\documentclass[a4paper,11pt]{article}
\pdfoutput=1 

\usepackage{jheppub} 
            
\usepackage{bm,amssymb,slashed,graphicx,multirow,soul,mathtools,xspace,array}  
\usepackage{float}                   
\allowdisplaybreaks
\usepackage{ bbold }
\usepackage{subfigure}

\newcommand{\todo}[1]{{\color{red} \ifmmode\else[todo]\fi #1}}
\usepackage[usenames,dvipsnames]{xcolor}
     \definecolor{hgreen}{rgb}{0,.3,0}
      \definecolor{darkgreen}{rgb}{0.3,.8,0.2}
     \definecolor{hred}{rgb}{.3,0,0}
     \definecolor{hblue}{rgb}{0,0,.3}
     \definecolor{LightGray}{gray}{0.95}
     
\usepackage{hyperref}
\usepackage[normalem]{ulem}

\newcommand{\ord}[1]{\mathcal{O}(#1)}     
\newcommand{\eps}{\epsilon}
\newcommand{\eV}{{\, \rm eV}}
\newcommand{\GeV}{{\, \rm GeV}}
\newcommand{\MeV}{{\, \rm MeV}}
\newcommand{\TeV}{{\, \rm TeV}}

\newcommand{\mueV}{{\, \mu {\rm eV}}}

\def\beq{\begin{equation}}
\def\eeq{\end{equation}}

\DeclareMathOperator{\Tr}{Tr}

\preprint{P3H-20-024, TTP20-025}

\title{Looking forward to Lepton-flavor-violating ALPs}

\author[a]{Lorenzo Calibbi,}
\author[b,c]{Diego Redigolo,}
\author[d]{Robert Ziegler,}
\author[e]{Jure Zupan,}

\affiliation[a]{School of Physics, Nankai University, Tianjin 300071, China}
\affiliation[b]{INFN Sezione di Firenze, Via G. Sansone 1, I-50019 Sesto Fiorentino, Italy and
Department of Physics and Astronomy, University of Florence, Italy}
\affiliation[c]{CERN, Theory Division, CH-1211 Geneva 23, Switzerland}
\affiliation[d]{Institute for Theoretical Particle Physics, Karlsruhe Institute of Technology, \\ Engesserstrasse 7, D-76128 Karlsruhe, Germany
}
\affiliation[e]{Department of Physics, University of Cincinnati, Cincinnati, Ohio 45221, USA}

\emailAdd{calibbi@nankai.edu.cn}
\emailAdd{d.redigolo@gmail.com}
\emailAdd{robert.ziegler@kit.edu}
\emailAdd{zupanje@ucmail.uc.edu}

\date{\today}

\abstract{We assess the status of past and future experiments on lepton flavor violating (LFV) muon and tau decays into a light, invisible, axion-like particle (ALP), $a$. We propose a new experimental setup for MEG II, the \emph{MEGII-fwd}, with a forward calorimeter placed downstream from the muon stopping target. Searching for $\mu \to e a$ decays MEGII-fwd is maximally sensitive to LFV ALPs, if these have nonzero couplings to right-handed leptons. The experimental set-up suppresses the (left-handed) Standard Model background in the forward direction by controlling the polarization purity of the muon beam. The reach of MEGII-fwd is compared with the present constraints, the reach of Mu3e and the Belle-II reach from $\tau \to \ell a$ decays. 
We show that a dedicated experimental campaign for LFV muon decays into ALPs at MEG II and Mu3e will be able to probe the ALP parameter space in an unexplored region well beyond the existing astrophysical constraints.  We study the implications of these searches for representative LFV ALP models, where the presence of a light ALP is motivated by neutrino masses, the strong CP problem and/or the SM flavor puzzle. To this extent we discuss the majoron in low-scale seesaw setups and introduce the LFV QCD axion, the LFV axiflavon and the leptonic familon, paying particular attention to the cases where the LFV ALPs constitute cold dark matter.}

\begin{document}

 \maketitle

\section{Introduction}

Probes of the Standard Model (SM) based on rare processes with charged leptons are set to improve substantially in the next decade. 
The muon beam experiments MEG II \cite{Baldini:2018nnn}, Mu3e \cite{Berger:2014vba,Blondel:2013ia}, COMET~\cite{Adamov:2018vin} and Mu2e~\cite{Bartoszek:2014mya} will collect unprecedented datasets using $\mathcal{O}(10^{15}-10^{17})$ muons each. Similarly,  Belle-II is expected to collect roughly $5\times10^{10}$ $\tau^+\tau^-$ pairs~\cite{Perez:2019cdy}, exceeding  by more than an order of magnitude  the datasets at Belle and BaBar. 
The standard New Physics (NP) targets for these experiments are rare lepton flavor violating (LFV) transitions\footnote{These are $\mu\to e\gamma$, $\mu\to e e e$ and $\mu\to e$ conversion for the muon and $\tau\to\ell\gamma$, $\tau\to \ell\ell\ell$, $\tau\to \ell\rho$, $\tau\to \ell\pi$ for the tau, where $\ell=e,\mu$. A more complete list of the LFV transitions and a theory summary can be found in Ref. \cite{Calibbi:2017uvl}.} induced by dimension-6 NP operators with SM particles on the external legs. The NP operators are suppressed by the heavy NP scale $\Lambda$ so that the corresponding LFV branching ratios scale as ${\rm BR}\propto 1/\Lambda^4$. Assuming $\mathcal{O}(1)$ Wilson coefficients for the dimension-6 operators the reach on the scale $\Lambda$ is expected to exceed $10^8$ GeV during the ongoing experimental campaign \cite{Calibbi:2017uvl}.  

\begin{figure}[t]
	\centering
	\includegraphics[width=0.95\linewidth]{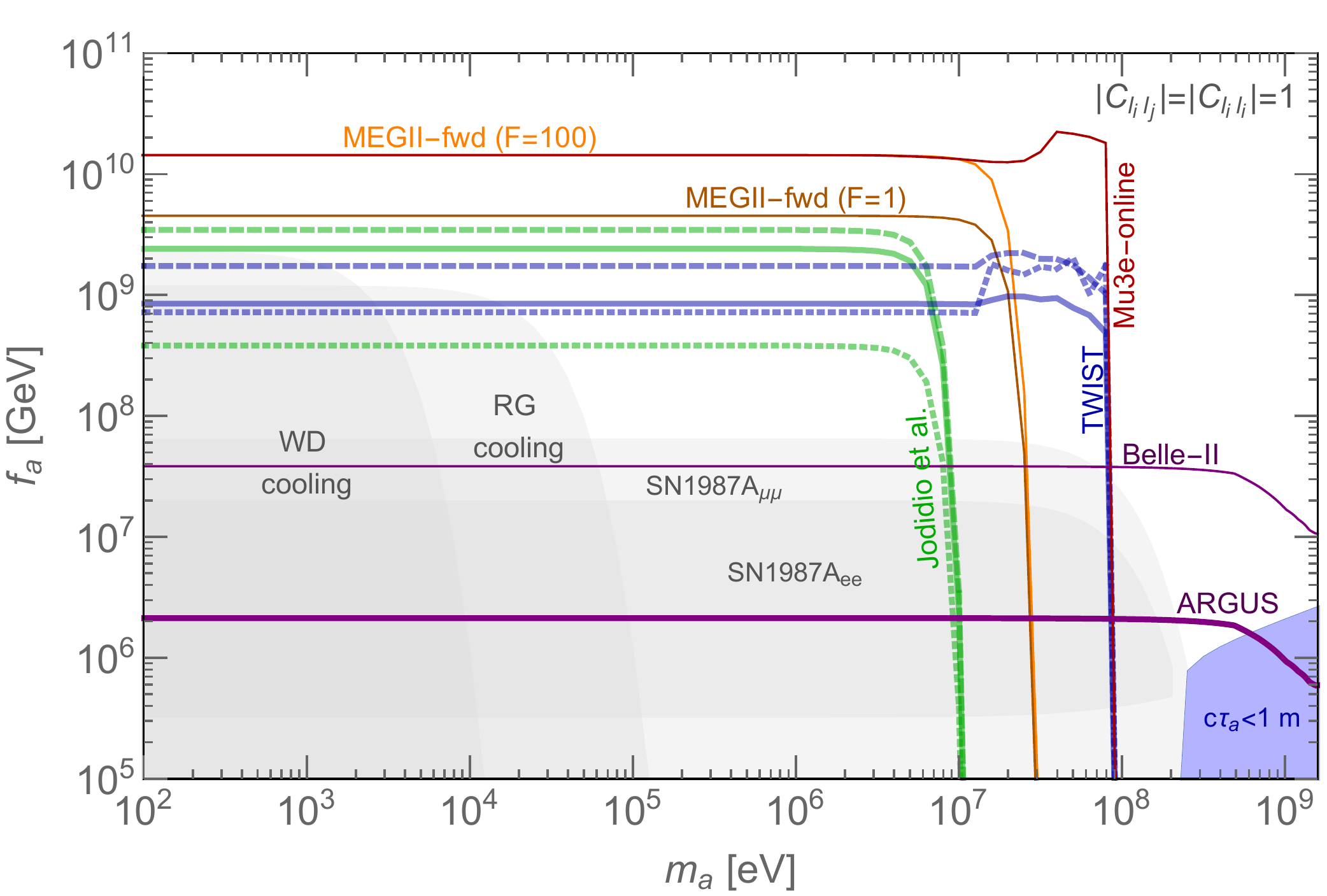}
	\caption{Summary of the present bounds and future projections for an ALP with generic couplings to leptons, i.e., we set $C_{\ell \ell'}^V = C_{\ell \ell'}^A =1$ for all the couplings in Eq.~\eqref{couplings}. For the isotropic case we set $C^V_{\mu e}=0$ and $C^A_{\mu e}=1$ (the opposite choice leads to the same results). In the $\text{V}\pm\text{A}$ case we set $C^V_{\mu e}=\pm C^A_{\mu e}=1$. The {\bf gray shaded} regions are excluded by the astrophysical bounds from star cooling due to $C_{ee}^A$ and  by SN1987A due to $C_{ee}^A$ and $C_{\mu\mu}^A$, see Sec.~\ref{sec:astro}. We present these bounds for the isotropic case. The {\bf blue shaded} region corresponds to a prompt/displaced ALP. The {\bf green solid} line is the exclusion due to the bound on $\mu^+\to e^+ a$ by Jodidio et al., assuming an isotropic ALP~\cite{Jodidio:1986mz}. The {\bf green dotted (dashed)} line is our recast of this bound for the $V-A$ ($V+A$) case. The sensitivity in the $V-A$ case is worse since then the signal is suppressed in the forward direction as much as the background. The {\bf blue solid (dotted, dashed)} lines are the bounds from the TWIST experiment on  isotropic ($V-A$, $V+A$) ALP~\cite{Bayes:2014lxz}. The {\bf dark orange thin solid} line is the MEGII-fwd projection for an isotropic ALP with no magnetic focusing while for the {\bf  orange thin solid} line we assumed that focusing increases the luminosity in the forward direction by a factor of 100, cf.~Sec.~\ref{sec:futuremutoe} for details. The {\bf dark red thin solid} line is the Mu3e projection  from~\cite{Perrevoort:2018okj}, for the isotropic ALP. The sensitivity for the other chiral structures is expected to be similar since there is no background suppression in this setup. The {\bf purple solid} line is the bound from the {$\tau\to  \mu a$} search by the ARGUS collaboration~\cite{Albrecht:1995ht}, and does not dependent on the chirality of the ALP couplings. The {\bf purple thin} line is the projected reach at Belle-II, see Sec.~\ref{sec:taudecays} for details. The bound on $\mu^+\to e^+ a \gamma$ from Crystal Box is subdominant, see  Sec.~\ref{sec:mutoegamma}, and is not displayed.}
\label{fig:money}	
\end{figure}

This can be contrasted with LFV decays into a light axion-like particle (ALP), $a$, in which case the LFV experiments probe much higher NP scales. The $\mu\to e a$, $ \tau\to \mu a$ or $ \tau\to e a$ decays are induced by dimension-5 operators so that the LFV branching ratios scale as ${\rm BR}\propto1/f_a^2$, where $f_a$ is the ALP decay constant. The projected bounds on LFV decays then translate to a reach on $f_a$ that, as we will show below, could exceed $10^{10}$~GeV, assuming $\mathcal{O}(1)$ flavor violating couplings. These scales are among the highest we could probe with ground-based experiments and are well above the present astrophysical constraints induced by the coupling of the light ALP to electrons. This conclusion is the main result of our paper and is presented in Fig.~\ref{fig:money}.

More broadly, in this paper we summarize the status of LFV ALP searches. The experimental difficulty is that the $\ell\to \ell' a$ decays look very similar to the SM decays, resulting in a single visible object plus missing energy (since the ALP is long-lived on detector scales in a large region of the allowed parameter space). As a consequence, the  $\ell\to \ell' a$ decays are not covered by the standard LFV searches and require dedicated experimental strategies/setups. We show that the experimental strategies to improve the coverage of the LFV ALP parameter space depends crucially i) on  the ALP mass, and  ii) on the chiral structure of the ALP couplings to the SM. For this reason we explore the full range of ALP masses from an effectively massless ALP with $m_a\ll m_e$ to a massive one up to $m_a\lesssim m_\tau$, as well as all the possible chiral structures of ALP couplings to the SM leptons. 

At present, the best bounds on $f_a$ from the $\mu\to ea$ decays give $f_a \gtrsim 10^9$ GeV \cite{Feng:1997tn}. The reach can be 
substantially improved with the next generation of experiments, mainly due to the increased integrated luminosities. 
For instance, a combination of the experiment by Jodidio et al. from 1986~\cite{Jodidio:1986mz} and the TWIST experiment from 2015~\cite{Bayes:2014lxz} gives the best present bounds on ${\rm BR}(\mu\to e a)$ based on merely $10^7-10^8$ stopped muons. This luminosity is at least seven orders of magnitude less than those expected at MEG II and Mu3e.

Taking full advantage of the available datasets will require adjusted experimental approaches. In Section \ref{sec:futuremutoe} we put forward a new experimental strategy to improve the bound on ${\rm BR}(\mu^+\to e^+ a)$ using MEG~II. The main idea is to mimic the 1986 experiment by Jodidio et al., utilizing that $\mu^+$ is polarized antiparallel to the beam direction, up to depolarization effects. We study the feasibility of a forward detector configuration in MEG II which we call \emph{MEGII-fwd} (``fwd'' for forward), where a calorimeter is placed in the forward direction relative to the muon beam. In this configuration the $\mu^+\to e^+a$ decay can be detected by searching for a positron of maximal energy emitted in the direction opposite  to the polarization of $\mu^+$.  For a highly polarized muon beam the SM background from $\mu^+\to e^+\nu \bar{\nu}$ is strongly suppressed in this part of the phase space, while the $\mu^+\to e^+a$ decay is allowed 
for an LFV ALP with nonzero right-handed couplings to the SM leptons.
We estimate the reach of this setup for two weeks of dedicated run at MEG II and for different configurations of the magnetic field, which will be crucial to control the polarization of the muon beam and the positron yield in the forward direction. We also compare the sensitivity of our proposal to the one that could be obtained at Mu3e by performing an online analysis of the positron spectrum obtained from triggerless data acquisition~\cite{Perrevoort:2018ttp}. Our conclusion is that the two experiments are complementary and, given their timelines, MEGII-fwd has a chance to explore new ALP parameter space several years before Mu3e.

Accessing this new portion of ALP parameter space could unveil connections between ALP Dark Matter (DM) and lepton flavor violation. Indeed, for ALP masses below 1-10 eV, the ALP is a viable DM  candidate, if it is produced non-thermally through the misalignment mechanism. The reach on LFV coupling from MEGII-fwd and Mu3e is then complemented by on-going and future experiments sensitive to the couplings of ALP DM to photons and electrons.

While the bulk of our analysis is based on a model-independent  parametrization of ALP couplings at low energies, we do explore the implications of the projected experimental sensitivities for a number of different NP scenarios, where the presence of a light ALP can be motivated by the strong CP problem, Dark Matter (DM), the SM flavor puzzle or neutrino masses.

First, we show how LFV decays arise naturally in QCD axion models of the DFSZ-type \cite{Zhitnitsky:1980tq,Dine:1981rt}. In these models the axion solves the strong CP problem which then connects the axion mass, $m_a$, and decay constant, $f_a$, giving $m_a\propto 1/f_a$. Furthermore, such a QCD axion can make up for the total amount of DM in the Universe~\cite{Preskill:1982cy,Abbott:1982af,Dine:1982ah,diCortona:2015ldu}. The future reach on $\mu^+\to e^+a$ decays could explore new axion parameter space in the mass range $30 \text{ meV}\gtrsim m_a\gtrsim0.4\text{ meV}$, where the upper bound is given by astrophyical constraints.  Interestingly, this high mass range requires either a non-standard cosmology~\cite{Hook:2019hdk} or special axion initial conditions~\cite{Arvanitaki:2019rax} in order for the axion to be the DM. The same mass range presents severe experimental challenges for axion DM detection through flavor diagonal couplings. While these could be possibly overcome by ongoing axion-mediated force experiments~\cite{Geraci:2017bmq}  or new experimental ideas \cite{Marsh:2018dlj,Mitridate:2020kly}, the QCD axion could also well be first observed through its flavor violating couplings. For related studies where flavor violating couplings are (only or also) in the quark sector, see \cite{MartinCamalich:2020dfe,Calibbi:2016hwq,Albrecht:2019zul,Bonnefoy:2019lsn,Gavela:2019wzg,Bjorkeroth:2018ipq,Alanne:2018fns,Ema:2018abj,Arias-Aragon:2017eww,Choi:2017gpf,Ema:2016ops}.

Second, we discuss the reach of LFV decays in the parameter space of the familon. This is the pseudo-Goldstone boson associated with the spontaneous breaking of the lepton flavor symmetry which could explain the charged lepton hierarchies via the Froggatt-Nielsen mechanism~\cite{Froggatt:1978nt}. In this setup we show how the strength of LFV decays is correlated with the texture of neutrino masses. A similar construction could also simultaneously address the strong CP problem and the flavor puzzle in the quark sector as proposed in Refs.~\cite{Calibbi:2016hwq,Ema:2016ops} (and was therefore dubbed as ``axiflavon''). In the axiflavon setup we show how the flavor violation in the quark sector could be naturally suppressed in a $U(2)$ flavor model similar to the one presented in Ref.~\cite{Linster:2018avp}. This would leave LFV decays as the main experimental signature to hunt for.  Similarly, our updated LFV sensitivities will probe the parameter space of  axion or relaxion models which try to address the flavor puzzle~\cite{Higaki:2019ojq,Carone:2019lfc,Davidi:2018sii,Bjorkeroth:2018dzu,Reig:2018ocz}.

As a final example of a well-motivated LFV ALP setup we discuss a class of majoron models where the breaking of lepton family number is decoupled from the spontaneous breaking of the lepton number~\cite{Ibarra:2011xn}. In this context the future reach of Mu3e and Belle~II will explore new parameter space beyond the present astrophysical bounds.

The paper is organized as follows. We start by setting up the notation in Sec.~\ref{sec:notation}, followed by the discussion of $\mu\to e a$ searches in Sec~\ref{sec:mutoe}, with section
Sec.~\ref{sec:pastmutoe} devoted to the review of past searches, while the future proposals, both our proposal for MEGII-fwd as well as the prospects at Mu3e, are presented in
Sec.~\ref{sec:futuremutoe}. Comments on $\mu\to e$ conversion are given in Sec. \ref{sec:mu2e}.
Sec.~\ref{sec:mutoegamma} contains a short summary of searches for $\mu\to e \gamma a$ decays, while in  Sec.~\ref{sec:taudecays} we summarize the bounds and prospects for tau decays. In Sec.~\ref{sec:astro}  we discuss the astrophysical constraints coming from star cooling and SN1987a, comparing these with the reach of LFV decays. In Sec.~\ref{sec:models} we discuss several models where LFV violation arises naturally: in Sec.~\ref{sec:DFSZQCDaxion} the LFV DFSZ axion, in Sec.~\ref{sec:axiflavon} the LFV axiflavon, in Sec.~\ref{sec:familon} the leptonic familon, and in Sec.~\ref{sec:majoron} the majoron. Finally, our conclusions are presented in Sec.~\ref{sec:conclusions}.

\begin{figure}[t]
	\centering
	\includegraphics[width=0.48\linewidth]{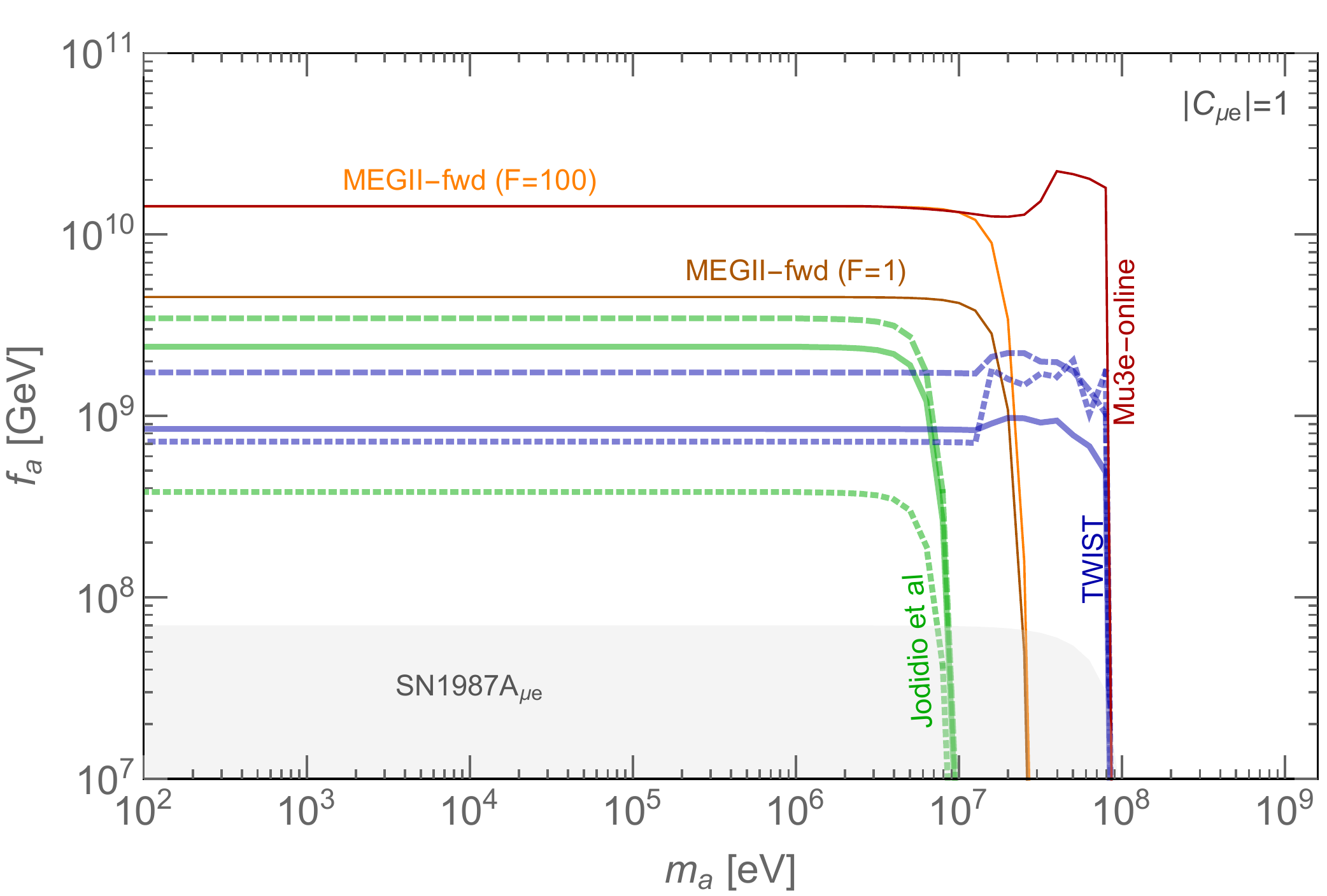}\quad
	\includegraphics[width=0.48\linewidth]{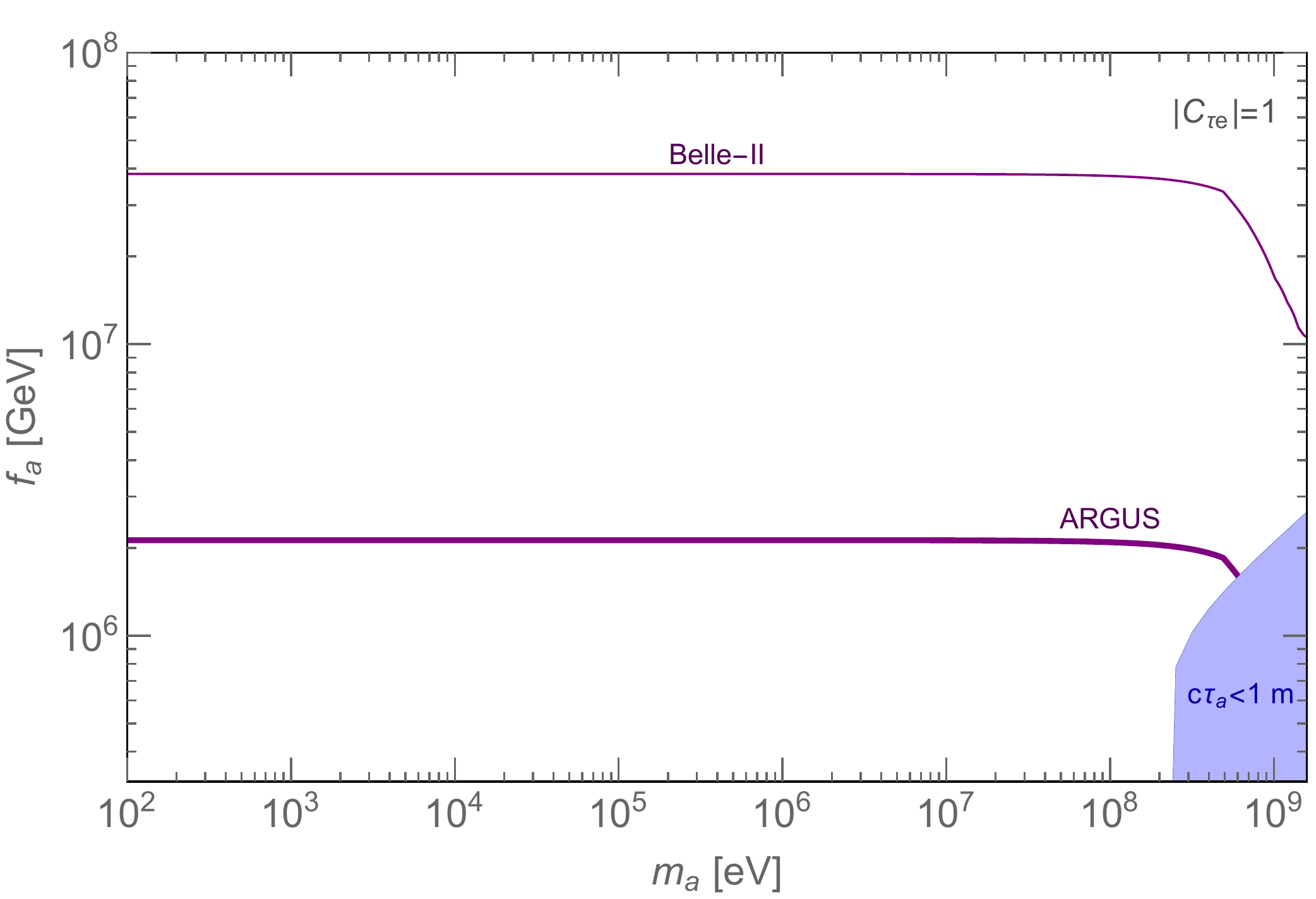}
	\caption{Summary of the bounds and future projections for LFV ALPs. The color coding is the same as in Fig~\ref{fig:money}. {\bf Left:} We set $C_{\mu e}=1$, while all the other couplings in Eq.~\eqref{couplings} are set to zero. For comparison we also show the bound on $\mu\to e a$ from SN1987A derived in Sec.~\ref{sec:astro} even though it is subdominant relative to the existing bounds from ground based experiments. {\bf Right:} The only nonzero coupling is $C_{\tau \mu}=1$. The plot for $C_{\tau e}=1$ is similar, and is not displayed for brevity.}
\label{fig:money2}
\end{figure}

\begin{table}[t]
\renewcommand{\arraystretch}{1.2}
\centering
\begin{tabular}{lcccc}
\hline\hline
 \multicolumn{5}{c}{Present best limits} \\
\hline
Process& BR Limit & Decay constant &  Bound (GeV) &  Experiment \\ \hline
Star cooling &  -- & $F_{ee}^A$ & $4.6 \times 10^9 $ &    WDs~\cite{Bertolami:2014wua}  \\
 &  -- & $F_{\mu\mu}^A$ & $1.3\times 10^8$ &    $\text{SN1987A}_{\mu\mu}$~\cite{Bollig:2020xdr,Croon:2020lrf}\\
 & $4 \times 10^{-3}$& $F_{\mu e}$ & $1.4\times 10^8$ &    $\text{SN1987A}_{\mu e}$ (Sec.~\ref{sec:astro})\\
\hline
 $\mu\to e\,a$ &  $2.6\times 10^{-6}$$^*$ & $F_{\mu e} ~(V\,\text{or}\,A)$ & $4.8 \times 10^9$ &  Jodidio at al.~\cite{Jodidio:1986mz}  \\
$\mu\to e\,a$  &  $2.5\times10^{-6}$$^*$ &  $F_{\mu e}~(V+A)$ & $4.9 \times 10^9$   & Jodidio et al.~\cite{Jodidio:1986mz}  \\
$\mu\to e\,a$  &  $5.8\times 10^{-5}$$^*$ & $F_{\mu e}~(V-A)$ & $1.0 \times 10^9 $ &   TWIST~\cite{Bayes:2014lxz} \\
 $\mu\to e\,a\,\gamma$ & $1.1\times 10^{-9}$$^*$    & $F_{\mu e}$ & $5.1 \times 10^{8 \#}$ &   Crystal Box~\cite{Bolton:1988af}  \\
\hline
$\tau\to e\,a$  &  $2.7\times 10^{-3}$$^{**}$ & $F_{\tau e}$ & $4.3 \times 10^6 $ &    ARGUS~\cite{Albrecht:1995ht}  \\
$\tau\to \mu\,a$  &  $4.5\times 10^{-3}$$^{**}$   & $F_{\tau \mu}$ & $3.3 \times 10^6$ &     ARGUS~\cite{Albrecht:1995ht}  \\
\hline
\hline
 \multicolumn{5}{c}{Expected future sensitivities} \\
 \hline
Process& BR Sens. & Decay constant &  Sens. (GeV) &  Experiment \\ 
\hline
$\mu\to e\,a$ &  $7.2\times 10^{-7}$$^{*}$ & $F_{\mu e} ~(V\,\text{or}\,A)$ & $9.2\times 10^9$ &  MEGII-fwd$^{\star}$ \\
$\mu\to e\,a$ &  $7.2\times 10^{-8}$$^{*}$ & $F_{\mu e} ~(V\,\text{or}\,A)$ & $2.9 \times 10^{10}$ &  MEGII-fwd$^{\star\star}$ \\
$\mu\to e\,a$  &  $7.3\times10^{-8}$$^{*}$ &  $F_{\mu e}~(V\,\text{or}\,A)$ & $2.9\times10^{10}$   & Mu3e~\cite{Perrevoort:2018okj}  \\
\hline
$\tau\to e\,a$  &  $8.3\times 10^{-6}$$^{**}$ & $F_{\tau e}$ & $7.7\times 10^7$ &    Belle II  \\
$\tau\to \mu\,a$  &  $2.0  \times10^{-5}$$^{**}$   & $F_{\tau \mu}$ & $4.9\times 10^7$ &     Belle II \\
\hline\hline
\end{tabular}
\caption{\label{tab:bounds} The present model independent $95\%$ C.L. best bounds  on leptonic ALP couplings $F_{\ell\ell'}^{V,A}$, Eq.~\eqref{eq:FVA},  are given in the upper part of the Table, with future projections listed in the lower part. The bounds assume $m_a$ below the mass resolution of the experiments considered here (see Fig.~\ref{fig:money} for modifications when $m_a$ is sizable). These follow from 
$90\%$ C.L. ($^*$) and $95\%$ C.L. ($^{**}$) bounds on  branching ratios in the 2nd column, rescaled using $Z_{95}/Z_{90} = 1.3$ when necessary.
The MEGII-fwd projections are obtained for two different sets of assumptions: MEGII-fwd$^\star$ assumes $\delta x_e=10^{-2}$ and $\langle P_\mu\rangle-1=10^{-2}$ with no focusing, while MEGII-fwd$^{\star\star}$ in contrast sets the focusing to $F=100$, roughly what was achieved in the 1986 experiment by Jodidio et al~\cite{Jodidio:1986mz}, cf.~Sec.~\ref{sec:futuremutoe} for details. The Belle II projection for $\tau\to\mu a$ is rescaled from the Belle MC simulation in Ref.~\cite{Yoshinobu:2017jti}, while the one for $\tau\to e a$ is rescaled directly from the ARGUS result~\cite{Albrecht:1995ht}. ($^{\#}$)~The Crystal Box bound on $F_{\mu e}$ can vary between $(5.1-8.3)\times 10^{8}\text{ GeV}$ depending on the assumed positron energy loss, cf. Eq.~\eqref{signalgamma}. }
\end{table}

\section{Notation and Summary}\label{sec:notation}
The ALP is a (pseudo-) Nambu-Goldstone boson (PNGB) and couples derivatively to the SM fermions. The interaction Lagrangian is thus given by 
\begin{align}
\label{couplings}
& \mathcal{L}_{\rm eff} = 
\sum_{i}\frac{\partial_\mu a}{2 f_a} \bar \ell_i C^A_{\ell_i \ell_i}\gamma^\mu \gamma_5 \ell_i+\sum_{i\ne j}\frac{\partial_\mu a}{2 f_a} \, \bar \ell_i \gamma^\mu (C^V_{\ell_i \ell_j} + C^A_{\ell_i \ell_j} \gamma_5 ) \ell_j  \, , 
\end{align}
where $C^A_{\ell_i\ell_i}$ is a real diagonal matrix\footnote{The diagonal vector couplings were set to zero, $C_{\ell_i\ell_i}^V=0$, via fermion field redefinitions that are anomalous only under $SU(2)_L$, and thus affect only the ALP couplings to electroweak gauge bosons.  These couplings can be constrained by their loop-induced contributions to $K \to \pi a$~\cite{Izaguirre:2016dfi}, which result for an invisible axion in the very weak constraint $f_a/C^V_{ee} \ge 0.7 \TeV $. }  and $C^{V,A}_{\ell_i \ell_j}$ are hermitian matrices in flavor space, while the summation is over $i,j=1,2,3$, and we ignore possible axion couplings to quarks. For ALP mass we take $m_a < m_{\tau}$, where $m_a$ could be well below the electron mass. The decay constant $f_a$ is related to the spontaneous breaking scale of the symmetry the ALP is associated with. 
We do not assume any relations between the couplings in Eq.~\eqref{couplings}, and discuss the experimental bounds and prospects separately. For these 6+3 couplings we also introduce the short-hand notation
\begin{align}
\label{eq:FVA}
F^{V,A}_{\ell_ i \ell_j} & = \frac{2 f_a}{C^{V,A}_{\ell_i \ell_j}} \, , & F_{\ell_ i \ell_j} & = \frac{2 f_a}{\sqrt{|C^{V}_{\ell_i \ell_j}|^2 + |C^{A}_{\ell_i \ell_j}|^2}} \, .  
\end{align}
When kinematically allowed, the couplings in Eq.~\eqref{couplings} give rise to LFV decays with the (invisible) ALP in the final state.\footnote{We note in passing, that while we do not study the phenomenology of the LFV neutrino decays, $\nu_i \to \nu_j a$, the typical decay time for this process is shorter than the age of the Universe for the ALP decay constants under consideration. This has interesting phenomenological consequences on neutrino cosmology that will be testable in future large scale structure surveys~\cite{Chacko:2019nej,Chacko:2020hmh}.} The corresponding total decay width is given by 
\begin{align}
\Gamma (\ell_i \to \ell_j \,a) = \frac{1}{16 \pi} \frac{m_{\ell_i}^3}{F^2_{\ell_i \ell_j}} \left( 1- \frac{m_a^2}{m_{\ell_i}^2} \right)^2 \, ,
\end{align}
where for simplicity we neglected the mass of the final-state lepton.
The differential decay rate reads (in the same  $m_{\ell_j} = 0$ limit) 
\begin{align}
\label{eq:diff}
\frac{\text{d} \Gamma(\ell_i  \to \ell_j  \, a)}{\text{d}\cos\theta} = \frac{m_{\ell_i}^3}{32 \pi F_{\ell_i  \ell_j }^2} \left( 1 - \frac{m_a^2}{m_{\ell_i}^2} \right)^2 \left[1 +2 P_{\ell_i} \cos \theta \frac{\text{Re}(C_{\ell_i  \ell_j }^V C_{\ell_i  \ell_j }^{A*})}{  \vert C_{\ell_i  \ell_j }^V\vert^2 + \vert C_{\ell_i  \ell_j }^A\vert^2 } \right] \, , 
\end{align}
where $\theta$ is the angle between the polarization vector, $\hat \eta$, of the decaying lepton $\ell_i$ and the momentum of the final state lepton $\ell_j$, while $P_{\ell_i}$ is the polarization of the decaying leptons. The convention used for $P_{\ell_i}$ is such that for the phenomenologically most important case of $\mu^+\to e^+ a$ decays we have $P_{\ell_i}=\hat{\eta}\cdot\hat{z}$, 
where $\hat z$ is the beam axis. The $\mu^+$ are predominantly polarized antiparallel to the beam direction, thus $P_{\ell_i}<0$ and $\theta$ is the angle between $-\hat z$ and the  momentum of the positron, cf. Fig. \ref{fig:summarymue} (left).

The total width of the ALP can be computed as a function of its mass by summing the different partial decay widths
\begin{equation}
\Gamma_{\text{tot}}(m_a)=\Gamma(a\to\gamma\gamma)+\sum_{i,j=1,2}\Gamma(a\to\ell_i \ell_j)+\sum_{i,j=1,2,3}\Gamma(a\to\nu_i \nu_j)\ .
\end{equation}
Since we restricted the ALP mass to $m_a< m_\tau$, only the decays to photons, neutrinos, electrons, and possibly muons are kinematically open. The corresponding partial decay widths are 
\begin{align}
&\Gamma(a\to\ell_i \ell_j)=\frac{m_a }{8\pi }\biggr[\biggr(\frac{m_{\ell_i}-m_{\ell_j}}{F_{ij}^V}\biggr)^2 z_+ +\biggr(\frac{m_{\ell_i}+m_{\ell_j}}{F_{ij}^A}\biggr)^2z_-\biggr]\,\sqrt{z_+z_-} \, ,
\\
&\Gamma(a\to\gamma\gamma)=\frac{\alpha_{\text{em}}^2E_{\text{eff}}^2}{64\pi^3}\frac{m_a^3}{f_a^2} \, ,
\end{align}
where $z_\pm=1-(m_{\ell_i}\pm m_{\ell_j})^2/m_a^2$, so that for $m_a \gg m_{\ell_i, \ell_j}$ we have $z_\pm\to1$. In the limit $m_i = m_j$ the result reduces to 
\begin{align}
\Gamma(a\to\ell_i \ell_i)=\frac{m_a }{2\pi } \biggr(\frac{m_{\ell_i}}{F_{ij}^A}\biggr)^2  \,\sqrt{1 - \frac{4 m_{\ell_i}^2}{m_a^2}} \, .
\end{align}
The ALP decays to neutrinos are often suppressed, so that in the bulk of the paper we set $\Gamma(a\to\nu_i \nu_j)=0$ (the majoron is an important exception, see Sec.~\ref{sec:majoron}).
The coupling to photons, $E_{\rm eff},$ depends on the UV physics as well as on the IR derivative couplings of ALP to 
the SM leptons running in the loop, 
\begin{equation}
\label{eq:EUV}
E_{\text{eff}}=E_{\text{UV}}+ \sum_{f}C_f^A B(\tau_f), \qquad B(\tau)=\tau\,\text{arctan}^2\frac{1}{\sqrt{\tau-1}}-1.
\end{equation}
Here, $\tau_f=4 m_f^2/m_a^2-i\epsilon$, and the summation is over the SM leptons, $f=e,\mu,\tau$. Note that the loop function in \eqref{eq:EUV} tends to $B(\infty)=0$ for heavy fermions, and thus the contributions due to the derivative ALP couplings  decouple in the heavy fermion limit.  
The anomaly contribution is encoded in the Wilson coefficient, $ E_{\rm UV}$,  and depends on the structure of the UV model. We use the the following normalization for the effective  ALP-photon Lagrangian\,\footnote{Our conventions are $\eps_{0123} = 1$ and $\tilde{F}_{\mu \nu } = 1/2 \, \eps_{\mu \nu \rho \sigma} F^{\rho \sigma}$.} 
\begin{align}
{\cal L}_{\rm eff} = E_{\rm UV} \frac{\alpha_\text{em}}{4 \pi} \frac{a}{f_a} F \tilde{F} \, ,
\end{align}
such that for the QCD axion $E_{\rm UV}=E/2N$, where $E$ and $N$ are the electromagnetic and color anomaly coefficients of the Peccei-Quinn symmetry.
For example, in the DFSZ-II model for the QCD axion~\cite{Zhitnitsky:1980tq,Dine:1981rt}, in which the charged leptons couple to the same Higgs as the up-quarks, one has $E_{\rm UV} = 1/3$. In Sec. \ref{sec:models} below, we give four explicit examples of LFV ALP models. For these we have 
$E_{\rm UV} = \{2/3,10/9,(10\div24),0\}$ 
for the LFV QCD axion (Sec.~\ref{sec:DFSZQCDaxion}), the LFV axiflavon (Sec.~\ref{sec:axiflavon}),
 the anarchic LFV familon (Sec.~\ref{sec:familon}) and the majoron, (Sec.~\ref{sec:majoron}), respectively.  
 If the ALP also couples to quarks and gluons, there are additional contributions to the effective photon coupling in Eq.~\eqref{eq:EUV}, both from heavy quarks as well as from pions running in the loop
 (see Ref.~\cite{Bauer:2017ris} for complete expressions). From now on we fix $E_{\text{UV}}=1$, unless specified otherwise, since its precise value does not affect most of the physics discussed in this paper. 
 \begin{figure}[t]
	\centering
	\includegraphics[width=0.95\linewidth]{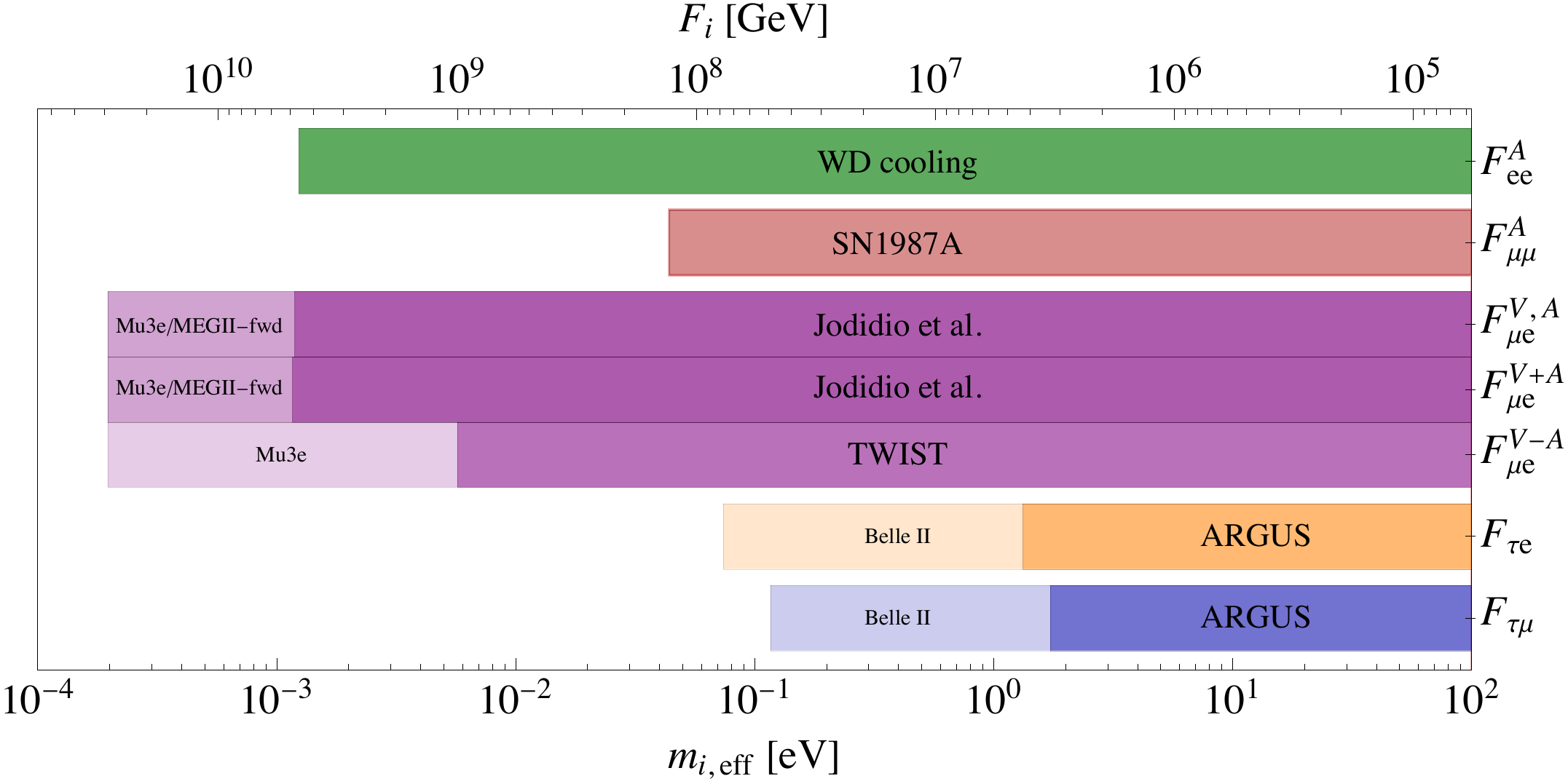}
	\caption{Current bounds and future projections for different couplings $F_i$ of an effectively massless ALP, also reported in Table~\ref{tab:bounds}. On the lower axis we indicate the corresponding values for the effective axion mass defined by $m_{i, {\rm eff}} = 4.7 \, \eV \times 10^6 \GeV/F_i$.}
	\label{fig:bounds-chart}
\end{figure}

As shown in Fig.~\ref{fig:money}, we focus in this paper on the region of parameter space where the ALP is long-lived on detector scales. As we will motivate extensively in Sec.~\ref{sec:models}, we believe that this region is the most appealing from a theoretical perspective. For the discussion of phenomenologically interesting decay channels in the displaced and prompt regions, we refer the reader to Refs.~\cite{Heeck:2017xmg,Bauer:2019gfk,Cornella:2019uxs} and to the recent MEG limit on LFV light particles decaying to two photons, $\mu\to e X,~X\to \gamma\gamma$ \cite{Baldini:2020okg}. ALP masses  in the range $2 m_e \le m_a \le m_\mu - m_e$ are constrained by the search for the $\mu \to 3 e$ decays at the SINDRUM experiment~\cite{SINDRUM}. Indeed, a tiny fraction of ALPs from prompt $\mu \to e a$ decays leads to $a\to e^+e^-$ decays 
within the detector, even for large ALP decay lengths. Assuming 1~cm for the effective decay length measurable in the instrumented volume, the SINDRUM upper bound of ${\rm BR} (\mu \to 3e) \le 10^{-12}$ is expected to constrain $f_a$ to be above $10^7 \GeV$ for  $C_{\mu e} = C_{ee} = 1$, which is much weaker than the other bounds discussed here. The bound on $f_a$ could be substantially improved  by the Mu3e experiment~\cite{refId0} given the larger luminosity. However, both the precise SINDRUM exclusion and the Mu3e projection would require to take into account the effect of further experimental cuts on the vertex quality and it is beyond the scope of this paper.  A complementary probe of LFV couplings in the region of heavier ALP masses is the muonium-antimuonium oscillation, which would be induced by $s$-- and $t$--channel exchanges of ALPs with couplings to $\mu e$. While the resulting bounds on $f_a$ are several order of magnitude below the ranges shown in Figs.~\ref{fig:money} and \ref{fig:money2},\footnote{For heavier ALPs, $m_a\gtrsim m_\mu$, we can integrate out the ALP to generate the muonium-antimuonium oscillation EFT operators. Translating the results of Ref.~\cite{Conlin:2020veq} to our notation gives
\begin{align}
\frac{1}{1.9{\rm~TeV}}&>\left|\frac{1}{F_{\mu e}^A}\pm \frac{1}{F_{\mu e}^V}\right|\left(\frac{m_\mu}{m_a}\right),\qquad 
&\frac{1}{3.8{\rm~TeV}}>\left|\frac{1}{(F_{\mu e}^A)^2}- \frac{1}{(F_{\mu e}^V)^2}\right|^{1/2}\left(\frac{m_\mu}{m_a}\right).
\end{align}
The constraints for light ALP, $m_a\lesssim m_\mu$, are obtained by taking $m_\mu/m_a\to 1$ in the two expressions above (see also similar results for heavy meson mixings in the limit of light ALP in Ref.~\cite{MartinCamalich:2020dfe}).  In the future these bounds could be improved for $m_a$ few GeV at Belle II by searching for $e^+ e^-\to ee\mu\mu$ events \cite{Endo:2020mev}.
}
 they are stringent enough to effectively rule out the LFV ALP explanations of possible deviations in $(g-2)_e$ and $(g-2)_\mu$~\cite{Bauer:2019gfk,Endo:2020mev,Cornella:2019uxs}.

In the numerical analyses throughout the paper all the axion couplings are assumed to be real  to simplify the discussion. 
The interpretations of the present LFV experimental results and future projections in terms of bounds on $F_{\ell_i \ell_j}$ are summarized in Fig.~\ref{fig:money}, assuming all the lepton couplings in Eq.~\eqref{couplings} to be $\mathcal{O}(1)$. Fig.~\ref{fig:money2} shows instead the same constraints for the case when only a single LFV coupling is taken to be nonzero. In Figs.~\ref{fig:money} and \ref{fig:money2} we also show the typical reach of astrophysical bounds on the ALP decay constant coming from star cooling and SN1987A observations (see Sec.~\ref{sec:astro} for details). In Table~\ref{tab:bounds} and Fig.~\ref{fig:bounds-chart} we summarize the current best bounds and future projections for an effectively massless $m_a$, i.e.~lighter than the typical mass resolution of the experiments considered here. This is the ALP mass range that applies to most of the concrete models discussed in Sec.~\ref{sec:models}. 
In the subsequent sections we discuss in detail the observables and the experiments
from which these constraints were derived.

\section{ALPs in $\mu^+ \to e^+ + {\rm invis.}$ decays}
\label{sec:mutoe}

We first summarize the status and prospects to search for the two body $\mu^+ \to e^+a$ rare decays, where  $a$ is invisible, i.e.,~it decays outside the detector. The challenge of this measurement is to distinguish $\mu^+\to e^+a$ from the background distribution of the SM $\mu^+\to e^+\,\nu\,\bar{\nu}$ decay. 

\begin{figure}[t]
	\centering
	\includegraphics[width=0.4\linewidth]{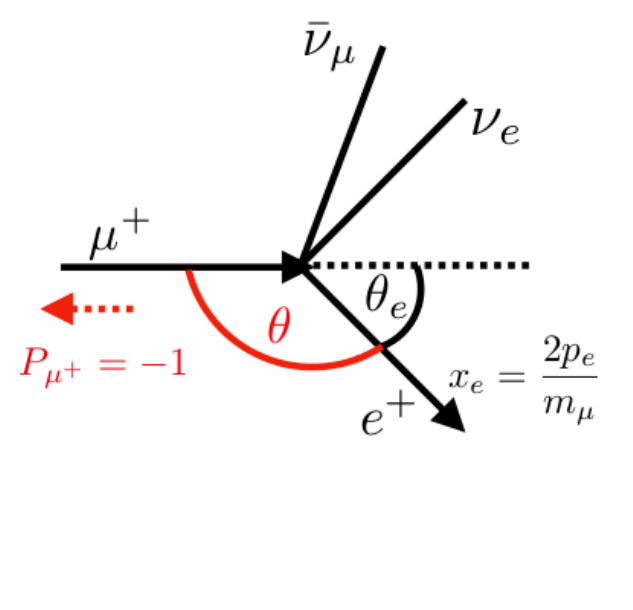}
	\includegraphics[width=0.58\linewidth]{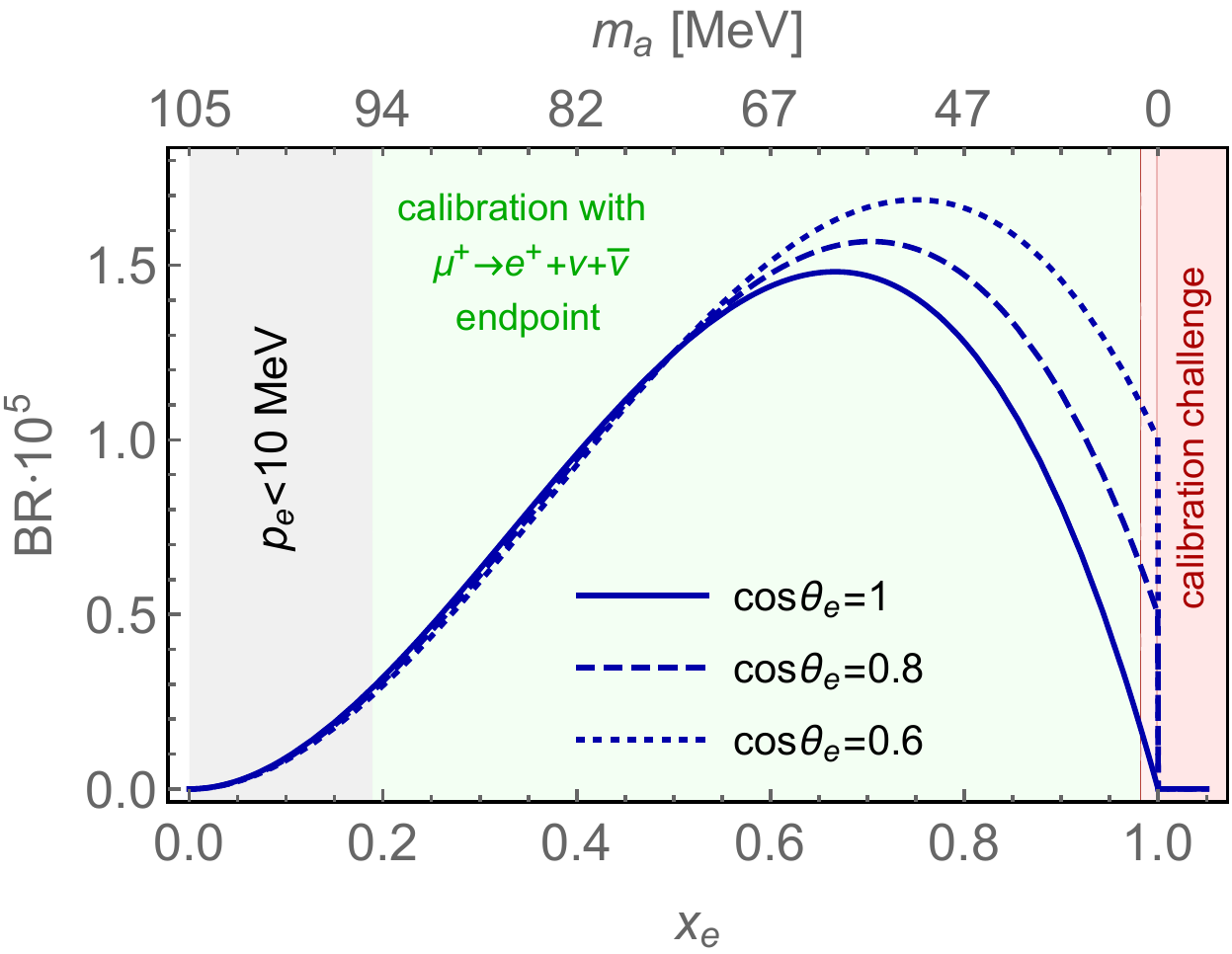}
	\caption{{\bf Left:} Cartoon summarizing the relevant kinematical variables in the SM decay of a polarized $\mu^+$. {\bf Right:} Summary of the experimental strategies to hunt for $\mu^+\to e^+a$. The three blue lines show the distribution of the positron spectrum in the SM decay $\mu^+\to e^+\nu_e\bar{\nu}_\mu$ for fixed angle $\cos\bar{\theta}_e=1,0.8,0.6$ up to a typical angular resolution of $\delta\theta_e=\pm5\times10^{-3}$. The muon beam is assumed to be 100\% polarized $\langle P_\mu\rangle=-1$ and as a consequence the positron distribution drops after having a maximum at  $x_\text{max}$ (see Eq.~\eqref{eq:xmax}). The value of the branching ratio at the end point depends on the angular position and resolution (see Eq.~\eqref{eq:endpoint}). In the gray region the positron momentum is below 10 MeV and typically not recorded, in the green region $\mu^+\to e^+a$ can be searched as a spike on the Michel background. In the red region looking for the signal spike of $\mu^+\to e^+a$ requires to overcome challenges in the calibration of the instrument. The two different red lines correspond to two different momentum resolutions as in Eq.~\eqref{eq:massless}: 0.13\% in Jodidio et al \cite{Jodidio:1986mz} and 1.8\%  in TWIST \cite{Bayes:2014lxz} (see text for details).}
	\label{fig:summarymue}
\end{figure}

The $\mu^+ \to e^+a$ decay produces a monochromatic positron line in the muon rest frame at the positron momentum
\begin{equation}
\big|\vec p_{e}^{\rm\,\, line}\big|=\sqrt{ \left(\frac{m_\mu^2-m_a^2+m_e^2}{2 m_\mu}\right)^2-m_e^2 },
\label{eq:mean}
\end{equation}
or in terms of the positron energy, $E_e^{\rm line}\simeq m_\mu/2$ for $m_{a}\ll m_\mu$. 
The angular distribution of the positrons depends on the initial muon polarization and the chiral structure of the ALP interactions. We discuss three representative cases that lead to distinct angular distributions: 
\begin{itemize}
\item the \emph{isotropic ALP} has either $C^V_{\mu e}=0$ or $C^A_{\mu e}=0$. The angular distribution of the final state positrons 
is isotropic in the muon rest frame and independent of the muon polarization, cf. Eq.~\eqref{eq:diff}.
\item the \emph{left/right-handed ALP} couples only to the left/right-handed SM fermions, i.e.,  $C^V_{\mu e}=-C^A_{\mu e}$ for the left-handed and $C^V_{\mu e}=+C^V_{\mu e}$ for the right-handed ALP. The positron angular distribution is $\propto (1\mp P_\mu\cos\theta)$ for the left(right)-handed ALP.  
\end{itemize}
The SM $\mu^+\to e^+\,\nu_e\,\bar{\nu}_\mu$ three-body decay  proceeds through an off-shell $W^+$ and produces the so-called Michel spectrum  
\begin{equation}
\frac{\text{d}^2\Gamma(\mu^+\to e^+\,\nu_e\,\bar{\nu}_\mu)}{\text{d}x_e\, \text{d}\cos\theta}\simeq \Gamma_{\mu}\big[\left(3-2x_e\right)-P_\mu(2x_e-1)\cos\theta\big]x_e^2 \, , \label{eq:back}
\end{equation}
with $ \Gamma_{\mu}\simeq {m_\mu^5G_F^2}/{(192\pi^3)}=3\times 10^{-10}\text{ eV}$ the total muon decay width, and $\theta$ the angle between muon polarization vector and the positron momentum in the muon rest frame, see Fig. \ref{fig:summarymue} (left).
The positron energy fraction $x_e=2E_e/m_\mu$ takes the values $0\leq x_e\leq 1$ (when neglecting the positron mass).  In writing Eq.~\eqref{eq:back} the NP scale was taken to be well above the weak scale, as is the case for NP models we are interested in, so that the three-body muon decay is the SM one. For further details on the SM muon properties we refer the reader to the two excellent reviews, Refs.~\cite{Scheck:1977yg,Kuno:1999jp}. 

For an unpolarized muon beam, $\langle P_\mu\rangle=0$, the SM background in Eq.~\eqref{eq:back} peaks at $E_{e}^{\text{line}}=m_\mu/2$ which corresponds to $x_e=1$. That is, the peak of the SM background for unpolarized muons coincides with the $\mu^+\to e^+ a$ positron line  for a massless ALP.  Luckily, this is not the situation encountered at the intense muon facilities. The muon flux in low energy muon beamlines such as those at TRIUMF or PSI is dominated by the muons produced from pions decaying at rest, at the surface of the production target. These muons are $100\%$ polarized in the direction opposite to the muon momentum, i.e., $\langle P_\mu\rangle=-1$ in the notation of Eq.~\eqref{eq:back}.  As a consequence, muon facilities produce very intense $\mu^+$ fluxes of almost $100\%$ polarized muons. The final polarization at the stopping point varies between $80\%$ and $100\%$ depending on the  size of the depolarization effects at the production point, during the propagation and at the stopping target. 

Polarization of the muon can significantly reduce the SM background. This is illustrated in Fig.~\ref{fig:summarymue} (right), which shows the Michel spectra  as functions of $x_e$ for fixed $\langle P_\mu\rangle=-1$ and three representative values of $\cos\bar{\theta}_e=1,0.8,0.6$, where $\theta_e\equiv \pi-\theta$ is the angle between the positron momentum and the muon beamline, see Fig.~\ref{fig:summarymue} (left).\footnote{Technically, the Michel spectra were integrated over small angular bins, $\cos\bar{\theta}_e\pm\delta\!\cos\theta_e$,
where $\delta\theta_e=5\times10^{-3}$, which is the typical angular resolution of these experiments.} For $0<\cos\bar{\theta}_e\leq 1$ the non-zero polarization moves the position of the maximum of the Michel spectrum  to 
\begin{equation}
x_{\text{max}}=\frac{3-\langle P_\mu\rangle \cos\bar{\theta}_e}{3(1-\langle P_\mu \rangle \cos\bar{\theta}_e)}\overset{\langle P_\mu \rangle =-1}{=}\frac{3+\cos\bar{\theta}_e}{3(1+\cos\bar{\theta}_e)}\ ,\label{eq:xmax}
\end{equation}
i.e., away from the massless ALP positron line, 
 while for $\cos\bar{\theta}_e\leq 0$ it remains at $x_{\text{max}}=1$. 
The SM decay rate at the position of the massless ALP positron line, $x_e=1$, is

\begin{equation}
\frac{d}{dx_e}\text{BR}(\mu\to e\,\nu_e\,\bar{\nu}_\mu)\vert_{x_e=1,\theta_e=\bar \theta_e}=2\delta\!\cos\theta_e\,(1+\langle P_\mu \rangle\cos\bar{\theta}_e)
\overset{\langle P_\mu \rangle =-1}{=}2\delta\!\cos\theta_e\,(1-\cos\bar{\theta}_e)\ . \label{eq:endpoint}
\end{equation}  
The SM background for fully polarized muons is exactly zero for $\cos\bar{\theta}_e=1$, i.e., for positrons emitted in the forward direction relative to the muon beam, up to  terms quadratic in the angular resolution. 
However, in order to have appreciable signal rates, one cannot work in the  the exact forward limit, $\cos\bar{\theta}_e=1$, but need to accept events in some range around $\bar \theta_e=0$, which also makes the SM background nonzero. Furthermore, any suppression in the average muon polarization increases the SM background linearly with $\delta P_\mu= \langle P_\mu\rangle+1$.  For the ALP positron line the background suppression is linear in the momentum uncertainty, $\delta x_e$, at least within the naive assumption of Gaussian smearing, making this uncertainty dominant compared to the one on the angle, $\delta \theta_e$.

With judicious choice of cuts one can optimize the signal to background ratios.  The only two discriminants between the SM background and the ALP signal are the momentum of the final state positron, $p_e$, and the angle $\theta_e$. The reach of different $\mu^+\to e^+ a$ searches can be  understood schematically in the cut-and-count scheme, giving 
\begin{equation}
\text{BR}(\mu^+\to e^+ a)\lesssim Z\sqrt{\frac{a_{\text{bkd}}}{a_{\text{sig}}^2 N_{\mu^+}}+\epsilon_{\text{sys}}^2 } \ ,\label{eq:reach}
\end{equation} 
where $Z=1.28 (1.64)$ for $90\%\,(95\%)$ C.L. intervals, taking the limit of Gaussian statistics and using limits for one-sided intervals. Here, $N_{\mu^+}$ is the number of $\mu^+$ at a given experiment, $a_{\text{bkd}}$ ($a_{\text{sig}}$)  is the background acceptance (signal efficiency), while $\epsilon_{\text{sys}}$ encodes the systematic uncertainties. The $a_{\text{bkd}}$ depends linearly on the momentum resolution, assuming the background is roughly constant in a given momentum bin. Whenever $\epsilon_{\text{sys}}\gtrsim 1/\sqrt{N_{\mu^+}}$ the reach in the branching ratio saturates independently on the muon luminosity.

Fig.~\ref{fig:summarymue} (right) summarizes the experimentally accessible regions that can be used for  $\mu^+\to e^+a$ searches. Due to the momentum threshold for soft positrons the ALPs with masses above $m_a \gtrsim 95$ MeV are unaccessible. For massive ALPs lighter than 95 MeV a standard bump hunt over the Michel spectrum can be performed, after the instrument is calibrated with the Michel spectrum endpoint. For an almost massless ALP, i.e., for masses below the momentum resolution of the experiment 
 \begin{equation}
 m_a\lesssim \sqrt{\frac{\delta p_e}{p_e}}\cdot m_\mu\ ,\label{eq:massless}
 \end{equation}
the positron line gets close to the endpoint of the Michel spectrum. As a consequence, a shift in the endpoint of the SM spectrum due to the momentum resolution or other uncertainties on the experimental setup would result in a spike which would be impossible to distinguish from the signal. Reducing the large systematics on the endpoint of the Michel spectrum then requires alternative ways to calibrate the instrument in this mass region. In Fig.~\ref{fig:summarymue} we show the two mass regions where this is  required for the momentum resolutions of the two TRIUMF experiments we describe here.

In Sec.~\ref{sec:pastmutoe} we first review the two experimental searches which were performed at TRIUMF: i) the 1986 experiment by Jodidio et al.~\cite{Jodidio:1986mz} and ii) the 2015 TWIST experiment~\cite{Bayes:2014lxz}. The two searches had very different philosophies. The setup by Jodidio et al. tried to suppress the SM background as much as possible, so that $a_{\text{bkd}}\ll a_{\text{sig}}$ in Eq.~\eqref{eq:reach}. This was achieved by using a highly polarized muon beam and measuring in the forward region with a tight angular cut and excellent momentum resolution. Crucially, magnetic focusing of the positrons improves the signal acceptance above the naive geometric one, and the sensitivity in Eq.~\eqref{eq:reach} is maximized. The TWIST experiment relies instead on  larger integrated luminosity and uses wide angular acceptance. As we show below, combining the results  of the two experiments gives complete coverage of different ALP masses and chiral structures of couplings for leptons. 

In Sec.~\ref{sec:futuremutoe} we discuss the future prospects for $\mu^+\to e^+ a$ at PSI: at MEG II and Mu3e \cite{refId0}. MEG II is set to soon start its physics run, with the primary goal to improve the sensitivity to $\mu^+\to e^+\gamma$ decays. The primary goal of Mu3e is to improve on $\mu^+\to e^+e^- e^+$,  with data taking scheduled to start in the relatively near future. Both experiments use the $\pi E5$ beamline which provides roughly $10^8 \mu/\text{sec}$ with muon momenta of $28\text{ MeV}$, and can also be used for $\mu^+\to e^+ a$ searches. The two experimental proposals to hunt for $\mu^+\to e^+a$, MEGII-fwd that is part of this paper, and Mu3e-online, are very different in spirit and could be complementary given that Mu3e will presumably start the physics run after MEG II will have collected the planned muon luminosity.

\subsection{Past Searches at TRIUMF: Jodidio et al. and TWIST}\label{sec:pastmutoe}
\noindent{\bf The 1986 experiment by Jodidio et al.}~\cite{Jodidio:1986mz} used two datasets, the ``spin held'' sample of $1.8\times10^7\, \mu^+$, and ``spin precessed'' sample of $1.4\times10^7\, \mu^+$. The two datasets both passed the trigger requirements, but differed in the magnetic field configurations that were used in the experiment. The  spin held sample had much higher purity of  polarized muons and was used to perform the $\mu^+\to e^+a$ search. The spin precessed sample was used to calibrate the end point of the Michel spectrum in order to reduce the systematic uncertainty. The extremely high purity of polarized muons  in the spin held sample was achieved through the use of high purity metal foils as targets, which highly suppressed the muonium production, while the strong magnetic field of $1.1\text{ T}$ parallel to the muon beam line suppressed the muon spin precession. The measured averaged muon polarization at the stopping point was 
\begin{equation}
\langle P_\mu\rangle=-0.99863\pm0.00088\, ,
\end{equation} 
where we assumed the SM values for the muon decaying parameters and combined in quadrature the statistical and systematic uncertainties. 

The $\mu^+\to e^+a$ analysis searched for a positron line in the ``spin held'' data. The cuts 
\begin{align}
&\cos\theta_e>0.975\ ,
&x_e>0.97\ ,\label{eq:cutsTRUMF}
\end{align}
 selected the region of phase space in which the SM three-body decay is heavily suppressed for polarized muons, see Eq.~\eqref{eq:endpoint}. 
 The positrons emitted in the beam direction, i.e., at $\cos\theta_e\simeq 1$, were measured downstream in the spectrometer after they bent by more than $90^\circ$ by a magnetic field. 
 
First, we  check how well the simple cut-and-count scheme reproduces the bound derived by the experimental collaboration for the massless isotropic ALP, 
\begin{equation}\label{eq:inv}
C^V_{\mu e}=0\text{ or } C^A_{\mu e}=0:\qquad \text{BR}(\mu\to e\, a)< 2.6\times10^{-6}~~(90\%~{\rm CL})\text{~\cite{Jodidio:1986mz}}.
\end{equation}
This will prove useful when deducing the projected sensitivities in the next section.  
Applying the cuts ~\eqref{eq:cutsTRUMF} on the Michel spectrum, Eq.~\eqref{eq:back}, gives $a_{\text{geo}}=5.3\times10^{-5}$  for the geometric acceptance of the background.
 Moreover, for a massless ALP, the number of background events in a $\pm 2\sigma$ band around $x_e=1$ can be estimated by integrating the distribution provided in Ref. \cite{Jodidio:1986mz}, accounting for the quoted momentum resolution  
\begin{equation}
\label{eq:momres}
\frac{\delta p_e}{p_e}=0.13\% .
\end{equation}
This leads to a background efficiency of $\epsilon_{\text{bkd}}=1.7\times 10^{-2}$, i.e., the fraction of  background events satisfying  \eqref{eq:cutsTRUMF} that are in the signal region $x_e\in [1-2\delta x_e,1+2\delta x_e]$. The background acceptance in the signal region of the initial spin held sample is thus $a_{\text{bkd}}=a_{\text{geo}}\cdot\epsilon_{\text{bkd}}= 8.7\times 10^{-7}$. Applying instead the $2\sigma$ band cut around $x_e\simeq1$ directly on the Michel spectrum, Eq.~\eqref{eq:back}, gives a similar estimate for the background acceptance, $a_{\text{bkd}}\simeq 10^{-6}$. This shows that using the simple analytical Michel spectrum is good enough for the purposes of our estimates in the next section. 

 In order to obtain the correct limits we still need to take into account the effect of magnetic focusing. After magnetic focusing and the cuts in Eq.~\eqref{eq:cutsTRUMF} the total number of observed events  in the spin held sample is $\simeq 7.4\times10^4$ (obtained by integrating the distribution published in Ref.~\cite{Jodidio:1986mz}).  Assuming that these events are entirely due to the SM, and neglecting the focusing, would require 
 $N_{\mu^{+}}\simeq1.4\times 10^9$ in the initial spin held sample, while in reality $N_{\mu^+}=1.8\times10^7$. The magnetic focusing therefore leads to effectively larger geometric acceptance,\footnote{We thank Angela Papa and Giovanni Signorelli for illuminating discussions about this point and refer to~\cite{Reiser:1995jm,Kumar:2009jva} for a detailed derivation of the focusing power of a solenoid lens.} 
 \beq
 \label{eq:focusing}
 F\equiv \frac{a_{\text{geo}+\text{focus}}}{a_{\text{geo}}}
 =77.8.
 \eeq

 The efficiency for the ALP signal in this experimental setup depends very much on the helicity structure of the ALP couplings to the SM current and on the ALP mass. For a massless isotropic ALP, i.e., $m_a\ll m_e$ and $C^V=0$ or $C^A=0$,  we find $a^{\text{ISO}}_{\text{geo}}=1.25\times 10^{-2}$ after the angular cut in Eq.~\eqref{eq:cutsTRUMF}. Assuming that the focusing lens act similarly on background and signal we find  $a^{\text{ISO}}_{\text{geo}+\text{focus}}=0.97$. The experimental analysis also assigns $\epsilon_{\text{sys}}=0.9\times 10^{-6}$ systematic uncertainty on the branching ratio  for an isotropic ALP. 
 Given the numbers above, Eq.~\eqref{eq:reach} then gives our recast bound of  $2.8\times 10^{-6}$, which agrees quite well with the result of the fit in Ref.~\cite{Jodidio:1986mz}. The agreement makes us confident that we understand the main features of the experimental setup. The above exercise also highlights  that the systematic uncertainty in the experiment by Jodidio et al. was smaller than the statistical one, and thus the bound on ${\rm BR}(\mu\to e a)$ would have benefitted from a factor of 10 bigger muon luminosity before hitting the bottleneck of systematics.

Next, we recast the result by Jodidio et al.~for different chiral structures of ALP couplings and for higher ALP masses. We assume that there are no changes in systematic uncertainties and the solenoid focusing effect. The ratios of signal acceptances are then obtained  by simply applying the angular cut in Eq.~\eqref{eq:cutsTRUMF} to the angular distribution predicted by the modified  ALP coupling and mass, and compare it to the baseline massless isotropic ALP. 
We obtain $a^{\text{RH}}_{\text{geo}}/a^{\text{ISO}}_{\text{geo}}\simeq 2$ and $a^{\text{LH}}_{\text{geo}}/a^{\text{ISO}}_{\text{geo}}\simeq0.012$ for the signal acceptances of right-handed (RH) ALP and left-handed (LH) ALP relative to the isotropic case, respectively. 
Rescaling the bound in Eq.~\eqref{eq:inv} gives for the massless right-handed ALP 
\begin{equation}
\label{eq:V+Abound}
C^V_{\mu e}=C^A_{\mu e}:\qquad \text{BR}(\mu\to e\, a)< 2.5\times10^{-6}~~(90\%~{\rm CL})\, .
\end{equation}
The improvement in the bound is marginal compared to the isotropic case because the magnetic focusing effect is already giving a signal acceptance very close to one. If focusing were a tunable parameter, the setup by Jodidio et al.~would have benefitted from slight defocusing which would have decreased the SM background but kept the signal almost unchanged, giving the best bound on the $V+A$ ALP.  

For massive ALP the measurement by Jodidio et al.~still translates into a bound since the tail of the signal distribution leaks into the experimental signal region due to the finite experimental resolution on the positron momentum, Eq.~\eqref{eq:momres}. For $p_{e}^{\text{line}}\simeq m_\mu/2$ the momentum spread is up to 70 keV. We recast the Jodidio et al.~bound to $m_a\gtrsim m_e$ by taking the signal to be a Gaussian in $p_e$ with the mean given by Eq.~\eqref{eq:mean} and the width set by the momentum resolution in Eq.~\eqref{eq:momres}. The overlap between the Gaussian smearing of the signal and the flat bin centered around $p_{e}^{\text{line}}=m_\mu/2$ with a width of 140 keV sets the efficiency of the signal. The result of this extrapolation is shown in Fig.~\ref{fig:money} and as expected the search strategy loses sensitivity very slowly for $m_a\gtrsim m_e$. The efficiency drops significantly only when $m_a\simeq m_\mu/10$, in which case the ALP mass starts to significantly shift the position of the positron line in Eq.~\eqref{eq:mean}.  

For pure left-handed ALP the systematic uncertainties related to the determination of the background endpoint are expected to grow significantly. The reason is that in this case the signal is suppressed more in the spin held dataset than in the spin precessed one. However, in the analysis \cite{Jodidio:1986mz} the latter was assumed to not be affected by the NP signal, in order to calibrate the instrument. Assuming that this is indeed still the case gives the green dashed line in Fig.~\ref{fig:money}, which should be viewed as only roughly indicative of what the correct bound for the left-handed ALP is.

\medskip\noindent {\bf The 2015 TWIST experiment}~\cite{Bayes:2014lxz} collected $5.8\times 10^8$ muons after the selection cuts were applied. The experimental collaboration studied the $\mu^+\to e^+a$ decay, varying both the mass of the ALP and the chirality of its couplings to the SM. Their results are summarized with the blue lines in Fig.~\ref{fig:money}. 

The experimental concept of TWIST is fundamentally  different from the previously discussed experiment by Jodidio et al. TWIST detected positrons using a spectrometer with an approximate cylindrical symmetry surrounding the muon beam line,
and momentum resolution at $x_e\simeq 1$ given by, 
\begin{equation}
\frac{\delta p_e}{p_e}\simeq \frac{1.1\%}{\vert \sin\theta_e\vert}\ ,\label{eq:momres:TWIST}
\end{equation}
which correspond to $\delta p_e=572\text{ keV}/\sin\theta_e $ at $p_e=52 \text{ MeV}$. Since the momentum resolution deteriorates in the forward direction (where $ \sin\theta_e\to0$), the experimental strategy of Jodidio et al.~cannot be implemented, even though the TWIST muon beam is highly polarized. The momentum resolution in Eq.~\eqref{eq:momres} translates into an upper bound below which the ALP can be considered effectively massless  
\begin{equation}
m_a \lesssim \sqrt{ 2 m_\mu \cdot \, 1 \text{MeV}}\simeq 14 \text{ MeV}\ .
\end{equation} 
The $1\text{ MeV}$ was obtained by setting $\cos\theta_e=0.8$ in Eq.~\eqref{eq:momres}, which is the positron angle for which TWIST had the widest momentum acceptance. 
 
As shown in Fig.~\ref{fig:money}, for a massive ALP the TWIST search covers a region that was left unexplored by the 1986 Jodidio et al. result, and extends the coverage up to an ALP mass of $86.6\text{ MeV}$. For higher masses the positron becomes too soft to be efficiently triggered on at TWIST. In this region older searches by Derenzo et al.~\cite{Derenzo:1969za}  and Bilger et al.~\cite{Bilger:1998rp} and the recent PIENU result~\cite{Aguilar-Arevalo:2020ljq} complement the high mass coverage. We refer to the PIENU paper~\cite{Aguilar-Arevalo:2020ljq} for a summary plot focused on the high mass region.

TWIST becomes less sensitive for a massless ALP  
due to the systematic uncertainties related to the calibration of the $x_e=1$ endpoint, which limited the sensitivity. Unlike Jodidio et al., 
TWIST collaboration had to rely on Monte Carlo modeling to calibrate the endpoint. The quoted systematic on the momentum edge represented an irreducible bottleneck to improving sensitivity of the massless ALP search. 
The  90\% C.L. upper bounds for massless ALP by TWIST, extracted using the Feldman-Cousins method \cite{Feldman:1997qc},
are
\begin{align}
&C^V_{\mu e}=0\text{ or } C^A_{\mu e}=0: &\text{BR}(\mu\to e\, a)< 2.1\times10^{-5}~~(90\%~{\rm CL}),\ \\
&C^V_{\mu e}=C^A_{\mu e}: &\text{BR}(\mu\to e\, a)< 1\times10^{-5}~~(90\%~{\rm CL}),~~\ \\
\label{eq:twistV-A}&C^V_{\mu e}=-C^A_{\mu e}: &\text{BR}(\mu\to e\, a)< 5.8\times10^{-5}~~(90\%~{\rm CL}).\ 
\end{align}  
Only the last bound, on the left-handed ALP, is stronger than our recast of the previous bound from Jodidio et al.~experiment, which, as discussed before, was less sensitive to left-handed ALPs. 
The gain of the TWIST experiment is due to the much larger muon luminosity compared to the one available to Jodidio et al., which compensates for the worse momentum resolution at TWIST.

\subsection{New searches at PSI: MEGII-fwd and Mu3e-online}
\label{sec:futuremutoe}
\begin{figure}[t]
	\centering
	\includegraphics[width=0.6\linewidth]{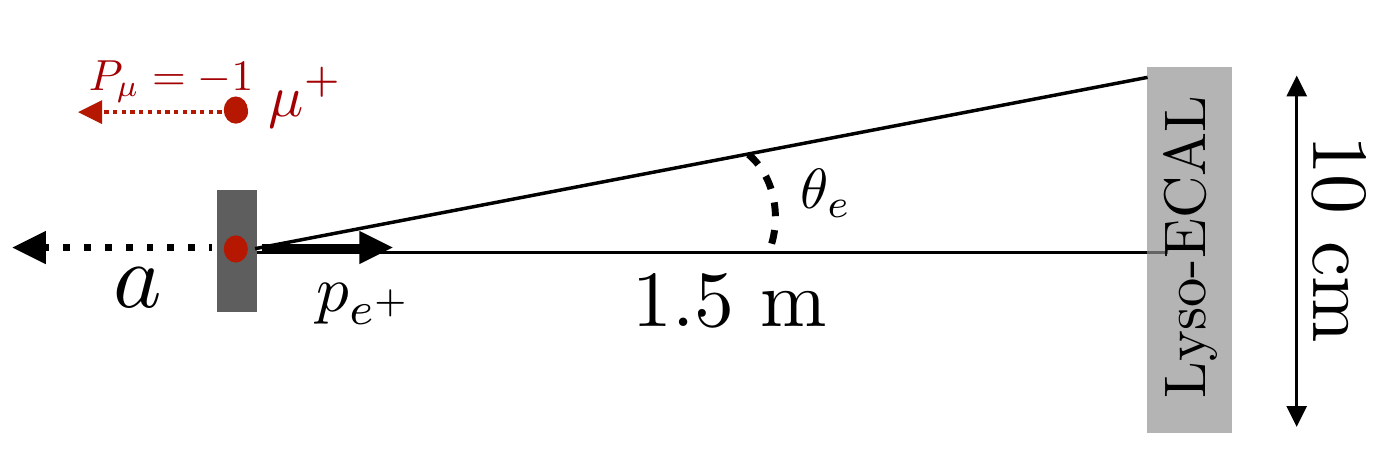}
	\caption{The proposed MEGII-fwd set-up. A Lyso-ECAL detector of $10\text{ cm}$ in diameter is placed along the muon beam line $1.5\text{ m}$ downstream from the stopping point. The muon polarization $P_\mu$ is in the opposite direction than the detected positron. 
 }
	\label{fig:cartoon}
\end{figure}

\noindent{\bf A proposal for MEGII-fwd.}
The MEG II experiment is expected to start its physics runs soon after completion of the engineering run that started in 2019. The goal of our proposal is to achieve at MEG II a configuration similar to the one used in the experiment by Jodidio et al.~\cite{Jodidio:1986mz}, so that the $\mu^+\to e^+ a$ decays of polarized muons are detected in the forward region where the SM background is suppressed, see Fig.~\ref{fig:summarymue} and discussion in Sec.~\ref{sec:pastmutoe}. Such a set-up requires: 
\begin{itemize}
\item A forward detector to collect energetic forward positrons. The final sensitivity to $\mu^+\to e^+ a$ decays will depend on the energy resolution of this detector. We assume that a Lyso calorimeter (Lyso-ECAL) with a 10 cm diameter and high enough threshold for positron energies can be installed in the forward direction roughly 1.5 meters downstream from the muon stopping target,  see Fig.~\ref{fig:cartoon}. The resulting angular coverage is $\theta\simeq1.91^{\circ}
$. We vary the positron momentum resolution at $p_{e^+}=m_\mu/2$ in the range\footnote{We are aware that the use of such a calorimeter for calibration purposes was discussed inside the MEG II collaboration. A momentum resolution $\delta p_{e^+}/p_{e^+}=10^{-2}$ at  $p_{e^+}=m_\mu/2$ is realistic for this ECAL~\cite{Atanov:2018ich}.  We thank Angela Papa for private communications regarding this.}   
\begin{equation}
\delta x_e\equiv\delta p_{e^+}/m_e=10^{-3}-10^{-1}\ .
\label{eq:momres:vary}
\end{equation}
 At a similar position the MEG II collaboration  planned to put a Radiative Decay Counter (RDC), with the aim to reduce the accidental background for the search of $\mu^+\to e^+\gamma$. However, RDC is designed to detect low-momenta positrons and would not be useful for the $\mu^+\to e^+a$ search.  
\item A new configuration of the MEG~II magnetic field in order to suppress depolarization effects and keep the $\mu^+$ 
antiparallel to the outgoing positron. The two main sources of depolarization of the muon beam are the so called ``halo muons'', emitted from pions decay in flight, and the angular divergence of the beam. How close the polarization can be kept to the maximal one, $\langle P_\mu\rangle=-1$ is crucial here, as this controls the suppression of the SM background, see \eqref{eq:endpoint}, which  directly affects the sensitivity to $\mu^+\to e^+ a$.
In what follows, we vary the depolarization in the range, 
\begin{equation}
\Delta P_\mu\equiv\langle P_\mu\rangle+1=10^{-3}-10^{-1}\ .\label{eq:polarization}
\end{equation}
In the predecessor experiment, the MEG experiment, the muon polarization at the stopping target was $\langle P_\mu\rangle=-0.86\pm 0.06$ \cite{Baldini:2015lwl}, which corresponds to the upper limit of the above range. The lower limit of the range assumes that the depolarization strategy similar to the one used in the 1998 experiment by Jodidio et al., Ref.~\cite{Jodidio:1986mz}, can be put in place. Collecting in addition a less pure sample of polarized muons would help in calibrating the endpoint of the Michel spectrum, $x_e=1$, exactly as it was done in Ref.~\cite{Jodidio:1986mz}.

\item A focusing lens to increase the positron luminosity in the forward direction. In the experiment by Jodidio et al.,  a solenoid lens was used to maximize the signal acceptance at the price of a higher SM background, cf. Sec.~\ref{sec:pastmutoe}. Focusing enlarges the effective size of the forward detector. The angular coverage of the Lyso-ECAL is very small so that without focusing the geometric acceptance of the signal is only,
\begin{equation}
a_{\text{geo}}^{\text{ISO}}=5.6\times 10^{-4}\quad\ ,\quad a_{\text{geo}}^{\text{V+A}}=1.2\times 10^{-3}\ ,\label{eq:focus}
\end{equation} 
for isotropic and $V+A$ ALP, respectively. The reach on the branching ratio scales with the focusing factor $F$ as $\text{BR}(\mu^+\to e^+ a)\sim1/\sqrt{F}$ as long as $F\times a_{\text{geo}}<1$ and the systematics uncertainties are subdominant. 
In the projections we leave $F$ as a free parameter, noting that $F\sim {\mathcal O}(10^2)$ was reached in the 1986 experiment by Jodidio et al., cf. Eq.~\eqref{eq:focusing}.

\item Sufficient time devoted to the physics run in the $\mu^+\to e^+a$ search configuration. 
As a reference point we will use 
\begin{equation}
N_\mu=1.2\times 10^{14}\mu^{+}\ ,
\end{equation}
which corresponds to a 2 week run at an instantaneous luminosity of $10^8\mu^+/\text{sec}$.\footnote{MEG II plans to run with the reduced instantaneous luminosity of $3\times 10^7 \mu^+/\text{sec}$ in order to decrease accidental coincidences, which is the dominant background in the $\mu^+\to e^+\gamma$ measurement. For the $\mu^+\to e^+a$ search one should attempt to use the full  instantaneous luminosity of $10^8 \mu^+/\text{sec}$ available at the $\pi E5$ beamline at PSI.}
This is seven orders of magnitude greater than the dataset collected by Jodidio et al. Crucially, the $\mu \to ea$ run could be performed at the very end of the $\mu\to e\gamma$ data taking at MEGII. This would allow to work on the necessary modifications and extend the physics purpose of MEG II before Mu3e starts. 
\end{itemize}

\begin{figure}[t]
	\centering
	\includegraphics[width=0.48\linewidth]{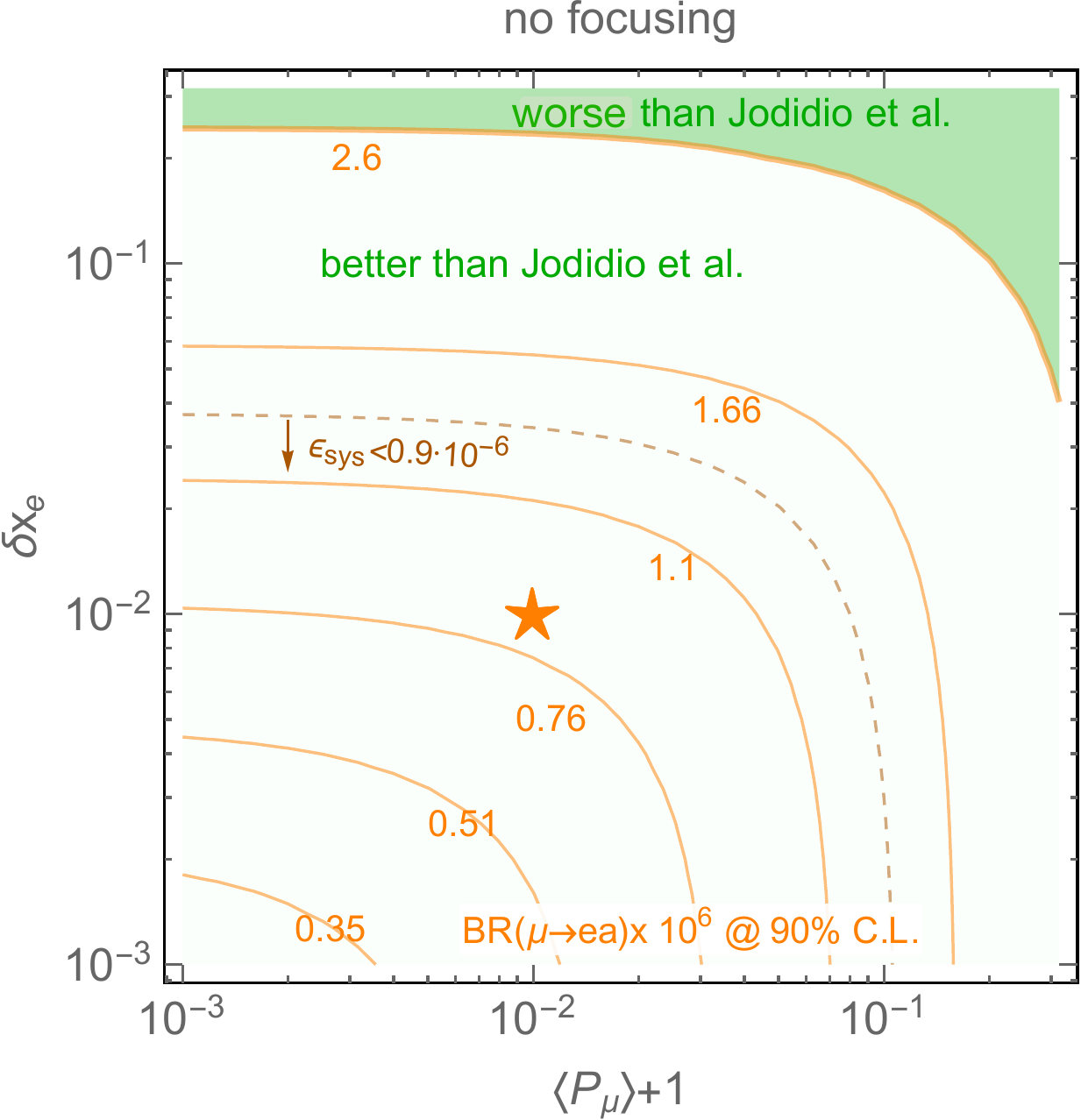}\quad
	\includegraphics[width=0.48\linewidth]{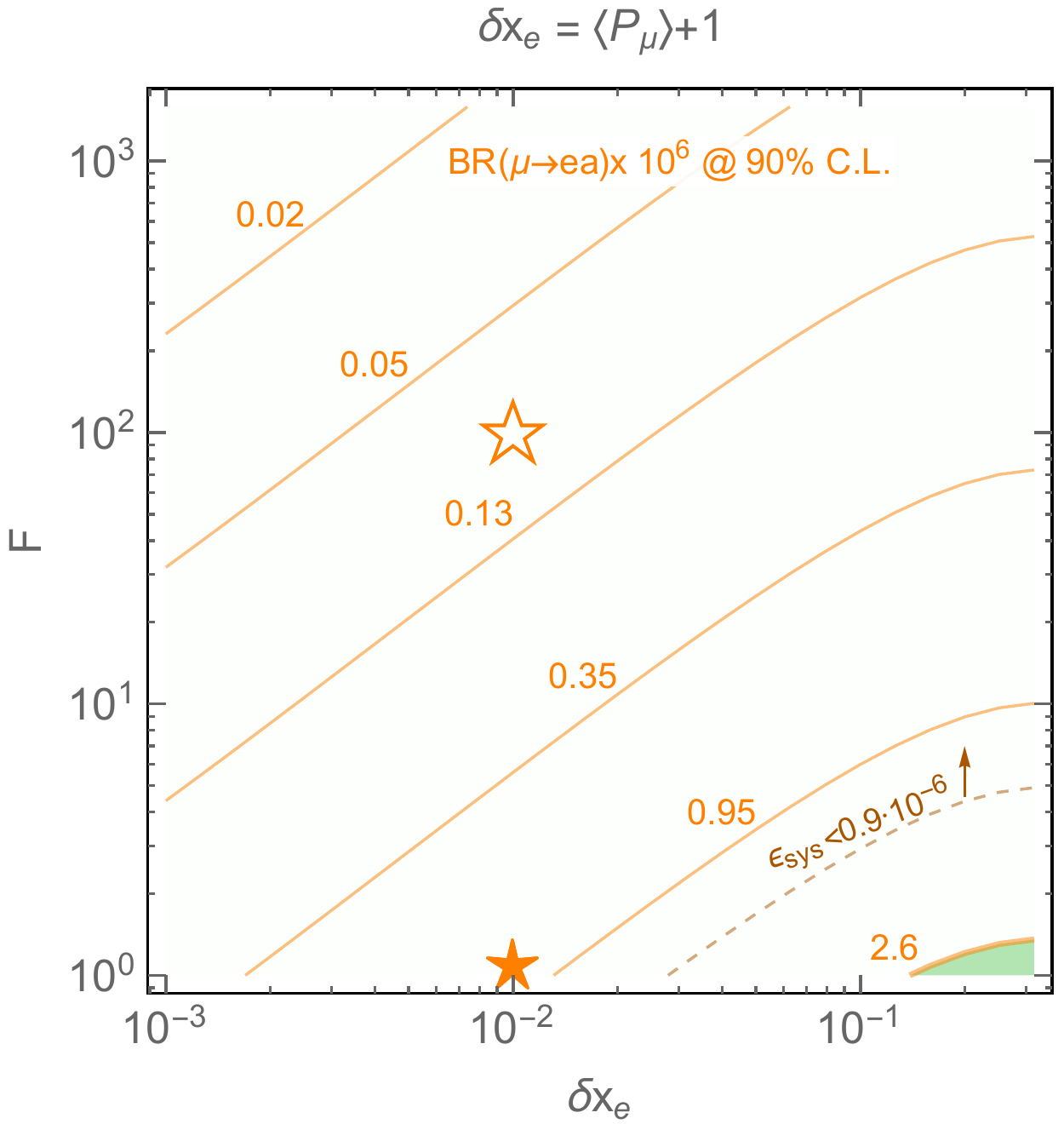}
	\caption{{\bf Left:} The {\bf orange} contours show the 90\% C.L. expected sensitivity on $\text{BR}(\mu^+\to e^+a)$ at MEGII-fwd  for a massless isotropic ALP (either $C^V_{\mu e}\neq0$ or $C^A_{\mu e}\neq0$) as a function of the momentum resolution $\delta x_e$, and the deviation of average polarization $\langle P_\mu \rangle$ from $-1$, assuming there is no magnetic focusing.  {\bf Right:} The expected sensitivity at MEGII-fwd ({\bf orange} contours) as a function of the momentum resolution $\delta x_e$ and focusing $F$, setting muon polarization to $\langle P_\mu\rangle =-1+\delta x_e$. The angular resolution is assumed to be subdominant and systematic uncertainties below  the statistical one. The {\bf darker green} region are excluded by the 1986 Jodidio et al. experiment~\cite{Jodidio:1986mz}, cf. Table~\ref{tab:bounds}. The {\bf dashed dark orange} line indicates where our future projections assume that systematic uncertainties can be lowered compared to the ones in Jodidio et al.~$\epsilon_{\text{sys}}=0.9\times 10^{-6}$. }
	\label{fig:reachMEGforward}
\end{figure}

In the left panel of Fig.~\ref{fig:reachMEGforward} we show how the reach on the branching ratio $\text{BR}(\mu^+\to e^+ a)$ for a massless ALP depends on the average polarization and the angular resolution. Interestingly, from the shape of the contours one sees that augmenting the polarization purity of the muon beam should go together with an increase of the momentum resolution in order to lead to a better experimental reach. The orange star in the plot is one of our benchmark configurations, where we assume no focusing in the forward direction and set $\delta x_e=\langle P_\mu\rangle+1=10^{-2}$. Already in these suboptimal conditions, MEGII-fwd could improve on the present bound from Jodidio et al. 

In the right panel we set $\delta x_e=\langle P_\mu\rangle+1$ and investigate the importance of the focusing. To be conservative we take as the benchmark $F=100$, which corresponds to an effective diameter of the forward calorimeter roughly 10 times bigger than its actual size. This is of the same order of magnitude as the focusing achieved in Ref.~\cite{Jodidio:1986mz}. However, a larger focusing would always be beneficial until $F\times a_{\rm geo}$ 
would be of order unity.

In Tab.~\ref{tab:bounds} and Fig.~\ref{fig:money} we show the estimated reach of MEGII-fwd for the two benchmark configurations described above. The reach for higher $m_a$, shown in Fig.~\ref{fig:money}, is computed accounting for the amount of signal which will overlap with the massless hypothesis given the experimental resolution. Due to worse momentum resolution the drop in the $f_a$ reach at high masses is slower at MEGII-fwd  compared to the Jodidio et al.~experiment.

\medskip\noindent{\bf Mu3e-online.}
The primary goal of the planned Mu3e experiment at PSI is to search for $\mu^+\to e^+e^-e^+$ with the unprecedented sensitivity of $10^{-16}$.
The key feature of Mu3e is that it will operate without a hardware trigger. 
The full detector read-out will be streamed to the filter farm at $80\text{ GB}/\text{sec}$, where the $\mu^+\to e^+e^-e^+$ events will be identified and eventually stored on disk. 

Recently, Ref. \cite{Perrevoort:2018ttp} performed a preliminary study of the Mu3e sensitivity to $\text{BR}(\mu^+\to e^+ a)$, based on a more detailed simulation of the $\mu^+\to e^+ a$ channel at the  phase~I of the Mu3e experiment~\cite{Perrevoort:2018okj}. The dark thin red line in Fig.~\ref{fig:money} shows the $95\%$ C.L. limit  on $f_a$ for isotropic LFV ALP that Mu3e is projected to achieve with a  physics run of 300 days. The $\sim 10^4$ improvement in $f_a$ reach over the TWIST experiment is mostly driven by the seven orders of magnitude larger dataset, which, however, does not come without challenges.

For $\mu^+\to e^+ a$ search the Mu3e will be able to reconstruct online all the single positron events corresponding to ``short tracks'', i.e., events with only four hits on the four detector layers. The online reconstruction of ``short tracks'' in the filter farm has been shown to reduce by a factor of a 100 
the data rate \cite{vomBruch:2017fqw}, making it possible to process all the short tracks events and store positron three-momenta, $\vec p_e$. 
A search for $\text{BR}(\mu^+\to e^+ a)$ positron line can be done as a bump hunt on the smooth SM $|\vec p_e|$ distribution, assuming that every positron event corresponds to a single track. 

The tracks in Mu3e will gyrate around the magnetic field of roughly $1 \, {\rm T}$. The typical $e^+$ gyroradius will be much larger than the radius of the Mu3e instrumented region -- a cylinder with radius of  around 6\text{ cm}. Encountering the detector material the positron will loose momentum and eventually stop. The positron will typically stop after half a turn, i.e., after having encountered at least four detector layers: 2 central layers + 2 external layers. This justifies the assumption of one positron per one track. Positrons emitted perpendicularly to the muon beam will instead perform many turns in the central layers without being stopped in the detector material. Enforcing an angular cut will minimize the impact of these re-curling tracks. As a result, the angular acceptance of the Mu3e analysis is expected to be reduced to the region $\theta_e<\pi/2- 0.1$. 

The momentum resolution for short tracks will be in the $\delta p_\text{short}=0.5-3\text{ MeV}$ range, roughly comparable to TWIST. In principle, a momentum resolution down to $\delta p_\text{long}=0.1-0.45\text{ MeV}$ could be achieved, if an experimental upgrade allowed to also process online the ``long tracks'', characterized by 6 or 8 hits in the detector layers. The Mu3e sensitivity  on  $\text{BR}(\mu^+\to e^+ a)$ extracted from the long track analysis would then improve by a factor of $\sqrt{\delta p_\text{long}/\delta p_\text{short}}\simeq0.4$~\cite{Perrevoort:2018okj}.  

Finally, we comment on the challenges of calibrating the instrument. As already mentioned, the current concept foresees to use the endpoint of the positron momentum spectrum to calibrate the online reconstructed tracks. This method will not allow to efficiently search for $ \mu^+\to e^+ a$ decays when $m_a\lesssim 10-25\text{ MeV}$, depending on the precise positron momentum resolution. Given the large amount of single positron data collected at Mu3e, Ref.~\cite{Perrevoort:2018okj} showed that, within a given mass assumption for the signal, one could in principle use the same momentum spectrum to simultaneously calibrate the apparatus and to perform the peak search. In the calibration fit  one would remove from the calibration dataset the signal region, which is defined as a $2\delta p_\text{short}$ band below and above the expected momentum of the positron corresponding to the ALP mass. Given that for $m_a\lesssim 25\text{ MeV}$ the signal region includes the Michel edge the determination of the scaling parameter $x_e=1$ deteriorates, resulting in a limited sensitivity at low masses. Including this effect the expected sensitivites for a massless isotropic ALP are
\begin{equation}
C^V_{\mu e}=0\text{ or } C^A_{\mu e}=0:\qquad \text{BR}(\mu\to e\, a)< 7.3\times 10^{-8}~~(90\%~{\rm CL})\ .
\end{equation}  
Notice that the deteriotation computed in Ref.~\cite{Perrevoort:2018okj} includes only the broadening of the $x_e$ statistical distribution estimated from a toy Monte Carlo sample of $10^9\mu^+$ (the root mean square error grows from $7\times 10^{-6}$ for $m_a=60\text{ MeV}$ to $3.8\times 10^{-5}$ for $m_a=10\text{ MeV}$). A similar effect is expected for both the left-handed and the right-handed ALPs so that the reach shown in Fig.~\ref{fig:money} is a realistic estimate of the Mu3e reach for all chiral structures.
It is possible, however, that alternative calibration strategies such as the one proposed in Ref.~\cite{Rutar:2016ozg} could improve the Mu3e reach at low masses.

\subsection{The potential at $\mu^-\to e^-$ conversion experiments}
\label{sec:mu2e}
 The $\mu\to e$ conversion in nuclei experiments, COMET at J-PARC \cite{Adamov:2018vin} and Mu2e at FNAL \cite{Bartoszek:2014mya}, are designed around very large muon fluxes, with over $10^9 \mu^-/s$ and $10^{10}\mu^-/s$ at the two respective experiments. One may hope that these could also be used for $\mu\to e a$ searches. However, in order to be able to deal with the vast data-streams the two experiments will not measure the full Michel spectrum but rather focus on the endpoint of the $e^-$ energy, which for the $\mu^-+(A,Z)\to e^- +(A,Z)$ process is at $E_e^{\rm end}=m_\mu-E_b-E_{\rm rec}$, with $E_b$ the binding energy of the muonic atom and $E_{\rm rec}$ the nuclear recoil energy. In Aluminum $E_e^{\rm end}\simeq 105 $ MeV, while in gold $E_e^{\rm end}\simeq 95 $ MeV. 
 
 The conversion with the emission of the ALP, $\mu^-+(A,Z)\to e^- +a+(A,Z)$, peaks instead  at the electron energy of around $E_e\simeq m_\mu/2$, but with the tails of the electron energy distribution that go all the way up to $E_e^{\rm end}$.  Nominally, searches for the $\mu\to e a$ process at COMET and Mu2e would also rely on these tails of the distribution, which poses a challenge for obtaining a competitive reach. For instance, for $E_e>100$ MeV only a small fraction of $\mu\to e a$ decays, about $2\times 10^{-10}$, would be inside the signal region for $\mu\to e$ conversion on Al \cite{GarciaiTormo:2011et} (see also \cite{Uesaka:2020okd}). Even with $10^{18}$ muons a competitive search would thus likely require relaxing the lower bound on $E_e$ and developing techniques to distinguish between the smooth shapes of the signal,  $\mu\to e a$, and SM background, $\mu \to e \nu\bar \nu$, decays in orbit.

\section{ALPs in $\mu \to e +\gamma+{\rm invis.}$ decays}\label{sec:mutoegamma}
The  $\mu^+ \to e^+ \gamma a$ decay offers a complementary probe of the LFV ALP which is less dependent on the chiral structure of the ALP couplings than the experimental searches for $\mu\to ea$. In Sec. \ref{sec:CB} we first discuss the searches performed at the Crystal Box experiment with a total number of $8.15\times10^{11}$ stopped muons~\cite{Goldman:1987hy,Bolton:1988af}. In Sec.~\ref{sec:futuremuegammaa} we discuss possible improvements in the reach were a similar search to be implemented  at MEG II. A more detailed analysis of a dedicated trigger at MEG II for this channel is left for future work.

\subsection{Past Searches at Crystal Box}\label{sec:CB}

\begin{figure}[t]
	\centering
	\includegraphics[width=0.48\linewidth]{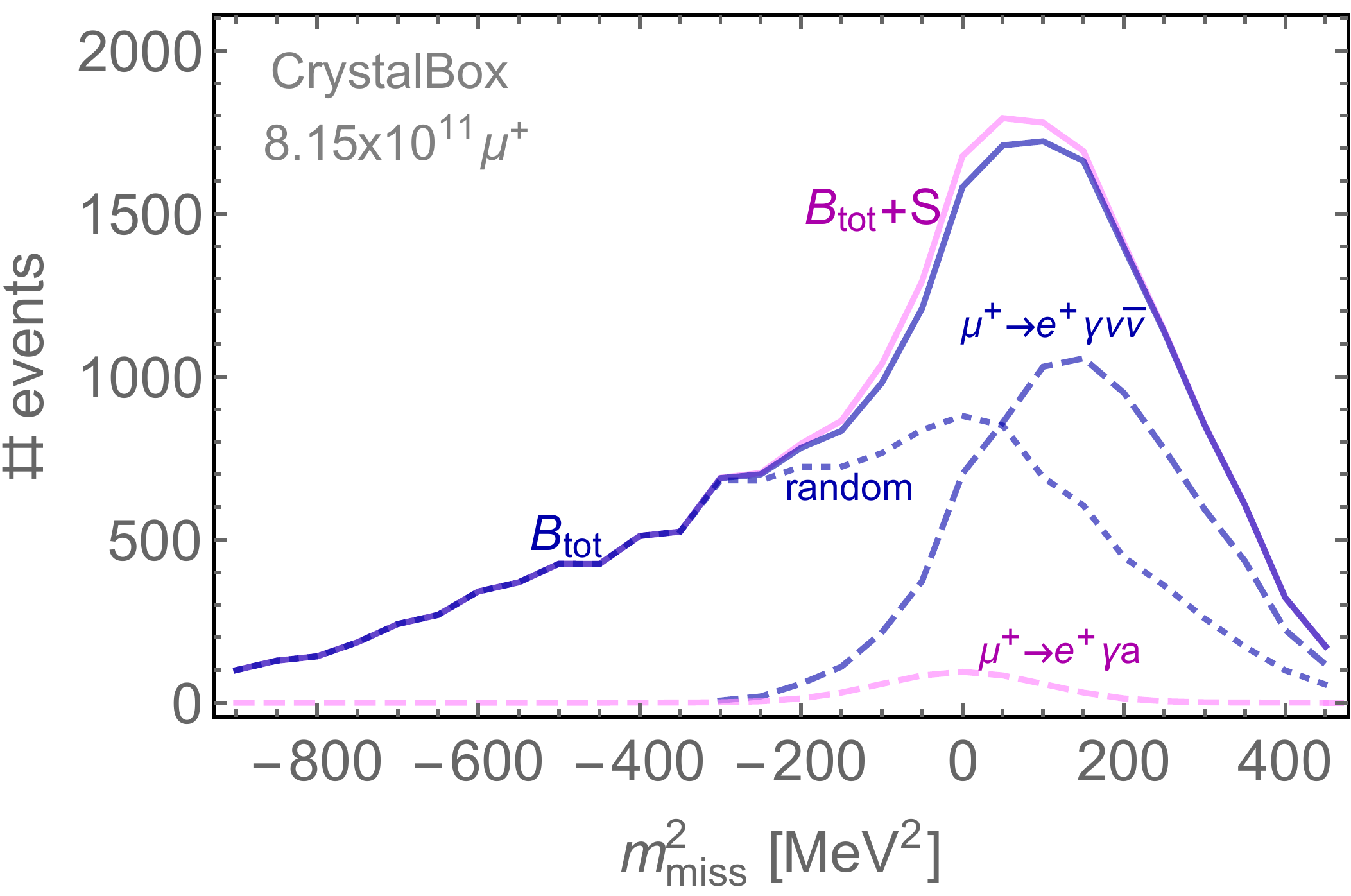}\quad
	\includegraphics[width=0.48\linewidth]{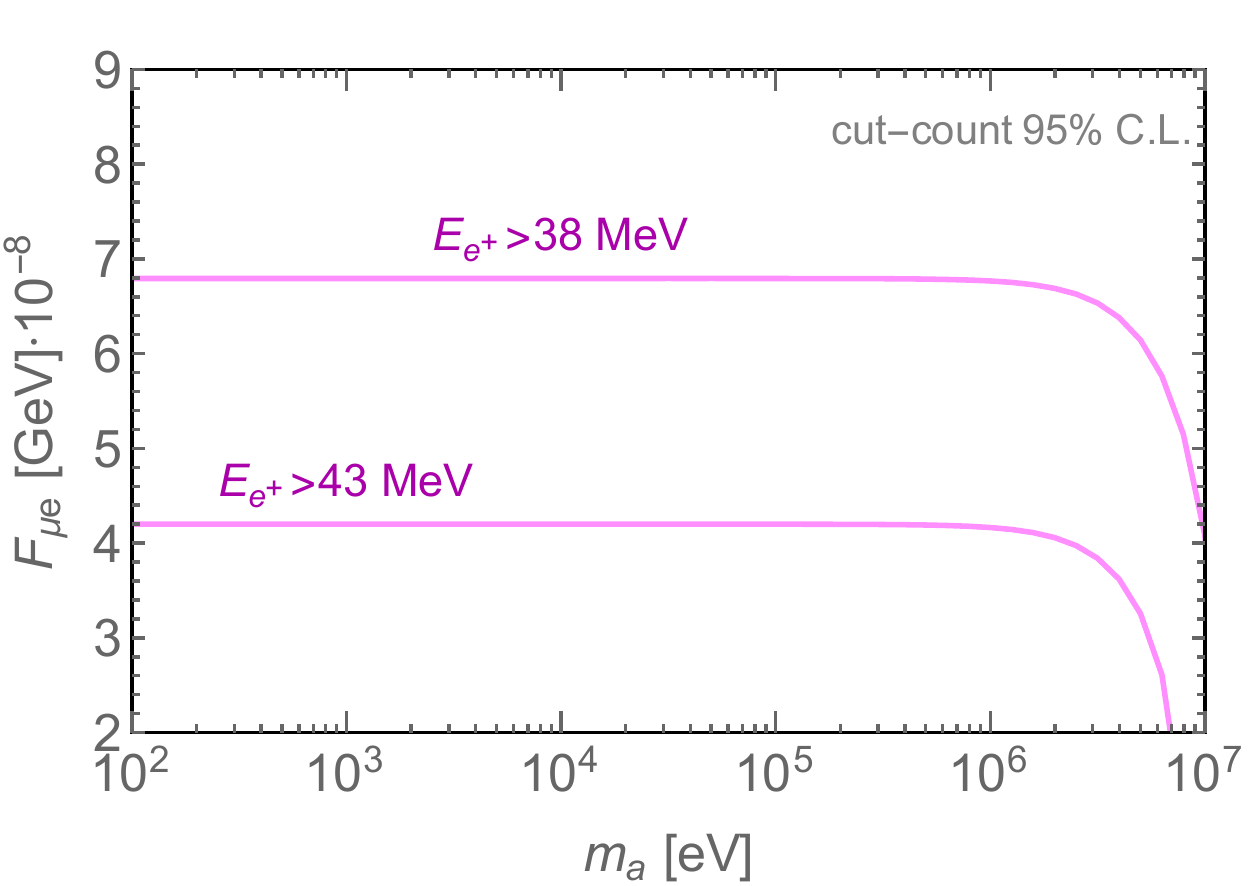}
	\caption{{\bf Left:} Missing mass distribution of background and signal events in Crystal Box after the cuts in Eqs.~\eqref{eq:cutsCB}, \eqref{eq:cutsCB2} are implemented and experimental efficiencies taken into account. {\bf Solid blue} is the total background composed by the {\bf dashed blue} distribution of $\mu^+ \to e^+ \gamma \nu\bar{\nu}$ events and the {\bf dotted blue} distribution of events where a hard photon from  $\mu^+ \to e^+ \gamma \nu\bar{\nu}$ and a positron from $\mu^+ \to e^+ \nu\bar{\nu}$ are randomly coincident. {\bf Solid magenta} shows the combined background plus signal distribution. The {\bf dashed magenta} is the signal shape smeared by the experimental resolution for $m_a=0$ and $\text{BR}(\mu^+\to e^+\gamma a)=3\times 10^{-9}$. {\bf Right:} the 95\% C.L. sensitivity of Crystal Box that we obtain using the simple cut and count scheme as described in the main text.  The two lines indicate how the final bound on $F_{\mu e}$ depends on the positron energy loss before it reaches the detector. The actual experimental bound reaches a 30\% higher value of $F_{\mu e}$ and is reported in Eq.~\eqref{signalgamma}.}\label{fig:CBplots} 
\end{figure}

In order to lower the trigger rate the Crystal Box required at the trigger level a hard positron and a photon of similar energy~\cite{Goldman:1987hy,Bolton:1988af}. The search for the three-body $\mu^+ \to e^+ \gamma a$ decay is then a search for a bump in the missing mass distribution in the collected data. The signal would be centered at $m_{\text{miss}}=m_a^2$ and spread by the  photon and positron energy resolutions and the resolution on the angle between the two. The SM background has two main components: the four-body $\mu^+ \to e^+ \gamma \nu\bar{\nu}$ decays, and the combinatorics background due to coincident $\mu^+ \to e^+ \nu\bar{\nu}$ and $\mu^+ \to e^+ \gamma\nu\bar{\nu}$ events. For the latter, a sufficiently hard positron from the $\mu^+ \to e^+ \nu\bar{\nu}$ decay is detected within a 1.5 ns time window together with a hard photon from the $\mu^+ \to e^+ \gamma\nu\bar{\nu}$ decay, while the soft positron is left undetected. The background and signal shapes at Crystal Box are shown in Fig.~\ref{fig:CBplots} (left). 

The rate of the three-body decay $\Gamma(\mu\to e\,a\,\gamma)$ for an ALP of mass $m_a$ is given by (in the limit $m_e \ll m_\mu$)
\begin{align}
\Gamma(\mu\to e\,a\,\gamma) \approx \frac{\alpha_{\text{em}} }{32\pi^2} \frac{m_\mu^3}{F_{\mu e}^2} \mathcal{I} (x_{\text{min}},y_{\text{min}},\eta_a) \, , \label{eq:width3body}
\end{align} 
with the phase space integral given by
\begin{equation}
\mathcal{I}(x_{\text{min}},y_{\text{min}},\eta_a)= \int_{y_{\rm min}}^{1-\eta_a}\hspace{-3mm} dy \hspace{-1mm}\int_{\max ( x_{\rm min}, 1-y - \eta_a )}^{ \frac{1- y - \eta_a}{1-y}}\hspace{-7mm} dx \,  \frac{y (1-x^2 - \eta^2_a) - 2 (1-\eta_a)(1-x-\eta_a)}{y^2 (1-x-y - \eta_a)} \, , 
\label{Iminmax}
\end{equation}
where 
\begin{equation}
x=2 E_e/m_\mu,\qquad y=2 E_\gamma/m_\mu,\qquad \eta_a= m_a^2/m_\mu^2.
\end{equation}
These kinematic variables are related to the angle $\theta_{e \gamma}$ between the electron and photon momenta, 
\begin{align}
\cos \theta_{e \gamma} = 1 + \frac{2  (1- x- y - \eta_a)}{x y} \, .
\end{align}
The branching ratio for the three-body decay, $\text{BR}(\mu\to e\,a\,\gamma)$,  is related to the  branching ratio for the 2-body decay, $\text{BR}(\mu\to e\,a)$, 
\begin{align}
\label{eq:Br:relation}
\text{BR}(\mu\to e\,a\,\gamma) \approx \frac{\alpha_{\text{em}}}{2\pi (1-\eta_a)^2}\mathcal{I}(x_{\text{min}},y_{\text{min}},\eta_a)\text{BR}(\mu\to e\,a)
\, . 
\end{align} 
where in the expressions for both $\text{BR}(\mu\to e\,a\,\gamma)$ and $\text{BR}(\mu\to e\,a)$ the mass of electron should be neglected. 

The infrared and collinear divergences (we are working in the limit $m_e \to 0$) are regulated by the experimental cuts on photon and positron energies  $x_{\rm min}, y_{\rm min}$. The experimental cuts in the Cristal Box search were \cite{Bolton:1988af}
\begin{align}
E_e>38-43\text{ MeV}\ , E_\gamma>38\text{ MeV} & \quad \Rightarrow\quad x_{\text{min}}=0.72-0.81\, ,~y_{\text{min}}=0.72 \, .\label{eq:cutsCB}\\
\theta_{e \gamma}> 140^\circ \quad & \Rightarrow \quad\cos \theta_{e \gamma} < -0.77 \, .
\label{eq:cutsCB2}
\end{align}
Imposing the $\theta_{e\gamma}$ cut in the phase space integral, Eq.~\eqref{Iminmax}, reduces $\mathcal{I}(x_{\text{min}},y_{\text{min}},\eta_a)$ by $\sim 10\%$ for a massless ALP, and has negligible effect for a massive ALP. The positron energy cut, $E_e> 38$ MeV, reported by the Crystal Box collaboration, refers to the positron energy measured by the scintillation crystals. When translating this cut to $x_{\rm min}$ in the phase space integral, Eq.~\eqref{Iminmax}, one needs to account for the positron's energy loss during the propagation to the detector. The Crystal Box collaboration indicated that the positron could loose up to 5 MeV before reaching the detector. In our analysis we therefore vary the cut on the positron energy in the range from 38 to 43 MeV.

In the special case of vanishing axion mass, $\eta_a \to 0$, the above expressions, Eq.~\eqref{eq:width3body}-\eqref{eq:Br:relation}, reduce to the ones obtained in  Ref.~\cite{Hirsch:2009ee}. Notice that the expression used by the experimental collaboration in Ref.~\cite{Goldman:1987hy}  appears to have a typographical error (the roles of $x$ and $y$ in the phase space integral in Eq.~\eqref{Iminmax} were exchanged). However, this does not affect the analysis in Ref.~\cite{Bolton:1988af} because of  symmetric cuts on photon and positron energies (up to the positron energy losses), cf. Eq.~\eqref{eq:cutsCB}.

The  90\% C.L. bound on the $\mu \to e a \gamma$ branching ratio for $m_a\simeq0$, obtained by the Crystal Box experiment, is
\begin{align}
\label{eq:extragamma}
\text{BR}(\mu\to e\,a\,\gamma) < 1.1\times10^{-9}~~(90\%~{\rm CL})\, ,
\end{align}
to be compared with the theory prediction, 
\begin{equation}
\text{BR}(\mu\to e\,a\,\gamma)\approx {(3.7-9.7) \times10^{-10} }\left( \frac{10^9 \GeV }{F_{\mu e} } \right)^2 \, , 
\label{signalgamma}
\end{equation}
where the range is due to the variation in the phase space integral,  $\mathcal{I}(x_{\text{min}},y_{\text{min}},\eta_a)\approx (0.004-0.011)$, obtained by varying the cut on the upstream positron energy between 38 MeV and 43 MeV. The 95\% C.L. bound  is then $F_{\mu e}\gtrsim (5.1-8.3)\times 10^8\text{ GeV}$, and  is always subdominant compared to the combination of the bounds obtained from $\mu^+\to e^+ a$ by Jodidio et al.~\cite{Jodidio:1986mz} and TWIST~\cite{Bayes:2014lxz}, cf. Table \ref{tab:bounds}. 

For completeness, we discuss a simple procedure that we use to roughly reproduce the Crystal Box result and extend it to $m_a>0$. The results of this recasting procedure are shown in the right panel of Fig.~\ref{fig:CBplots}. We first extract the experimental efficiency for the $\mu^+\to e^+ \gamma a$ signal from the bound on the branching ratio, $\text{BR}_{\text{excl.}}$, in Eq.~\eqref{eq:extragamma}, which corresponds to $N_{\rm excl.}=165$ signal events~\cite{Bolton:1988af}. Given that the total number of stopped muons is $N_{\mu^+}=8.15\times 10^{11}$ we get
\begin{equation}
\epsilon_S= \frac{N_{\text{excl.}}}{N_{\mu^+}\times \text{BR}_{\text{excl.}}}\simeq0.18\ .
\end{equation}
The signal shape can be extracted from the binned MC sample given in Ref.~\cite{Goldman:1987hy},
and is well described by a Gaussian centered at $m_{\text{miss}}^2=0$ with a variance of $\sigma_{S}=100 \text{ MeV}^2$. 

For a given mass $m_a$ one can then estimate the sensitivity of Crystal Box on $\text{BR}(\mu\to e\,a\,\gamma)$  by considering the expected background and signal in a missing mass window, $(m_a^2-\sigma_S,m_a^2+\sigma_S)$ and using the asymptotic formula at $90\%$ C.L. For $m_a=0$, this procedure gives $\text{BR}(\mu\to e\,a\,\gamma)<1.6\times 10^{-9}$ which is weaker than the experimental bound in Eq.~\eqref{signalgamma} only by about 30\%. This is not surprising since the full likelihood analysis has better discriminating power than the simple cut and count analysis we are performing. The bound on the branching ratio gets stronger for $m_{\text{miss}}^2>100\text{ MeV}^2$ because the background is suppressed. However, this effect is compensated by the phase space suppression of the signal, Eq.~\eqref{eq:width3body}. All in all, the bound on $F_{\mu e}$ is constant up to $m_a\simeq 10\text{ MeV}$ after which the signal phase space starts to shrink significantly with increasing ALP mass, given the strong energy cuts on the photon and positron energies.

\subsection{Possibilities for future search at MEG II}\label{sec:futuremuegammaa}

We now briefly comment on the possibility for MEG II to improve on the Crystal Box result discussed above. Naively, the MEG II luminosity will exceed the one of Crystal Box by at least 3 orders of magnitude. In optimal conditions, this could lead to an increase of sensitivity on $F_{\mu e}$ of more than a factor of 5 with respect to the current bound in Table.~\ref{tab:bounds}. With such an optimistic gain MEG II could start probing new parameter space beyond the current TWIST bound for the $V-A$ ALPs. An improvement on $\mu\to e a \gamma$ would complement the MEGII-fwd proposal, Sec.~\ref{sec:futuremutoe}, since this is only sensitive to ALPs with some amount of right handed couplings to the SM leptons.  The combination of the two searches, $\mu\to e a \gamma$ and $\mu\to e a $ at MEGII-fwd would then fully cover the possible chiral structures of the ALP couplings. 

Clearly, the above naive estimate for the improvement on the reach is far from guaranteed. A more realistic estimate  of the reach would require a dedicated trigger study. The current energy and angular cuts of the MEG II trigger dedicated to the $\mu\to e \gamma$ search are designed to select a very energetic positron and a photon exactly back to back~\cite{TheMEG:2016wtm}. Keeping these cuts will greatly suppress the signal rate of $\mu\to e \gamma a$, and make it impossible to perform the search. An interesting  possibility would be to relax the energy and angular cuts of the trigger down to similar values than the ones in Crystal Box. The final expected sensitivity at MEG II will strongly depend on the signal vs. background efficiencies whose detailed determination goes beyond the scope of this paper and is left for future investigations.

\section{ALPs in $\tau$ decays}\label{sec:taudecays}
The search strategies for $\tau\to \mu a$ and $\tau\to e a$ decays are qualitatively different from the $\mu\to e a$ searches. The main differences can be traced to the fact that $\tau$ has a much shorter life-time ($3\times 10^{-13}$ s vs. $2\times 10^{-6}$ s for the muon), that it has many more decay channels, and that from the $\tau$ production it is not possible to unambiguously reconstruct the $\tau$ rest frame. 

\subsection{Past Searches at ARGUS}
\noindent{\bf The ARGUS experiment in 1995} derived bounds on tauonic LFV ALP couplings~\cite{Albrecht:1995ht}.  The tau data sample was produced from $e^+e^-$ collisions in the DORIS II storage ring at DESY at a center of mass energy varying between 9.4 and 10.6 GeV with a total integrated luminosity of $472\text{ pb}^{-1}$. 

The challenge in $\tau\to \ell_i a$ search is to disentangle the signal decay from the SM $\tau\to \ell_i\, \nu\bar{\nu}$ decays. The search would be easier in the tau rest frame, since then the lepton from $\tau\to \ell_i a$ is monochromatic and one can do a line search on top of the smooth $\tau\to \ell_i\, \nu\bar{\nu}$ background. Unfortunately, the tau is pair-produced in $e^+e^-\to \tau^+\tau^-$ collisions. Each of the tau's decays into a final state with at least one invisible particle, making exact reconstructions of the tau rest frames impossible. 
Instead, the ARGUS analysis used the ``pseudo-rest frame'' technique.
The idea is to require one side of the $\tau^+\tau^-$ pair to decay into a three prong hadronic mode. The direction of the $\tau$ momentum  is then approximated by the direction of the combined momentum of the three prong decay products. In the center of mass of $e^+e^-$ collision the tau energy equals the beam energy, while the two taus are back to back. This gives enough constraints so that one can boost the leptonically decaying tau to its, approximate  ``pseudo-rest frame''. The crucial property of this frame is that the sensitivity on LFV two body tau decays into ALPs does not depend much on the ALP mass (see~\cite{Albrecht:1995ht} for further details). 

For a massless ALP ARGUS obtained~\cite{Albrecht:1995ht} 
\begin{align}
&\text{BR}(\tau\to e \,a)< 2.7\times10^{-3}~~(95\%~{\rm C.L.})\quad \Rightarrow\quad F_{\tau e}\gtrsim 4.3\times 10^6\text{ GeV}\, ,\\
&\text{BR}(\tau\to \mu \,a)< 4.5\times10^{-3}~~(95\%~{\rm C.L.})\quad \Rightarrow\quad F_{\tau \mu}\gtrsim 3.3\times 10^6\text{ GeV} \,. \label{ARGUSmubound}
\end{align}
The  bound on $\text{BR}(\tau\to \mu \,a)$ is less stringent than $\text{BR}(\tau\to e \,a)$ at low masses while they become comparable for higher masses. 
The final bound on the ALP decay constant from $\text{BR}(\tau\to e a)$ is shown in Fig.~\ref{fig:money}. The mass dependence of the bound is predominantly due to the phase space suppression of the two-body decay for heavier ALPs.

\subsection{Future Searches at Belle-II}
\noindent{\bf Belle and Belle-II.}
While Belle and Babar collected $\approx2000$ times larger datasets of $\tau$'s than ARGUS, no experimental searches for $\tau\to \ell_i a$ were performed yet. However, a recent simulation of the expected limit at the Belle experiment  with integrated luminosity of 
$1020\text{ fb}^{-1}$ was performed in  Ref.~\cite{Yoshinobu:2017jti}, and obtained  for a massless ALP  at 90\% CL)
\begin{align}
&\text{Belle (1/ab) prospect:}\quad\text{BR}(\tau\to \mu\,a)< 1.1\times10^{-4} &\Rightarrow\quad F_{\tau \mu}\gtrsim 2.1\times 10^7\text{ GeV}\ .
\end{align}
Notice that this bound is almost exactly a factor of $\sqrt{2000}$ more stringent than the present one from ARGUS, Eq.~\eqref{ARGUSmubound}. Using the same simple rescaling with the luminosity, we obtain for the expected 95\% CL limit for a massless ALP for  Belle II with $50\text{ ab}^{-1}$,
\begin{align}
&\text{Belle-II (50/ab) prospect:}\quad\text{BR}(\tau\to \mu\,a)< 2.0 \times10^{-5} &\Rightarrow~ F_{\tau \mu}\gtrsim 4.9 \times 10^7\text{ GeV}.
\label{eq:belle2}
\end{align}
The limit as a function of ALP mass is shown with the purple line in Fig.~\ref{fig:money}. In the absence of MC analysis of Belle or Belle II reach for $\tau\to e a$, we estimate the Belle II sensitivity by performing the naive rescaling of the ARGUS result with luminosity, which gives
\begin{align}
&\text{Belle-II (50/ab) prospect:}\quad\text{BR}(\tau\to e\,a)< 8.3 \times 10^{-6}&\Rightarrow F_{\tau e}\gtrsim 7.7\times 10^7\text{ GeV}\ .
\end{align}

Belle II may improve the ARGUS searches for $\tau\to e a$ and $\tau\to\mu a$ transitions beyond mere increase in statistics. First of all, it could be interesting to explore the reach on $\text{BR}(\tau\to \mu\,a\, \gamma)$ and $\text{BR}(\tau\to e\,a\, \gamma)$, especially since for muons $\mu\to e a\gamma$  gives  constraints that are not that far from the two-body $\mu \to e a$ decay (see \cite{Heeck:2016xkh} for similar comments in the context of a light $Z'$). Secondly, further improvements of  $\tau\to \ell_i a$ searches may be possible. A possibly interesting direction, while still using the tau  ``pseudo-rest frame'',  is to employ in addition variables that tag the tau polarization, such as the directions of pions in two prong tau decays. 
If successful, this could allow to further suppress the SM $\tau\to \ell_i\, \nu\bar{\nu}$ background, similarly to what was done for $\mu\to e a$ decays in the experiment by Jodidio et al.~\cite{Jodidio:1986mz}. 

\section{Bounds from Astrophysics and Cosmology}
\label{sec:astro:cosmo}
The bounds on ALP couplings from the astrophysical observations and from cosmology fall into two categories, depending on whether the ALP is assumed to constitute the observed DM relic abundance or not.  In Section \ref{sec:astro}  we first discuss the constraints from stellar cooling, which do not depend on whether or not ALP is the DM. In Sec. \ref{sec:ALP:DM}  we then explore in which parts of the parameter space the LFV ALP could explain the observed DM abundance. 

\subsection{Bounds from stellar cooling}
\label{sec:astro} 
The emission of light particles inside stars can alter stellar evolution to an extent that is in conflict with observations. This leads to powerful constraints on the interactions of such light particles with matter and radiation~\cite{Raffelt:1996wa}. Our primary interest here are the ALP couplings to leptons. In this context,  ALP couplings to electrons can lead to efficient energy loss mechanisms in stars. For massless ALP (such as the QCD axion) the studies of red giants (RG)~\cite{Raffelt:1994ry,Viaux:2013lha} and white dwarfs (WD)~\cite{Bertolami:2014wua},  give roughly comparable bounds,
which at 95\% C.L. are\footnote{Interestingly, several stellar systems exhibit hints of non-standard energy losses. The global fit performed in Ref.~\cite{Giannotti:2017hny} finds that an axion/ALP solution 
to these anomalies with a coupling 
$5.4 \times 10^9 \GeV~\lesssim~ F^A_{ee}~ \lesssim~ 8.1 \times 10^9 \GeV \, ~~(\text{1$\sigma$ range})$
is preferred at the $3\sigma$ level over the case of only the standard energy loss through neutrinos.}
\begin{align}
\label{eq:stars}
F^A_{ee} & \gtrsim 4.6 \times 10^9 \GeV  \, \, {\rm (WD)}  \,, & F^A_{ee} & \gtrsim 2.4 \times 10^9 \GeV  \, \, {\rm (RG)} \, .
\end{align}
In both cases the dominant cooling mechanism is ALP bremsstrahlung in electron--nucleus scattering,  $e^- + N \to e^-+ N + a$. This dominates over the Compton process, $\gamma+e^-\to e^-+ a$, and electron-electron bremsstrahlung, $e^- + e^- \to e^-+ e^- + a$, which are relevant only when electrons are non-degenerate~\cite{Raffelt:1996wa}.

For non-negligible ALP masses, the cooling rates are expected to be Boltzmann-suppressed. Following Ref.~\cite{Raffelt:1990yz}, we estimate the resulting constraints on massive ALPs by rescaling the energy loss rates with  the ratio $R (m_a, T)$ of ALP energy densities ${\cal E}_a$ for the massive and massless case
\begin{align}
R (m_a, T) & \equiv {\cal E}_a (m_a, T)/{\cal E}_a (0, T) \, . 
\label{Rdef}
\end{align}
The energy densities are given by
\begin{align}
\label{eq:E:dependence}
 {\cal E}_a (m_a, T) = \frac{1}{2 \pi^2} \int_{m_a}^\infty  \frac{E^2 \sqrt{E^2-m_a^2}}{e^{E/T} -1} \, dE = 
\begin{cases} 
\frac{\pi^2}{30} T^4 & m_a \ll T,
 \\
\frac{1}{(2 \pi)^{3/2}} T^4 \left( \frac{m_a}{T} \right)^{5/2}  e^{- m_a /T} & m_a \gg T. 
\end{cases} \, 
\end{align}
Since the cooling rates scale with $(F^A_{ee})^{-2}$, the constraints on massive ALPs in Fig.~\ref{fig:money} are obtained from the bounds on $F_{ee}^A$ for the massless ALP by rescaling them with the factor $\sqrt{R (m_a,T)}$. Because of the Boltzmann suppression  the star cooling bounds rapidly shut off for ALP masses above $m_a \approx 2 T$, where $T_{\rm RG} \approx 10^8 \rm K \approx 8.6 \, {\rm keV}$ and  $T_{\rm WD} \approx 10^7 \rm K \approx 0.8 \, {\rm keV}$. 

For heavier ALPs the relevant astrophysics constraints are due to neutron star cooling in supernova (SN) explosions, since the nascent proto-neutron star (PNS) reaches a temperature of order $30 \MeV$ a few  seconds after the start of the supernova explosion~\cite{Burrows:1986me, Bethe:1990mw}. In order to estimate the SN bound on $F_{aa}^A$ we use the expression for the energy loss rate per unit mass, $\eps$,  through electron-nucleon bremsstrahlung under highly degenerate conditions~\cite{Raffelt:1989zt, Raffelt:1996wa}, 
\begin{align}
\eps = \frac{\pi}{15} \alpha_{\rm em}^2 \frac{T^4}{m_n (F^A_{ee})^2} Y_p\, \mathcal{I} \, . 
\label{eps}
\end{align}
Here, $Y_p = n_p/n_B$ is the number density of protons relative to the baryon number density, while $ \mathcal{I}$ is the angular integral that includes plasma screening effects,
\begin{align}
\label{eq:F:angular}
\mathcal{I}& = \int \frac{d \Omega_2}{4 \pi}\frac{d \Omega_a}{4 \pi} \frac{\left( 1 -\beta_{\rm F}^2 \right) \left[ 2 \left( 1- c_{12} \right) - \left( c_{1a} - c_{2a} \right)^2 \right] }{\left( 1- c_{1a} \beta_{\rm F} \right) \left( 1- c_{2a} \beta_{\rm F} \right) \left( 1- c_{12}  \right)  \left( 1- c_{12} + \kappa^2_{\rm DH} \right)} \, .
\end{align}
It depends on the electron velocity at the Fermi surface, $\beta_{\rm F} = p_{\rm F}/E_{\rm F} = p_{\rm F}/\sqrt{p_{\rm F}^2 + m_e^2}  \approx 1$, while  $c_{12}, c_{1a}, c_{2a}$ are the cosines of the angles between the 3-momenta ${\bf p}_1$ (${\bf p}_2$) of the incoming (outgoing) electron  and the ALP 3-momentum ${\bf p}_a$, respectively.  The screening effects enter through the parameter $\kappa_{\rm DH}^2 = k_{\rm DH}^2/2 p_{\rm F}^2$, where $k_{\rm DH} $ is the Debye screening scale. Note that the PNS with temperature $T_{\rm NS} \approx 30 \MeV$ and mass density $\rho \approx \rho_{\rm nuclear}$ can be treated as composed of weakly coupled degenerate plasmas. The electron screening can then be neglected with respect to the proton screening, giving $k_{\rm DH}^2 = 4 \pi \alpha_\text{em} n_p/T_{\rm NS}$. Since the main contribution to the angular integral \eqref{eq:F:angular} comes from the forward direction, $c_{12} = 1$ and $c_{1a} = c_{2a}$, the integral is well approximated by the simplified form obtained by setting $c_{2a} = c_{1a}$ in the denominator. 
In the ultra-relativistic case then~\cite{Raffelt:1990yz} 
\begin{align}
\mathcal{I} = \frac{2 + \kappa^2_{\rm DH}}{2} \log \frac{2 + \kappa^2_{\rm DH}}{\kappa^2_{\rm DH}}  - 1\, .
\label{F}
\end{align}
For the numerical evaluation we use $n_B = 1.4 \times 10^6 \MeV^3,~Y_p = 0.2,~p_F =  204 \MeV$, so that $\kappa_{\rm DH} = 0.10$ and $ \mathcal{I}= 4.3$. This implies for the energy loss rate
\begin{align}
\eps = 1.2 \times 10^{20} \, \frac{{\rm erg}}{{\rm g \, s}} \left( \frac{10^7 \GeV}{F^A_{ee}} \right)^2 \, .
\end{align}
Imposing the crude bound on the energy loss of $\eps \lesssim 10^{19} \, {\rm erg} \, {\rm g}^{-1} \, {\rm s}^{-1}$~\cite{Raffelt:1990yz}, leads in the case of a massless ALP to the bound
\begin{align}
\label{eq:SNbound}
F^A_{ee} & \gtrsim 3.4 \times 10^7 \GeV  \, \, {\rm (SN1987A)}  \, .
\end{align}
For massive ALP we rescale this bound by the factor $R (m_a, T)$ in Eq.~\eqref{Rdef} with $T \to T_{\rm NS} \approx 30 \, {\rm MeV}$. 
Note that for heavy ALPs the $m_a^{5/2}$ dependence in the energy loss rate \eqref{eq:E:dependence} appreciably counteracts the $\exp(-m_a/T)$ suppression, so that the SN bounds on $F_{aa}^A$ are important up to $m_a\approx 200$ MeV, cf.~Fig.~\ref{fig:money}.

 A qualitatively different regime is obtained for small values of $F_{ee}^A$. For small enough $F_{ee}^A$ the ALP interactions  become so strong that the ALP remains trapped within the stellar material, in which case there are no bounds on $F_{ee}^A$ from the stellar cooling constraints. Following Ref.~\cite{Raffelt:1990yz}, we estimate this limiting values of $F_{ee}^A$ for RG, WD and NS systems by requiring that the mean free path  $\lambda$ of the ALP is smaller than the corresponding stellar effective radius $R_0$, cf.~Table~\ref{tab:astroinput}. The mean free path is calculated from ALP decay and absorption rates, 
\begin{align}
\lambda^{-1} = \beta^{-1}   \Gamma_{\rm abs} + (\beta \gamma)^{-1} \Gamma_{\rm decay} \, .
\end{align}
The absorption rate is approximately given by the total energy loss rate per volume, $\rho \epsilon$, divided by the ALP energy density, ${\cal E}_a$ (even if ALPs are not in thermal equilibrium) 
\begin{align}
\Gamma_{\rm abs} = \rho \eps/{\cal E}_a \, . 
\end{align}
As this estimate just follows from the principle of detailed balance, the absorption rate is independent of the ALP mass. We can therefore use for the energy loss per unit mass, $\eps$, the result in Eq.~\eqref{eps}, while ${\cal E}_a = {\cal E}_a (0,T)$ in Eq.~\eqref{Rdef}.  The dependence of $\lambda$ on the ALP mass dominantly enters through the kinematical factors $\beta$ and $\gamma$ which are given by the integrals 
\begin{align} 
\beta & = \frac{\left\langle(E^2 - m_a^2)\right\rangle_T}{ \left\langle E \sqrt{ E^2 - m_a^2}\right\rangle_T}\quad\ ,\quad \beta \gamma=\frac{\left\langle E/m_a\left(E^2 - m_a^2\right) \right\rangle_T}{\left\langle E \sqrt{ E^2 - m_a^2}\right\rangle_T}\ ,
\end{align}
where $\langle\dots \rangle_T$ denotes a thermal average using the Bose-Einstein thermal distribution of ALPs. In the  limit of large masses, $m_a \gg T$, one finds $ \beta \approx \gamma \approx  \sqrt{8/\pi} \sqrt{T/m_a}$. 

For the decay rates, $\Gamma_{\rm decay}$, we only consider decays to electrons and photons induced by $F^A_{ee}$. The resulting decay widths are given in Sec.~\ref{sec:notation} (we set $E_{\rm UV}=1$ in these expressions, for definiteness). The contributions from $\gamma\gamma$ decay channel are always subdominant -- for  low ALP masses, where the decays to photons dominate $\Gamma_{\rm decay}$, the absorption rate is more important in determining $\lambda$, while for higher ALP masses the decays to electrons or even to muons dominate over decays into photons. 

We have used the inputs in Table~\ref{tab:astroinput} (taken from Ref.~\cite{Raffelt:1996wa}) to derive the bounds in the trapping regime in Fig.~\ref{fig:money}. Note that for WDs the plasma is typically strongly coupled, so that the Debye screening given in Eq.~\eqref{F} is not an appropriate description. In these case we simply use $ \mathcal{I}\approx 1$, which provides a good approximation~\cite{Raffelt:1996wa, Nakagawa:1987pga, Nakagawa:1988rhp}.

We have also included the recent bounds on ALP couplings to muons obtained in Ref.~\cite{Bollig:2020xdr} from SN1987A (see also Ref.~\cite{Croon:2020lrf}). In contrast to our rough description of the PNS, these authors have used dedicated simulations which lead to the robust (and conservative) bound of $F_{\mu \mu}^A \ge 1.3 \times 10^8 \GeV$ for a massless axion.  As for the case of the bounds on electron couplings from WDs and RGs, we simply rescale this bound by  $\sqrt{R (m_a,T)}$ defined in Eq.~\eqref{Rdef} (with $T \approx 30 \MeV$) in order to account for non-zero axion masses. Note that the bound on muons is stronger than the SN bound on electron couplings in Eq.~\eqref{eq:SNbound}, because energy loss through the Compton process for non-relativstic and non-degenerate muons is much more efficient than the same process for highly relativstic and degenerate electrons, see Refs.~\cite{Hatsuda:1987ba, Raffelt:1996wa}. 

\begin{table}[t]
\renewcommand{\arraystretch}{1.2}
\centering
\begin{tabular}{lcccccc}
\hline\hline
& $T$ & $\rho$ &  $p_F$ &  $Y_p$ &  $R_0$ \\ \hline
RG & $8.6 \, {\rm keV} $ & $4.3 \MeV^4$ & $409 \, {\rm keV} $ & $0.5$ & $10^4 \, {\rm km} $\\
WD & $0.8 \, {\rm keV} $ & $7.7 \MeV^4$ & $495 \, {\rm keV} $ & $0.5$ & $10^3 \, {\rm km} $\\
PNS (SN) & $30 \, {\rm MeV} $ & $1.3 \times 10^9 \MeV^4$ & $204 \, {\rm MeV} $ & $0.2$ & $10 \,\, {\rm km} $\\
\hline\hline
\end{tabular}
\caption{\label{tab:astroinput} Numerical inputs used in the evaluation of the stellar cooling bounds.}
\end{table}

Finally, we comment on the bound on BR($\mu \to e a)$ from SN1987A. This decay contributes to the cooling of the PNS, with a cooling rate that is  given by (see Ref.~\cite{MartinCamalich:2020dfe})
\beq
\begin{split}
\label{eq:SN}
\eps =&\frac{m_\mu^3\Gamma(\mu\to e a)}{\pi^2 \rho (m_\mu^2-m_n^2)} \int^\infty_0 p_\mu \,dp_\mu \int^{p_e^{\rm max}}_{p_e^{\rm min}}p_e \, dp_e \frac{E_\mu-E_n}{E_\mu E_e} f_\mu(E_\mu) (1-f_e({E_e})),
\end{split}
\eeq
where $p_e^{\rm max}$ ($p_e^{\rm min}$) is the maximal (minimal)  electron momentum in the $\mu \to e a$ decay, if $\mu$ has momentum $p_\mu=|\pmb{p}_\mu|$, and $E_i^2 = p_i^2 + m_i^2$ are the energies in the PNS's rest frame. Moreover, $f_i (E_i)=1/\big(1+\exp\big({\frac{E_i-\mu_i}{T}}\big)\big)$
are the Fermi distribution functions with $\mu_i$ the chemical potentials, which  is the only non-trivial input in Eq.~\eqref{eq:SN}. Using the numerical values $\mu_\mu \approx 41 \MeV$ and $\mu_e \approx 190 \MeV$  which we obtained as described below, the cooling rate becomes
\begin{align}
\eps \approx 10^{19} \, \frac{{\rm erg}}{{\rm g \, s}} \left( \frac{{\rm BR} (\mu \to e a)}{4 \times 10^{-3}} \right) \, .
\label{epsmu}
\end{align}
The resulting bound on ${\rm BR} (\mu \to e a)$  is therefore about three orders of magnitude weaker than the constraints from laboratory experiments. 

It is instructive to compare this result with the case where Pauli blocking in Eq.~\eqref{eq:SN} is neglected (i.e. $f_e(E_e) \to 1$), in which case the energy loss rate would be approximated by
\begin{align}
\eps  \simeq \frac{n_\mu}{\rho} \left( m_\mu - m_e \right) \Gamma(\mu \to ea) = \frac{Y_\mu}{m_n} \left( m_\mu - m_e \right) \Gamma(\mu \to ea)  \, . 
\end{align}
Using $Y_\mu \approx  2.9 \times 10^{-3} \simeq Y_e \, \exp{[-(m_\mu - m_e)/T]}$ (see below), the resulting bound on ${\rm BR} (\mu \to e a)$ is about a factor 50 larger than the exact result, indicating that Pauli blocking is an important effect. This can be  understood by the large Fermi energy of the electrons, $E_F \approx p_F \approx 200 \MeV$, which implies that muons at rest can hardly decay because almost all the relevant electron levels are filled. 

The above values of $\mu_\mu, \mu_e$ and $Y_\mu$ were obtained by treating the PNS  as a non-interacting Fermi gas, where all particles are in thermal and chemical equilibrium (see also Ref.~\cite{MartinCamalich:2020dfe}). At given temperature the number density of a given particle then depends only  on its chemical potential, which in turn can be written as a linear combination of the chemical potentials associated with the conserved quantum numbers, i.e., charge $Q$, baryon number $B$ and the individual lepton numbers $L_i$.  Furthermore, we assumed that neutrinos are trapped inside the PNS, which implies that the muon number density vanishes, $Y_{L_\mu} = Y_\mu + Y_{\nu_\mu} = 0$~\cite{Prakash:1996xs}. For the electron number density (relative to the baryon number density $n_B$) we used $Y_{L_e} = 0.3$~\cite{Raffelt:1996wa}. With these three input values and the constraint of charge neutrality, $n_Q = 0$, one can numerically solve for the 4 unknown chemical potentials $\mu_Q, \mu_B, \mu_{L_e}, \mu_{L_\mu}$ and thus obtain number densities $n_i$ and chemical potentials $\mu_i$ for all the involved particles. The tau leptons could be trivially included, taking $Y_{L_\tau} =0$. However, unlike $Y_\mu$, the tau number density is negligible, $Y_\tau \approx 0$. The  electron's Fermi energy of about $200 \MeV$ is just large enough to excite a relatively substantial muon population via electron scattering, while taus are simply too heavy. Note that this rather crude treatment of the PNS's thermodynamics appears to be consistent with the results from the dedicated simulations in Ref.~\cite{Bollig:2020xdr}, since we obtain similar bounds on $F^A_{\mu \mu}$ using the expressions for the energy loss rate through Compton (as appropriate for muons) with the above input.  

Finally, the derived SN bounds depend crucially on the  SN explosion mechanism.  If the SN1987A was not triggered by the canonical delayed neutrino mechanism but is rather due to the collapse-induced thermonuclear explosion, there would be no resulting bounds on the ALP couplings \cite{Bar:2019ifz}. However, this interpretation would be disfavored, if the possible ALMA detection of a compact object in the remnant of SN 1987A~\cite{Cigan:2019shp}, consistent with the neutron star~\cite{Page:2020gsx}, is confirmed.

\subsection{ALP Dark Matter}
\label{sec:ALP:DM} 
\begin{figure}[t]
	\centering
	\includegraphics[width=0.95\linewidth]{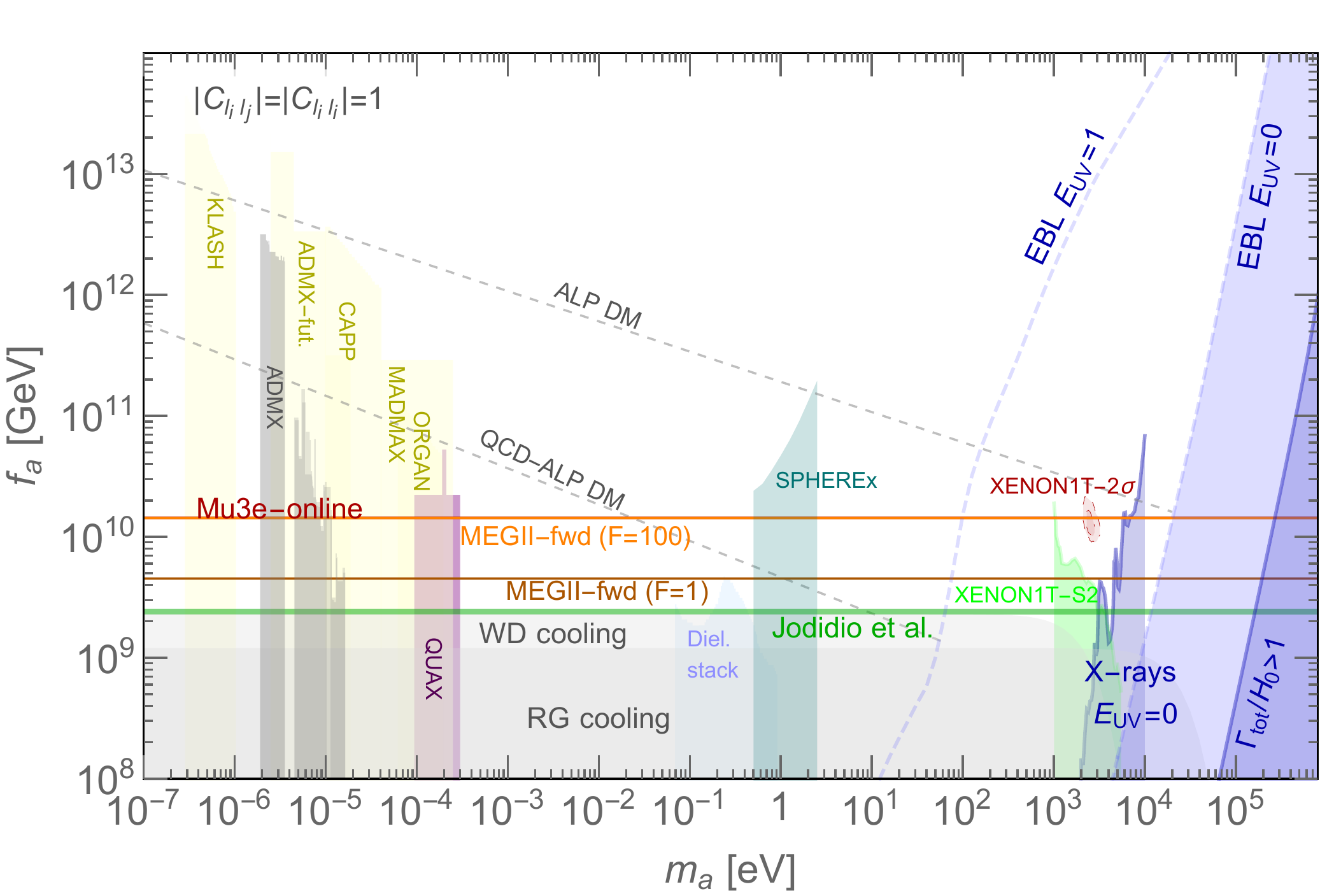}
	\caption{The impact of present and future $\mu\to e a$  searches compared to other light ALP DM searches, taking $E_{\text{UV}}=1,0$ as two representative examples. The {\bf green solid} line shows the current best bound on the isotropic LFV ALP~\cite{Jodidio:1986mz}, the {\bf (dark) orange thin} line gives our MEGII-fwd projection assuming $F=100$ focusing enhancement (no focusing). The {\bf dark red} line (overlapping with the orange thin line) shows the sensitivity of Mu3e-online analysis~\cite{Perrevoort:2018okj}. In the blue region enclosed by the {\bf solid blue} line the ALP decays within the present Hubble time, while the region to the right of the {\bf dashed blue} line is excluded by the extragalactic diffuse background light measurements for $E_{UV}=0,1$. We also show the X-rays constraints for $E_{\text{UV}}=0$~\cite{Boyarsky:2006hr,Figueroa-Feliciano:2015gwa}. The red blob indicates where ALP DM could explain the XENON1T anomaly~\cite{Bloch:2020uzh}. The {\bf dashed gray lines} denote two scenarios where the observed DM relic abundance is due to ALPs  produced trough the misalignment mechanism. On the upper line the ALP mass is temperature independent, cf. Eq.~\eqref{eq:relicmaconst}, while on the lower line the temperature dependence is parametrically  similar to the one for the QCD axion, cf. Eq.~\eqref{enhancement}. The {\bf gray shaded} regions are excluded by the star cooling bounds, and the ADMX data~\cite{Braine:2019fqb,Boutan:2018uoc,Du:2018uak}. The light green region is excluded by the S2 only analysis of XENON1T~\cite{XENON:2019gfn} and Panda-X~\cite{Fu:2017lfc}. The {\bf purple shaded} region shows the future reach of axion-magnon conversion experiments such as QUAX~\cite{Barbieri:1985cp,Barbieri:2016vwg,Chigusa:2020gfs}. Regarding the coupling to photons, the {\bf cyan band} shows the future sensitivity of SPHEREx estimated in Ref.~\cite{Creque-Sarbinowski:2018ebl}, assuming ALP decay exclusively to two photons, while the {\bf yellow bands} show the future sensitivities of resonant microwave cavities such as ADMX~\cite{Shokair:2014rna}, CAPP~\cite{Petrakou:2017epq}, KLASH~\cite{Gatti:2018ojx}, and ORGAN~\cite{McAllister:2017lkb}, dielectric haloscopes such as MADMAX~\cite{TheMADMAXWorkingGroup:2016hpc} and the reach of the dielectric stack proposal~\cite{Baryakhtar:2018doz} is shown with {\bf light blue}. }
	\label{fig:moneyDM}
\end{figure}

Next, we explore under what circumstances the LFV ALP is a viable DM candidate, with the correct DM relic abundance. This will lead us to two important conclusions. Firstly, since the expected sensitivity of future LFV experiments is $f_a\gtrsim 10^{10}\text{ GeV}$, well above the present astrophysical bounds,  the LFV experiments will explore a part of the parameter space of an ultralight ALP DM. Secondly, the LFV experiments cannot resolve ALP masses below 1 MeV (i.e. the typical experimental mass resolution). Below this mass range pinpointing the ALP mass will require experiments that search for ALP DM using other means, through its couplings to photons and/or electrons. 

The minimal requirement for ALP to be the DM is that it is stable on timescales longer than the lifetime of the Universe. Assuming the $a\to \gamma\gamma$ decay dominates, this translates into the following constraint, 
\begin{equation}
\frac{H_0}{\Gamma_{\text{tot}}}=H_0\tau_a>1,\quad\text{where}\quad H_0\tau_a\simeq 5.4\left(\frac{1}{E_{\text{eff}}^2}\right)^2\left(\frac{10\text{ keV}}{m_a}\right)^3\left(\frac{f_a}{10^{10}\text{ GeV}}\right)^2.
\end{equation}
Taking  $f_a=10^{10}\text{ GeV}$ as a reference value, this means that the ALP DM probed in LFV experiments must have a mass below 10 keV.  If other decay channels, such as $a\to \nu_i\bar \nu_j$, are appreciable, then the above bound on the ALP mass is correspondingly lowered (this is for instance the case for the majoron, see Sec. \ref{sec:majoron}). For the rest of this section we will assume that the ALP decay channels apart from $a\to \gamma\gamma$ can be neglected. 

The $a\to \gamma\gamma$ decays  contribute to the extragalactic background light (EBL) and may be bounded by EBL measurements. The ALP decay results in a line at frequency $\nu_a=1.2\times 10^{14}\, m_a/\text{eV} \text{ Hz}$ with intensity 
\begin{equation}
\nu_a I_{\nu_a}=\frac{5\times 10^{-3}\text{ W} \text{m}^{-2} \text{sr}^{-1}}{\tau_{\gamma\gamma}H_0}\simeq 6\times10^{-13}\text{ W} \text{m}^{-2} \text{sr}^{-1} \left(\frac{m_a}{1\text{ eV}}\right)^3\left(\frac{10^{10}E_{\text{eff}}\text{ GeV}}{f_a}\right)^2.\label{eq:intensity}
\end{equation}
A conservative bound on the decay width $\Gamma(a\to\gamma\gamma)$
is obtained by requiring that the line intensity in Eq.~\eqref{eq:intensity} is less than what is observed at that frequency~\cite{Creque-Sarbinowski:2018ebl}. An updated map of the EBL observations at different frequencies can be found in Ref.~\cite{Hill:2018trh}. For instance, the observed EBL intensity in the optical band is $10^{-8}\text{ W} \text{m}^{-2} \text{sr}^{-1}$, constraining the axion width well below $H_0$. Converting the EBL constraints to a bound on $f_a$ gives the light blue exclusion region in Fig.~\ref{fig:moneyDM} for for $E_{\rm UV}=1$ and $E_{\rm UV}=0$. Stronger bounds can be obtained from the measurements of the X-ray microcalorimeters in the XQC rocket~\cite{Boyarsky:2006hr,Figueroa-Feliciano:2015gwa}. These constraints will be further improved by future X-ray missions such as Athena~\cite{Barret:2013bna} or by future line-intensity mapping campaigns~\cite{Creque-Sarbinowski:2018ebl}. Particularly relevant for the ALP parameter space is the SPHEREx project~\cite{Dore:2014cca}, a funded  two-year mission by NASA with a planned launch in 2023, that will probe optical and near infrared frequencies corresponding to $m_a\sim \text{ eV}$. The SPHEREx reach is denoted as green region in Fig. \ref{fig:moneyDM}.

The ALP that satisfies the stability and decaying DM bounds could be a viable DM candidate. The main production mechanism in the allowed region of parameter space in Fig.~\ref{fig:moneyDM} is the misalignment mechanism, first discussed in the context of the QCD axion in Refs.~\cite{Preskill:1982cy,Abbott:1982af,Dine:1982ah} and then generalized to a generic ALP in Ref.~\cite{Arias:2012az}.\footnote{The production of hot ALPs through freeze-in via lepton annihilation, $\ell^+ \ell^-\to a \gamma$, or lepton-photon collisions, $\ell^{\pm}\gamma\to e^{\pm} a$, could lead to a too large contribution to $\Delta N_{\text{eff}}$. It is easy to check that this contribution is in fact negligible in the parameter space in Fig.~\ref{fig:moneyDM}, once the present bounds on the axion-electron couplings from stellar cooling are taken into account.  (A systematic study of hot ALPs production through muon or tau couplings in the case where couplings to electrons are switched off has been performed in Ref.~\cite{DEramo:2018vss}.)} If inflation occurred below the scale of ALP global symmetry breaking, the initial misallignment of the ALP, $a_0$, is frozen by the inflationary dynamics and acts as the initial condition. Conventionally, it is parametrized in terms of an angular variable, $a_0=f_a\theta_0$, where $\theta_0\in [0,\pi)$. As long as the Hubble expansion rate is large, $H>m_a$, the field is frozen at its initial value $a_0$. At temperature $T_{\rm osc}$, when $H(T_{\text{osc}})\simeq m_a$, the field starts to oscillate and produces the ALP number density  $n_a(T_{\text{osc}})=\tfrac{1}{2}m_a^2a_0^2$, which then expands adiabatically until the present time. For the case when ALP oscillations occur during the radiation dominated epoch the resulting ALP relic abundance is (see also \cite{Blinov:2019rhb}) 
\begin{equation}
\Omega_a^{T\text{-indep.}}h^2=0.12\times10^{-2} \sqrt{\frac{m_a}{\text{eV}}} \left(\frac{f_a}{10^{10}\text{GeV}}\right)^2\left(\frac{\theta_0}{\pi}\right)^2 \left(\frac{90}{g_*(T_{\text{osc}})}\right)^{1/4}.\label{eq:relicmaconst}
\end{equation} 
Since the ALP mass is bounded from above by EBL constraints, the future reach of  LFV searches (i.e. $f_a\approx 10^{10}$ GeV) will probe a region of parameter space where the production from misalignment does not suffice to obtain the total observed DM abundance ($\Omega_{\rm DM}h^2\simeq0.12$ \cite{Ade:2015xua}) with an ALP mass independent on temperature. The relation between $m_a$ and $f_a$ that leads to the observed DM abundance for $\theta_0\sim1$ for a temperature independent ALP mass is corresponds to the upper gray dashed line in Fig.~\ref{fig:moneyDM}. Below this line the ALP DM produced through the misalignment mechanism is under-abundant, while above this line a smaller value of $\theta_0$ is needed in order to obtain $\Omega_{\text{DM}}h^2=0.12$. 

An interesting alternative possibility that leads to enhanced misalignment production is if the ALP mass comes from a dynamical mechanism like the one of the QCD axion. At zero temperature the ALP mass is given by $m_a=\Lambda^2/f_a$, while at finite temperature the mass is suppressed, and is given by $m_a(T)=m_a(\Lambda/T)^b$, where $b=4$ in QCD (an expression for $b$ in a general gauge theory can be found in Ref.~\cite{Gross:1980br}). The relic density from misalignment for this case has been studied in Ref.~\cite{Arias:2012az}, and more recently in Ref.~\cite{Blinov:2019rhb}, and is given by  
\begin{equation}
\Omega_a^{T\text{-dep}}=\Omega_a^{T\text{-indep.}}\left(\frac{2+b}{2}\right)^{\frac{3+b}{2+b}}\left(\frac{5}{8\pi}\frac{M_{\text{Pl}}}{f_a}\sqrt{\frac{90}{g_*(T_{\text{osc}})}}\right)^{\frac{b}{4+2b}}.
\label{enhancement}
\end{equation}
The two additional factors on the r.h.s. enhance the relic abundance with respect to the  result in Eq.~\eqref{eq:relicmaconst}. 
In Fig~\ref{fig:moneyDM} we show that for $b=4$ this enhancement is large enough that the correct DM relic abundance is obtained in a large region of parameter space that will be tested by future LFV searches. The ALP abundance can be enhanced even further at small decay constants by tuning the ALP initial condition very close to $\theta_0=\pi$, such that non-linearities dominate the production~\cite{Arvanitaki:2019rax}, or by mixing of the ALP with a dark photon in a presence of a magnetic field in the early Universe~\cite{Hook:2019hdk}. In summary, these different production mechanisms can make the ALP abundance match the current DM abundance in the whole parameter space shown in Fig.~\ref{fig:moneyDM} which is not excluded by present constraints.

Parts of the LFV ALP parameter space will be probed by other means. In Fig.~\ref{fig:moneyDM} we show two types of such probes, based either on axion couplings to electrons or to photon. All of these probes require the ALP to be the DM, and assume that the ALP is responsible for the full DM relic abundance. Note that in this case, for the range of masses shown in Fig.~\ref{fig:moneyDM}, the description of ALP in terms of classical background field is justified in the early Universe, since there are many ALPs inside a single de Broglie volume.

The ALP couplings to electrons can then lead to interesting constraints from electron recoil experiments where the ALP energy gets absorbed in the detector. The energy threshold of the DM experiment translate directly to the lower ALP mass that these experiment can probe.  We show the current constraints from Panda-X~\cite{Fu:2017lfc} and from the S2-only analysis of XENON1T \cite{Aprile:2019xxb}. The fit of the XENON1T anomaly derived in \cite{Bloch:2020uzh} is also shown. Further improvements at lower ALP mass are expected from low threshold experiments like SENSEI~\cite{Bloch:2016sjj}. The efficient axion-magnon conversion in an experiment such as QUAX~\cite{Barbieri:1985cp,Barbieri:2016vwg} could probe a portion of the ALP parameter space in the $m_a\sim 10-50\,\mu\text{eV}$ window. The light purple region in Fig.~\ref{fig:moneyDM} shows the future reach of QUAX derived in Ref.~\cite{Chigusa:2020gfs} (the present sensitivity in Ref.~\cite{Crescini:2020cvl} is still outside the plotted range in $f_a$).  At higher masses the low threshold DM absorption experiments based on existing technology are generically weaker than the stellar cooling bounds, even if improvements can be foreseen with future technology under optimistic conditions~\cite{Marsh:2018dlj,Mitridate:2020kly}. 

If the ALP coupling  to photons is not suppressed, the standard searches for the QCD axion will cover significant parts of the parameter space, as seen in Fig.~\ref{fig:moneyDM} for the case of $E_{\rm UV}=1$. The gray region around $0.2\,\mu\text{eV}$ is excluded by  current ADMX data \cite{Braine:2019fqb,Boutan:2018uoc,Du:2018uak}. The future axion haloscope campaigns will explore the ALP mass region between $0.2-20\,\mu\text{eV}$. In Fig.~\ref{fig:moneyDM} we show in yellow the estimated sensitivities of CAPP \cite{Petrakou:2017epq}, KLASH \cite{Gatti:2018ojx}, ORGAN \cite{McAllister:2017lkb}, MADMAX~\cite{TheMADMAXWorkingGroup:2016hpc} and the ADMX upgrade \cite{Shokair:2014rna}. We also include the ``dieletric stack'' proposal which could have sensivity beyond the current stellar cooling bounds between 0.1 and 1 eV, depending on the value of $E_{\text{UV}}$ \cite{Baryakhtar:2018doz}. The sensitivities of large-scale helioscopes such as IAXO \cite{Irastorza:2013dav,Armengaud:2019uso}, and light-shining-through-wall experiments such as ALPS-II \cite{Bahre:2013ywa} lie below the current stellar cooling bounds for our choice of parameters and is not shown in Fig.~\ref{fig:moneyDM}.

\section{LFV ALP models}\label{sec:models}

So far we were concerned with the model independent bounds on LFV ALP couplings, Eq.~\eqref{couplings}. In the remainder of this paper we focus instead on  several  representative models of  ALPs with LFV couplings: the LFV QCD axion, the LFV axiflavon, the leptonic familon and the majoron. These examples are representative for broad classes of models, and illustrate how flavor-violating couplings to leptons can naturally arise for PNGBs of global symmetries addressing the strong CP problem, the SM flavor puzzle, or neutrino masses.
In the first three models the LFV couplings are generated at tree level via non-universal charges of the global symmetry under which the ALP transform non-linearly while in the last one the LFV comes from loop-induced couplings. For each model above we also present the parameter space where the ALP can be a viable DM candidate as discussed in Sec.~\ref{sec:ALP:DM}.

The LFV QCD axion (Sec.~\ref{sec:DFSZQCDaxion}) and the LFV axiflavon (Sec.~\ref{sec:axiflavon}) are two explicit realizations of  the QCD axion, which elegantly solves the strong CP problem in the SM via the spontaneous breaking of a $U(1)$ Peccei-Quinn (PQ) symmetry that is anomalous under QCD. In Sec.~\ref{sec:DFSZQCDaxion} we first show that in DFSZ-like  models~\cite{Zhitnitsky:1980tq,Dine:1981rt} the QCD axion can naturally have LFV couplings while keeping the couplings to quarks flavor diagonal.  In Sec.~\ref{sec:axiflavon} we go one step further and identify the PQ symmetry with a subgroup of the flavor symmetry that gives the hierarchical masses and mixings of the SM fermions. The LFV axiflavon is obtained for the $U(2)$ flavor group, since in this case the flavor violating couplings are parametrically larger in the leptonic than the quark sector.

The Leptonic familon (Sec.~\ref{sec:familon}) is the PNGB of a $U(1)$ flavor symmetry in the leptonic sector. The spontaneous breaking of this symmetry could explain the hierarchies among the charged leptons via the Froggatt-Nielsen mechanism~\cite{Froggatt:1978nt,Leurer:1992wg,Leurer:1993gy} (for recent variations see \cite{Smolkovic:2019jow,Bonnefoy:2019lsn}). 
In the LFV familon setup the strengths of the LFV couplings depend on the texture of the PMNS matrix, as we will see in detail below. 

Our final example, the majoron, is the PNGB associated with the spontaneous breaking of the lepton number~\cite{Chikashige:1980ui, Schechter:1981cv}. In Sec.~\ref{sec:majoron} we show that in a non-minimal class of seesaw models the majoron has parametrically enhanced LFV couplings. In these theories an approximate generalized lepton number suppresses the neutrino masses~\cite{Broncano:2002rw,Raidal:2004vt, Kersten:2007vk, Abada:2007ux,Gavela:2009cd,Ibarra:2010xw,Ibarra:2011xn,Dinh:2012bp,Cely:2012bz,Alonso:2012ji}, without suppressing the majoron couplings to the SM.

\subsection{The LFV  QCD Axion}
\label{sec:DFSZQCDaxion}
The mass of the QCD axion is entirely due to the QCD anomaly, and is given by~\cite{Gorghetto:2018ocs} 
\begin{align}
\label{eq:axion:mass}
m_a = 5.691(51) \mueV \left( \frac{10^{12} \GeV}{f_a} \right) \, .
\end{align}
The value of the axion decay constraint $f_a$ therefore completely determines the mass of the QCD axion, which for all the processes we consider is effectively massless. 

Astrophysical constraints require the axion to be very weakly coupled, with a lifetime larger than the age of the universe and a mass below $3\times 10^{-2}\text{ eV}$. In this range the QCD axion is a perfectly viable cold DM candidate in large parts of the parameter space. One of the simplest scenarios for axion production is the misalignment mechanism described in Sec.~\ref{sec:ALP:DM}. In the QCD axion case the observed DM abundance is obtained for misalignment angles of order unity $\theta_0\sim 1$ with an axion decay constants $f_ a \sim 10^{(11 \div 13)}$ GeV. For smaller decay constants, within the reach of LFV experiments, the axion relic from the standard misalignment contribution is under-abundant unless non-trivial dynamics or tuning are invoked (see discussion in Sec.~\ref{sec:ALP:DM}). 

The axion couplings to fermions in Eq.~\eqref{couplings} arise from rotating the PQ current to the fermion mass basis, with unitary rotations $V^{f}$ defined by $V_{L}^{f\,\dagger} y_f V_{R}^f= y_f^{\rm diag}$. Denoting the PQ charge matrices by $X_f$, one has 
\begin{align}
C_{f_i f_j}^{V,A} & = -\frac{1}{2N} \left( V_{R}^{f\,\dagger} X_{f_R} V_{R}^f \pm V_{L}^{f\,\dagger} X_{f_L} V_{L}^f \right)_{ij}  \, ,
\label{Cdef}
\end{align}
 where $2N$ is the domain wall number. This implies that off-diagonal couplings arise when the PQ charges are not diagonal in the same basis as the Yukawa couplings, $y_f$.  Their sizes depend on the misalignment between the two bases, which is parametrized by the unitary rotations $V_{R,L}^f$. 
 We focus on the situation where the PQ charges in the quark sector are universal, so that the QCD axion only has flavor violating couplings in the lepton sector. (This is of course not the most general case. If PQ charges in the quark sector are not universal, the results from Ref. \cite{MartinCamalich:2020dfe} apply, with the bound from $K^+ \to \pi^+ a $ leading to tight constraints on $f_a$.)

In the following, we specify a DFSZ-like model of the QCD axion with LFV couplings.  The field content of the theory consists of the SM fermions, two Higgs doublets, $H_{1,2}$, and a complex scalar $S$ that is a gauge singlet. 
The model contains an anomalous global $U(1)$ PQ symmetry under which the scalar fields carry charges $X_S=1$, $X_{H_2}=2+X_{H_1}$. As a consequence, the scalar potential contains the couplings $H_2^\dagger H_1 S^2$ and $(S^\dagger S)^2$, but not, for instance, $H_1^\dagger H_2 S^2$ or $S^4$. The fermionic $U(1)_{\rm PQ}$ charges are flavor universal in the quark sector, $X_{u_{Ri}}=-X_{H_1}$, $X_{d_{Ri}}=X_{H_2}$, $X_{q_{Li}}=0$, $i=1,2,3$, while they are generation dependent in the lepton sector, such that 
the Yukawa interaction Lagrangian takes the form (here $\tilde{H}_i = i \sigma_2 H_i^*$ and $a,b=2,3$)
\begin{align}
{\cal L} & =  y^{e}_{1a} \overline{\ell}_{L1} e_{Ra} \tilde{H}_{1} + y^{e}_{a1} \overline{\ell}_{La} e_{R1} \tilde{H}_{1} + y^{e}_{ab} \overline{\ell}_{La} e_{Rb} \tilde{H}_{2} - y^{u} \overline{q}_{L} u_{R} H_{1} + y^{d}  \overline{q}_{L} d_{R} \tilde{H}_{2}  + {\rm h.c.}\, .
\label{L1}
\end{align}
The first generation leptons carry charges $X_{e_{R1}}=X_{H_1}$, $X_{\ell_{L1}}= 2$ under $U(1)_{\rm PQ}$, while the 2nd and 3rd generation leptons have $X_{e_{Ra}}=X_{H_2}$, $X_{\ell_{La}}=0$, where $a=2,3$.
In Eq.~\eqref{L1} $y^{e}_{1a}$ and $ y^{e}_{a1}$ are complex 2-vectors, $ y^{e}_{ab}$ is a complex $2\times 2$ matrix, while for simplicity we do not display the flavor indices on $3\times 3$ complex Yukawa matrices in the quark sector, $y^u, y^d$.
The forms of $y^{e}_{1a}, y^{e}_{a1},y^{e}_{ab}$ and $y^{u,d}$ are not fixed by the $U(1)_{\rm PQ}$ symmetry and are thus external to the discussion (presumably there is additional flavor dynamics that gives their form and thus the required hierarchy of SM fermion masses). 

The form of the scalar potential is assumed to be suitable to induce vacuum expectation values for all scalars, with the ratio of Higgs vevs given by $v_2/v_1 = \tan\beta\equiv t_\beta$ and $\langle S \rangle\equiv v_{\rm PQ}/\sqrt2 \gg v_i$, while $v_1^2+v_2^2=v^2$ with $v=246$ GeV the electroweak vev. The axion $a$ is then mainly contained in $S$, i.e., $S=\langle S\rangle \exp\big(i  a/ v_{\rm PQ}\big)+\cdots$, and partially in the two Higgs doublets, $H_i=\langle H_i\rangle \exp\big(i  X_{H_i} a/ v_{\rm PQ}\big)+\cdots$  (here we only show the leading dependence on $a$). Requiring that the axion is orthogonal to the Goldstone boson eaten by the $Z$ completely fixes the embedding of $U(1)_{\rm PQ}$, i.e., it fixes the PQ charge of the two Higgs doublets to be $X_{H_1}=-2 s_\beta^2$, $X_{H_2}=2 c_\beta^2$, where we used the shortened notation $s_\beta\equiv \sin\beta$, $c_\beta\equiv \cos \beta$. 

It is conventional to remove the axion from the Yukawa interactions \eqref{L1} through phase redefinitions of the SM fermion fields. Working in this basis, the axion couples derivatively to the fermions as in Eq.~\eqref{couplings}, and in addition  has couplings to gluons and photons induced by the color and EM anomalies
\begin{align}
\label{eq:La:QCDaxion}
{\cal L}_a & = \  \frac{a}{f_a} \frac{\alpha_s}{8 \pi} G_{\mu \nu} \tilde{G}^{\mu \nu} + \frac{E}{N} \frac{a}{f_a} \frac{\alpha_{\rm em}}{8 \pi} F_{\mu \nu} \tilde{F}^{\mu \nu}  +\frac{\partial_\mu a}{2 f_a} \overline{f}_i \gamma^\mu \left[ C^V_{ij} + C^A_{ij} \gamma_5 \right] f_j \, .
\end{align}
The axion decay constant is related to the PQ breaking vev,  $\langle S \rangle = \sqrt2 N  f_a$. The anomaly coefficients are given by  $2 N =- 6$ and $E/N = 4/3$. 

The couplings to fermions are given by Eq.~\eqref{Cdef}. In the quark sector the PQ charges are flavor universal, and so are the axion couplings,
  \begin{align}
  \label{eq:Cuu:Cdd}
 C^{V}_{u_i u_j} & = C^{A}_{u_i u_j} = \frac{s_\beta^2}{3} \delta_{ij} \, , & 
  C^{V}_{d_i d_j} & = C^{A}_{d_i d_j} =  \frac{c_\beta^2}{3} \delta_{ij} \, . 
 \end{align} 
In contrast, in the charged lepton sector the PQ charges are not universal and therefore the axion couplings to charged leptons depend on the unitary rotations that diagonalize the Yukawas as in Eq.~\eqref{Cdef}. It is useful to introduce the hermitian matrices
\begin{align}
\label{eq:epsilon:eLR:QCDaxion}
\eps^{e_{L}}_{ij} & \equiv (V^e_{L})^*_{1i}  (V^e_{L})_{1j} \, , & 
\eps^{e_{R}}_{ij} & \equiv (V^e_{R})^*_{1i}  (V^e_{R})_{1j} \, , 
\end{align}
 which satisfy
 \begin{align}
 0 & \le \eps^{e_P}_{ii} \le 1 \, , & \sum_i \eps^{e_P}_{ii}  & = 1 \, , &  |\eps^{e_P}_{ij}| & = \sqrt{\eps^{e_P}_{ii} \eps^{e_P}_{jj}} \, ,
 \end{align}
for $P= L,R$. In terms of these parameters one has
\begin{align}
\label{eq:Cellell}
C^V_{\ell_i \ell_j} & = \frac{1}{3} \left[ c_\beta^2 \delta_{ij} - \eps^{e_{R}}_{ij} +\eps^{e_{L}}_{ij} \right] \, , & 
C^A_{\ell_i \ell_j} & = \frac{1}{3} \left[ c_\beta^2 \delta_{ij} - \eps^{e_{R}}_{ij} - \eps^{e_{L}}_{ij} \right] \, .
\end{align}
Note that the flavor diagonal parts of the vectorial couplings, $C_{f_if_j}^V$,  Eqs. \eqref{eq:Cuu:Cdd}, \eqref{eq:Cellell}, can be set to zero through fermion field redefinitions (these would introduce couplings to EW boson field strengths as in Eq. \eqref{eq:La:QCDaxion} that are, however, not relevant for our analyses). 

To show the impact of the experimental searches for LFV processes with muons, we construct three benchmarks for off-diagonal matrices $\epsilon_{ij}^{e_L}$ and $\epsilon_{ij}^{e_R}$. To do so we first choose a particular form of the leptonic Yukawa couplings, where we assume that in the basis in which $y_{ab}^e$ is diagonal the Yukawa 2-vectors  $y_{1a}^e$, $y_{1b}^e$ have zero couplings between the 1st and 3rd generation. Within these assumptions the charged lepton mass matrix is completely fixed, apart from a single continuous parameter, $\eta$,
\begin{align}
\label{eq:mije}
m^e_{ij} & = 
\begin{pmatrix} 
0 & \eta \, m_\mu & 0 \\ 
 -m_e/\eta & m_{\rm eff} & 0 \\
0 & 0 & m_\tau 
\end{pmatrix} \, , & m_{\rm eff} \equiv \sqrt{m_\mu^2 (1-\eta^2) + m_e^2 (1-1/\eta^2)}  
 \, .
\end{align}
The parameter $\eta$ controls the size of left- and right-handed rotations. We restrict its values to the range $m_e/m_\mu \le \eta \le 1$ such that there are no unnaturally large cancellations when diagonalizing the mass matrix. Choosing three representative values of $\eta$ gives 
\begin{align}
 (V^e_{L})_{12}  & \approx \begin{cases} \frac{1}{\sqrt{2}} & \eta = \frac{1}{\sqrt{2}},
 \\  
 \sqrt{\frac{m_e}{m_\mu}} & \eta = \sqrt{\frac{m_e}{m_\mu}},
    \\   
    \frac{m_e}{ m_\mu}& \eta = \sqrt{2} \frac{m_e}{m_\mu},
      \end{cases} &
  (V^e_{R})_{12}  & \approx \begin{cases}- \frac{m_e}{ m_\mu}  & \eta = \frac{1}{\sqrt{2}},
 \\ 
- \sqrt{\frac{m_e}{m_\mu}}  & \eta =\sqrt{\frac{m_e}{m_\mu}},
    \\
    -  \frac{1}{\sqrt{2}}& \eta = \sqrt{2} \frac{m_e}{m_\mu},
       \end{cases}
\end{align}
which we take as the three representative benchmarks: the ``$V-A$'', ``$V$'' and ``$V+A$''  scenarios, respectively. As per our assumptions, the only flavor violating couplings are between the 1st and the 2nd generation leptons. 

 \begin{table}[t]
\renewcommand{\arraystretch}{1.2}
\centering
\begin{tabular}{cccc}
\hline\hline
& \multicolumn{3}{c}{Bound on $f_a$ (in GeV)} \\
 & Benchmark ``$V$'' & Benchmark ``$V+A$'' & Benchmark ``$V-A$'' \\ 
\hline
SN1987A & $9.4 \times 10^{7}$ & $9.4 \times 10^{7}$ & $9.4 \times 10^{7}$  \\
WD cooling & $1.3 \times 10^{9}$  & $9.3 \times 10^{8}$ & $9.3 \times 10^{8}$ \\
$\mu \to e\,a$ & $1.1 \times 10^{8}$  & $8.0 \times 10^{8}$  & $1.2 \times 10^{8}$ \\
\hline\hline
\end{tabular}
\caption{Bounds on the axion decay constant $f_a$ (in GeV) for the three bechmarks of the LFV axion model choosing $\beta = 1$, cf. Eqs. \eqref{eq:BM:V}-\eqref{eq:BM:V-A}. 
}
\label{tab:LFVaxion}
\end{table}
\begin{figure}[t]
	\centering
	\includegraphics[width=0.8\linewidth]{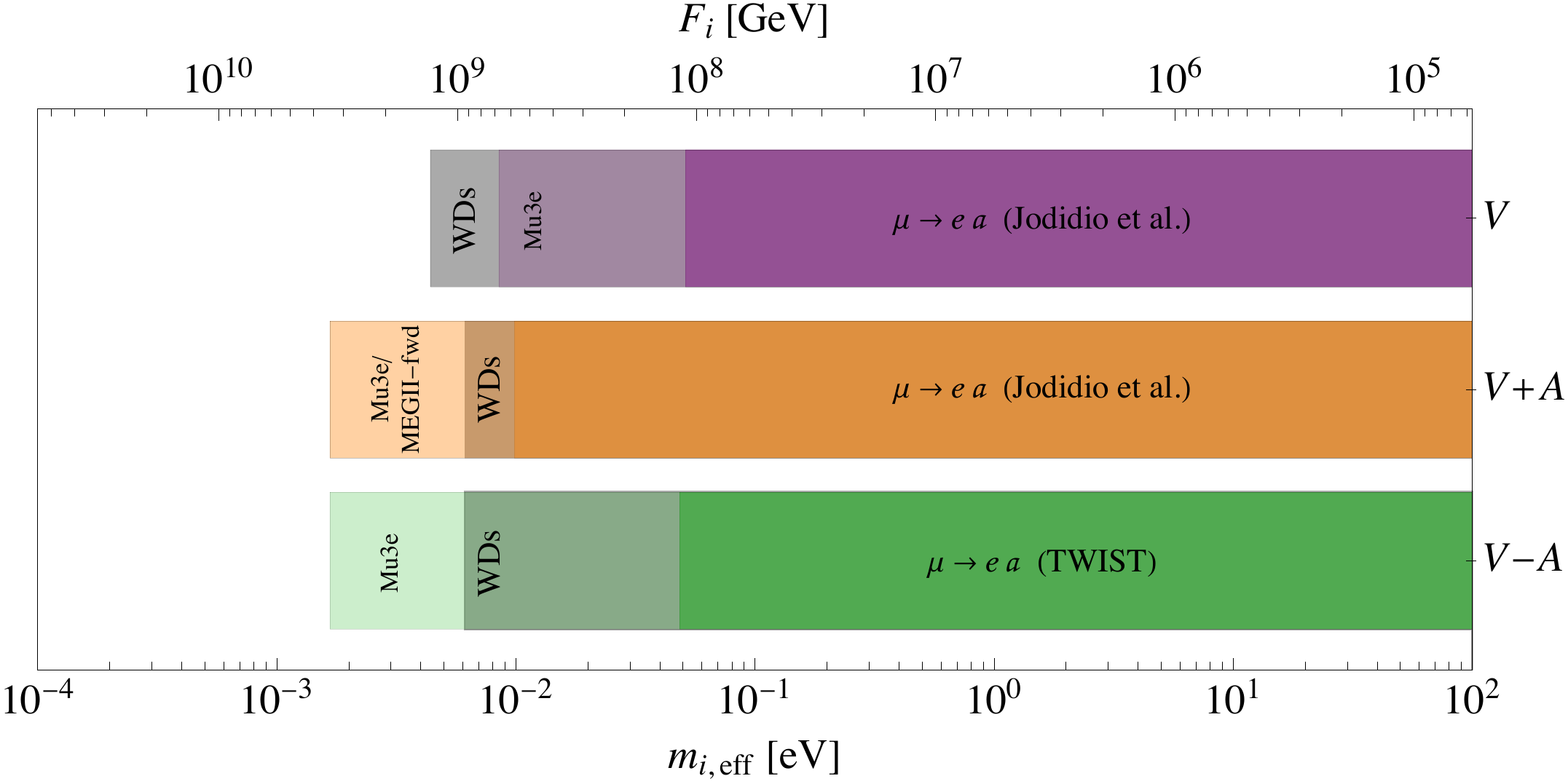}
	\caption{Present and expected future bounds on $f_a$ and $m_a$ for the LFV QCD axion in the three scenarios described in detail in the text, see also Table \ref{tab:LFVaxion} and Eqs. \eqref{eq:BM:V}-\eqref{eq:BM:V-A}. On the lower axis we indicate the corresponding values for the effective axion mass defined by $m_{i, {\rm eff}} = 4.7 \, \eV \times 10^6 \GeV/F_i$.}
	\label{fig:axion-chart}
\end{figure} 

More explicitly, the axion couplings in the three scenarios are,
\begin{itemize}
 \item   {\bf Benchmark ``$V$''} $(\eta = \sqrt{m_e/m_\mu})$ 
\begin{align}
\label{eq:BM:V}
C^V_{\mu e} & \approx  2/3 \sqrt{m_e/m_\mu}\, , & C^A_{\mu e} & = 0 \, , & C^A_{ee} \approx c_\beta^2/3 -2/3 \, , 
\end{align}
  \item  {\bf Benchmark ``$V+A$''}  ($\eta =\sqrt{2} m_e/m_\mu$) 
\begin{align}
C^V_{\mu e} & \approx C^A_{\mu e} \approx 1/6 \, , & C^A_{ee} \approx c_\beta^2/3 -1/2 \, ,
 \end{align}  
\item {\bf Benchmark ``$V-A$''} ($\eta = 1/\sqrt2$) 
 \begin{align} 
 \label{eq:BM:V-A}
 C^V_{\mu e} & \approx - C^A_{\mu e} \approx 1/6\, , & C^A_{ee} \approx c_\beta^2/3 -1/2  \, .  
 \end{align}
\end{itemize}
We can now reinterpret the model independent bounds on LFV ALPs, derived in Sections \ref{sec:mutoe}-\ref{sec:astro:cosmo}, for the three LFV QCD axion benchmarks (choosing $\beta = 1$ as a representative value). The resulting bounds on $f_a$ from $\mu\to e a$ and  from WD cooling, obtained by rescaling respectively the bounds on $F_{\mu e}^{A,V}$, $F_{ee}^{A}$ in Table \ref{tab:bounds} by the appropriate values of $C_{\mu e}^{V,A}$, $C_{ee}^A$ in the three benchmarks, are collected in Table \ref{tab:LFVaxion} and  presented graphically in Fig. \ref{fig:axion-chart}. 
\begin{figure}[t]
	\centering
	\includegraphics[width=0.49\linewidth]{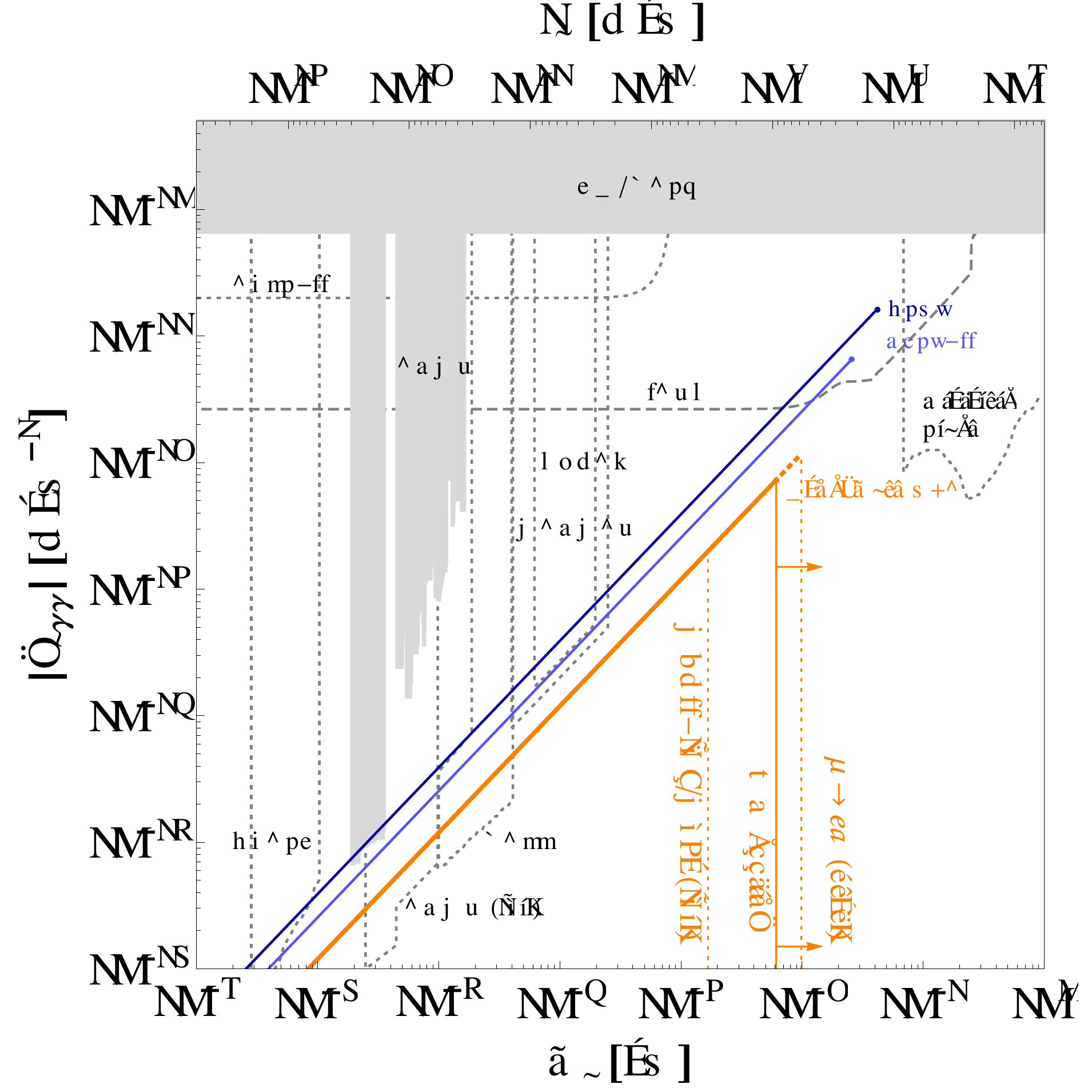}
	\hfill
	\includegraphics[width=0.49\linewidth]{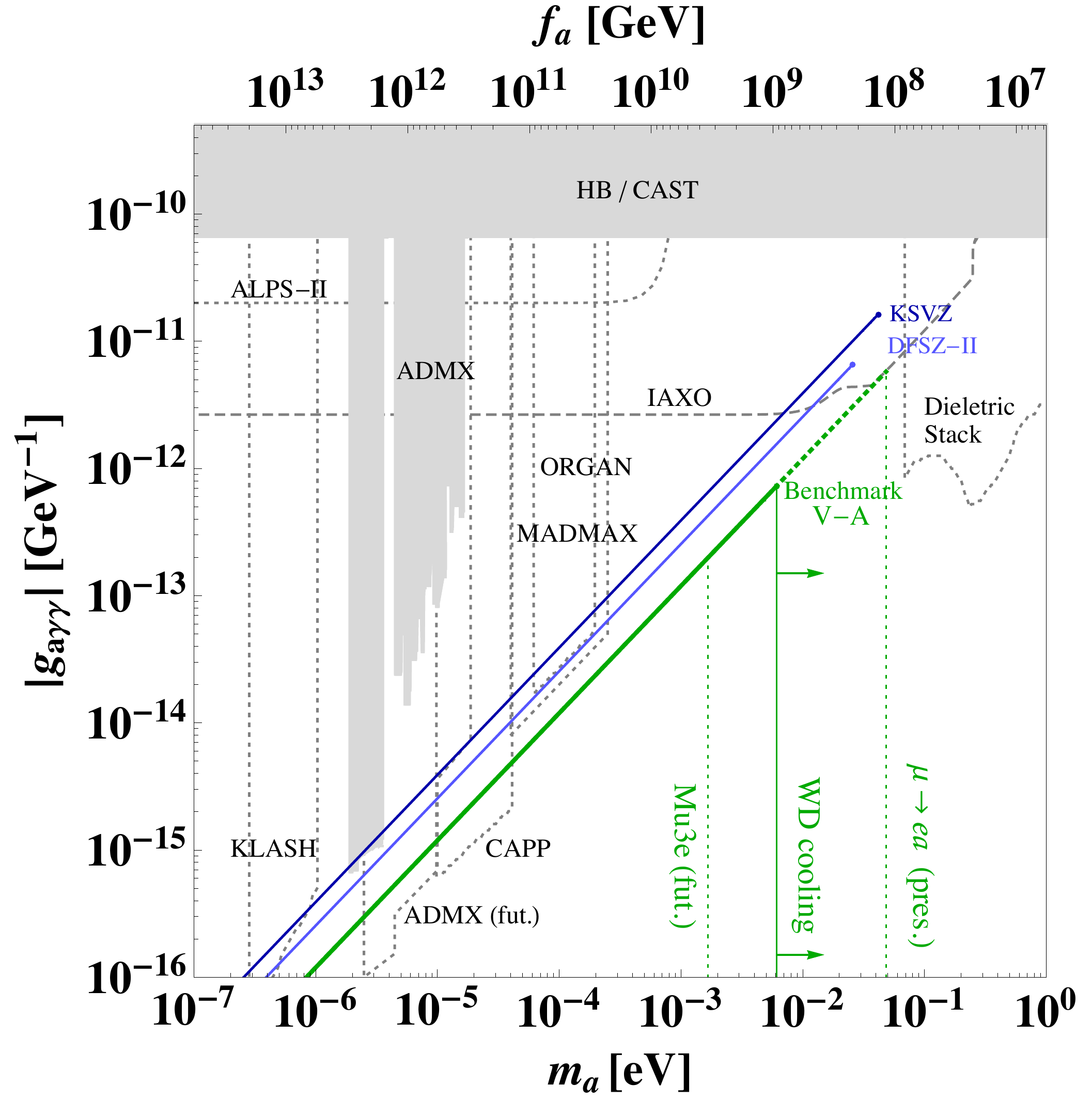}
	\caption{{\bf Left:} the ``{$V+A$}'' LFV axion corresponds to the  {\bf orange bold solid} line. {\bf Right:}  the ``{$V-A$}'' LFV axion corresponds to the {\bf green bold solid } line. The {\bf grey shaded} regions show the present bounds on LFV QCD axion couplings to photons as a function of axion mass for the two benchmarks. The {\bf grey dashed} lines denote future projected sensitivities on the photon coupling. The {\bf solid orange/green} vertical lines show the present upper bound on the axion mass from WD cooling in the two models. The {\bf dotted orange/green} vertical lines show the present bound and future reach on the   axion mass from LFV experiments. For a comparison we also show lines corresponding to the standard KSVZ ({\bf dark blue}) and DSFZ-II ({\bf blue}) models. The former one is limited by the SN1987A bound, the latter one by WD cooling. For both the DSFZ-II and our LFV QCD axion models we set $\beta=1$. See the main text for details.}
	\label{fig:axion-plot}
\end{figure}

The SN1987 bound is modified with respect to the one discussed in Sec.~\ref{sec:astro} due to the axion couplings to quarks and gluons that then result in the axion couplings to nucleons (due to smaller scattering cross sections the processes involving electrons lead only to subleading corrections). 
Adopting the treatment of Ref.~\cite{SNbound}, the relevant bound is on the effective coupling to nucleons, 
\begin{align}
C_N \equiv \sqrt{C_n^2 + 0.29 \, C_p^2 + 0.27 \, C_p C_n } \, , 
\end{align}
 where $C_{p,n}$ are the axion couplings to protons and neutrons, respectively. Using the expressions in Ref.~\cite{diCortona:2015ldu} along with the values of couplings to quarks and gluons in Eqs.~\eqref{eq:La:QCDaxion}, \eqref{eq:Cuu:Cdd}, we get for the LFV axion model, for all three benchmarks,
\beq
C_N =  \sqrt{0.042 - 0.084 \, c_\beta^2 + 0.18 \, c_\beta^4}.
\eeq
 For given $c_\beta$ the bound on $f_a$ follows from the  bound on the effective decay constant $F_N \equiv 2 f_a/C_N \ge 10^9 \GeV$~\cite{SNbound}. The resulting bound on $f_a$ is of the order $f_a \gtrsim 10^8 \GeV$, with mild dependce on $c_\beta$. 

In  Fig.~\ref{fig:axion-plot} we show the constraints on the axion-photon couplings
 \beq
 \label{eq:gagg}
g_{a \gamma \gamma} =\frac{1}{f_a} \frac{\alpha_{\rm em}}{2 \pi}  \big(E/N - 1.92\big),
 \eeq
  as a function of the axion mass, $m_a$. The orange (green) solid line in the left (right) panel  shows $g_{a \gamma \gamma}$ as a function of $m_a$ in the $V+A$ ($V-A)$ benchmark up to the exclusion by the WD cooling constraints (the dotted continuation of the line is excluded by WD cooling but is not excluded by the direct $\mu\to e a$ searches).  Note that for the LFV QCD axion, $E/N=4/3$, leading to smaller coupling to photons, $|g_{a\gamma\gamma}|\simeq 0.6 \times \alpha_{\rm em}/(2\pi f_a)$, than for the flavor universal original KSVZ model, for which $E/N=0$ and thus $|g_{a\gamma\gamma}|\simeq 1.9 \times \alpha_{\rm em}/(2\pi f_a)$ \cite{Kim:1979if,Shifman:1979if}, and DFSZ-II model~\cite{Zhitnitsky:1980tq,Dine:1981rt}, for which $E/N=2/3$ and $|g_{a\gamma\gamma}|\simeq 1.3 \times \alpha_{\rm em}/(2\pi f_a)$  (though large variations in this coupling are possible depending on the precise choices of the heavy fields and their charges \cite{DiLuzio:2020wdo,DiLuzio:2016sbl,DiLuzio:2017pfr}). The dotted orange (green) vertical lines show projected and present bounds from $\mu\to e a$ searches, as denoted. 
  
  The gray regions are excluded by other axion experiments:  
  CAST \cite{Anastassopoulos:2017ftl}, cooling of horizontal branch stars (HB) and ADMX \cite{Braine:2019fqb,Boutan:2018uoc,Du:2018uak}. The gray dashed lines denote future projections from different axion searches already discussed in Sec.~\ref{sec:ALP:DM}. We show the reach of the future ADMX upgrade \cite{Shokair:2014rna}, of CAPP \cite{Petrakou:2017epq}, KLASH \cite{Gatti:2018ojx}, and ORGAN \cite{McAllister:2017lkb}, MADMAX \cite{TheMADMAXWorkingGroup:2016hpc} and the ``dieletric stack'' proposal~\cite{Baryakhtar:2018doz}. We also include the reach of large-scale helioscopes such as IAXO \cite{Irastorza:2013dav,Armengaud:2019uso}, and light-shining-through-wall experiments such as ALPS-II \cite{Bahre:2013ywa}.

Fig.~\ref{fig:axion-plot} demonstrates the complementarity between this diverse experimental program based on axion couplings to photons and electrons, and  the reach of the LFV experiments MEG-fwd and Mu3e, in order to search for  LFV axions. In particular, a signal in a future LFV search would be incompatible with an axion lighter than a few $\text{meV}$. Such an LFV axion line will be challenging to test in axion haloscopes because of the infamous DFSZ accidental suppression of the photon coupling, see Eq.~\eqref{eq:gagg}. In contrast, the future ADMX experimental campaign and CAPP will probe the LFV axion for masses between few $\mu\text{eV}$ and few tens of $\mu\text{eV}$, which are inaccessible through LFV experiments.  

\subsection{The LFV Axiflavon}
\label{sec:axiflavon}
We discuss next the possibility that the PQ symmetry that solves the strong CP problem is also responsible for explaining the smallness of the SM Yukawas, i.e., that the QCD axion is the {\em axiflavon}. This framework naturally results in a QCD axion with flavor-violating couplings.

The simplest scenarios of this kind arise when the $U(1)_{\rm PQ}$ is responsible for explaining all the SM fermion mass hierarchies and mixings, along the lines of the Froggatt-Nielsen models~\cite{Froggatt:1978nt,Leurer:1992wg,Leurer:1993gy}, as in Refs.~\cite{Ema:2016ops,Calibbi:2016hwq}. In these constructions the strongest bound on the axion decay constant always arises from $K^+ \to \pi^+ a$ constraints \cite{Calibbi:2016hwq}, since the axion coupling to $sd$ is suppressed only by roughly the size of the Cabibbo angle, $V_{us}= \lambda \simeq 0.2$.  Indeed, for the $U(1)$ axiflavon the LH quark charges, $X^{q_L}_i$, are non-universal so that from Eq.~\eqref{Cdef},
\begin{align}
C^{V}_{sd} \sim V_{us} \, \frac{X^{q_L}_1 - X^{q_L}_2}{2N} \sim V_{us}. 
\end{align}

The $sd$ axiflavon couplings, on the other hand, can be strongly suppressed in $U(2)_F$ flavor models~\cite{Barbieri:1995uv, Barbieri:1997tu, Linster:2018avp}. In these classes of models the light generations form doublets of $U(2)_F$. The $U(1)_F$ factor acts as the PQ symmetry that gives rise to the QCD axion after spontaneous symmetry breaking. This scenario successfully explains the fermion mass hierarchies and mixings in terms of just two small parameters. In the model of Ref.~\cite{Linster:2018avp} both quark and lepton flavor violating couplings between the first two generations are equally suppressed because of the assumed structure that is compatible with $SU(5)$ unification. Here we present a variant of this model that leads instead to parametrically large $\mu e$ couplings (their enhancement can be traced to large PMNS mixing angles). This model is a successful model of flavor  and at the same time an example of a large class of flavored axion models where the LFV couplings are sizable while the FV couplings to quarks are suppressed. 

In the LFV axiflavon model the $U(2)_F = SU(2)_F \times U(1)_F$ quantum numbers of almost all the SM fermion are the same as in Ref.~\cite{Linster:2018avp}. In particular, all the fermions transform as ${\bf 2} + {\bf 1}$ under $SU(2)_F$. The only difference with respect to Ref.~\cite{Linster:2018avp} is that the $U(1)_F$ charge of the $SU(2)_F$ singlet left-handed lepton, $L_3$, is $-1$ instead of $1$.  Table~\ref{tab:charges} summarizes the complete field content and the transformation properties under the $U(2)_F$ flavor group. In addition to the SM fermions, the electroweak doublets $Q_i, L_i$, and singlets $U_i^c, D_i^c,E_i^c$, $i=1,2,3$ (in Table  ~\ref{tab:charges}, $a=1,2$), the model contains the SM Higgs doublet as well as two scalar spurions, $\phi_a$ and $\chi$. 

As in Ref.~\cite{Linster:2018avp}, the breaking of the flavor symmetry  is parametrized by two scalar spurions $\phi$ and $\chi$, which transform under $U(2)_F$ as  $\phi$ = ${\bf 2}_{-1}$ and $\chi$= ${\bf 1}_{-1}$.  
These fields acquire the following flavor symmetry breaking vevs
\begin{align}
\label{eq:vevs:phi}
\langle \phi \rangle & =  \begin{pmatrix} \eps_\phi \Lambda \\ 0 \end{pmatrix} \, , &
 \langle \chi \rangle & = \eps_\chi \Lambda \, ,
\end{align}
where $\Lambda$ is, up to ${\mathcal O}(1)$ factors, the typical mass of the heavy states present in the full UV model (their exact structure is not important for our effective low energy discussion as they are integrated out). The values of the two small parameters in the two spurions are 
fixed by the fit to quark masses and mixings to be about $\eps_{\phi} \sim \lambda^2$ and  $\eps_{\chi} \sim \lambda^3$.  
\begin{table}[t]
\centering
\begin{tabular}{c|cccccc|cccc|ccc}
\hline\hline
& $U^c_a$ & $D^c_a$ & $Q_a$ & $U^c_3$ & $D^c_3$ & $Q_3$ & $E^c_a$ & $L_a$ &$E^c_3$ & $L_3$  &  $H$ & $\phi_a$ & $\chi$ \\
\hline
$SU(2)_F$ & ${\bf 2}$ & ${\bf 2}$  & ${\bf 2}$ &  ${\bf 1}$ & ${\bf 1}$ & ${\bf 1}$ & ${\bf 2}$ & ${\bf 2}$  & ${\bf 1}$ & ${\bf 1}$ & ${\bf 1}$ & ${\bf 2}$ & ${\bf 1}$ \\    
$U(1)_F$ & $1$ & $1$ & $1$ & $0$ & $1$ & $0$ & $1$ & $1$ & $0$ & $-1$ & $0$ & $-1$ & $-1$    \\
\hline\hline
\end{tabular}
\caption{The $U(2)_F$ quantum numbers of the SM fermions and the scalars $H$, $\phi_a$ and $\chi$ in the example LFV axiflavon model. \label{tab:charges}}
\end{table}

Since the SM fermions are charged under $U(2)_F$, the Yukawa interactions between the SM fermions and the Higgs require insertions of the spurion fields in order to form invariants under $U(2)_F$. 
This leads to non-renormalizable interactions  suppressed by appropriate powers of $\Lambda$. After replacing spurions with their vevs, Eq.~\eqref{eq:vevs:phi}, the dependence on the heavy scale $\Lambda$ drops out. The  hierarchies in Yukawa matrices then arise from powers of the small parameters $\eps_{\phi, \chi}$, giving for the mass matrices  
\begin{align}
\label{eq:mf}
m_u  & \approx \frac{v}{\sqrt 2}
\begin{pmatrix}
\lambda_{11}^u \eps_\phi^2  \eps_\chi^4 & \lambda_{12}^u \eps_\chi^2  & \lambda_{13}^u \eps_\phi  \eps_\chi^2  \\ 
- \lambda_{12}^u \eps_\chi^2 & \lambda_{22}^u \eps_\phi^2  & \lambda_{23}^u \eps_\phi \\
\lambda_{31}^u \eps_\phi  \eps_\chi^2 & \lambda_{32}^u \eps_\phi & \lambda_{33}^u
\end{pmatrix} \, , & 
m_d  & \approx \frac{v}{\sqrt 2}
\begin{pmatrix}
\lambda_{11}^d \eps_\phi^2  \eps_\chi^4 & \lambda_{12}^d \eps_\chi^2  & \lambda_{13}^d \eps_\phi  \eps_\chi^3  \\ 
- \lambda_{12}^d \eps_\chi^2 & \lambda_{22}^d \eps_\phi^2  & \lambda_{23}^d \eps_\phi \eps_\chi \\
\lambda_{31}^d \eps_\phi  \eps_\chi^2 & \lambda_{32}^d \eps_\phi & \lambda_{33}^d \eps_\chi 
\end{pmatrix} \, , \\
m_e  & \approx \frac{v}{\sqrt 2}
\begin{pmatrix}
\lambda_{11}^e \eps_\phi^2  \eps_\chi^4 & \lambda_{12}^e \eps_\chi^2  & \lambda_{13}^e \eps_\phi  \eps_\chi^2  
 \\ 
- \lambda_{12}^e \eps_\chi^2 & \lambda_{22}^e \eps_\phi^2  & \lambda_{23}^e \eps_\phi  \\
\lambda_{31}^e \eps_\phi  \eps_\chi & \lambda_{32}^e \eps_\phi \eps_\chi & \lambda_{33}^e \eps_\chi 
\end{pmatrix} \, ,  
&
m_{\nu}  & \approx \frac{v^2}{2M} 
\begin{pmatrix}
 \lambda_{11}^\nu \eps_\phi^2 \epsilon_\chi^4 & \lambda_{12}^\nu \eps_\phi^2 \epsilon_\chi^2 &  \lambda_{13}^\nu  \eps_\phi \epsilon_\chi  \\
\lambda_{12}^\nu \eps_\phi^2 \epsilon_\chi^2 & \lambda_{22}^\nu \epsilon_\phi^2 & \lambda_{23}^\nu \epsilon_\phi \epsilon_\chi \\
 \lambda_{13}^\nu  \eps_\phi \epsilon_\chi & \lambda_{23}^\nu\epsilon_\phi \epsilon_\chi & \lambda_{33}^\nu \epsilon_\chi^2 
\end{pmatrix} \, ,
\end{align}
where we kept only the leading contributions in $\eps_{\phi, \chi}$.
The parameters $\lambda_{ij}^f$ are ${\cal O}(1)$ complex coefficients. The structure of the neutrino mass matrix $m_\nu$ was obtained under the assumption that the neutrinos are Majorana with the masses arising  from the Weinberg operator with the UV suppression scale $M$. Notice that the 1-2 entry in $m_\nu$ of order $\eps_\chi^2$ vanishes due to anti-symmetrization, so that the first nonzero entry is of ${\mathcal O}(\epsilon_\phi^2\epsilon_\chi^2)$. 

The quark sector of this model is identical to the one in Ref.~\cite{Linster:2018avp}, and so are the unitary rotations that diagonalize the quark masses. Their parametric structure is given by,
\begin{align}
\label{eq:VLu:LFVaxiflavon}
V^u_L & \sim V^u_R \sim \begin{pmatrix}
1  & \lambda   & \lambda^7  \\ 
 \lambda & 1 & \lambda^2  \\
\lambda^3 & \lambda^2 &1
\end{pmatrix} \, , &  
V^d_L &   \sim \begin{pmatrix}
1  & \lambda   & \lambda^3  \\ 
 \lambda & 1 & \lambda^2  \\
\lambda^3 & \lambda^2 &1
\end{pmatrix} \, , &  
V^d_R &    \sim \begin{pmatrix}
1  & \lambda   & \lambda^5  \\ 
 \lambda & 1 & 1  \\
\lambda & 1 &1
\end{pmatrix} \, ,
\end{align}
where $\lambda = V_{us}\sim 0.2$. The charged lepton mass matrix is parametrically the same as in Ref.~\cite{Linster:2018avp}, and so are the parametric sizes of the unitary rotations that diagonalize it,
\begin{align}
\label{eq:VLe:LFVaxiflavon}
V^e_R &   \sim \begin{pmatrix}
1  & \lambda   & \lambda^3  \\ 
 \lambda & 1 & \lambda^2  \\
\lambda^3 & \lambda^2 &1
\end{pmatrix} \, , &  
V^e_L &    \sim \begin{pmatrix}
1  & \lambda   & \lambda^5  \\ 
 \lambda & 1 & 1  \\
\lambda & 1 &1
\end{pmatrix} \, .
\end{align}
It is easy to check that the neutrino sector can reproduce the PMNS matrix and neutrino mass differences for $\lambda^\nu_{ij}$ that are ${\cal O}(1)$ for normal neutrino mass ordering.
\begin{figure}[t]
	\centering
	\includegraphics[width=0.55\linewidth]{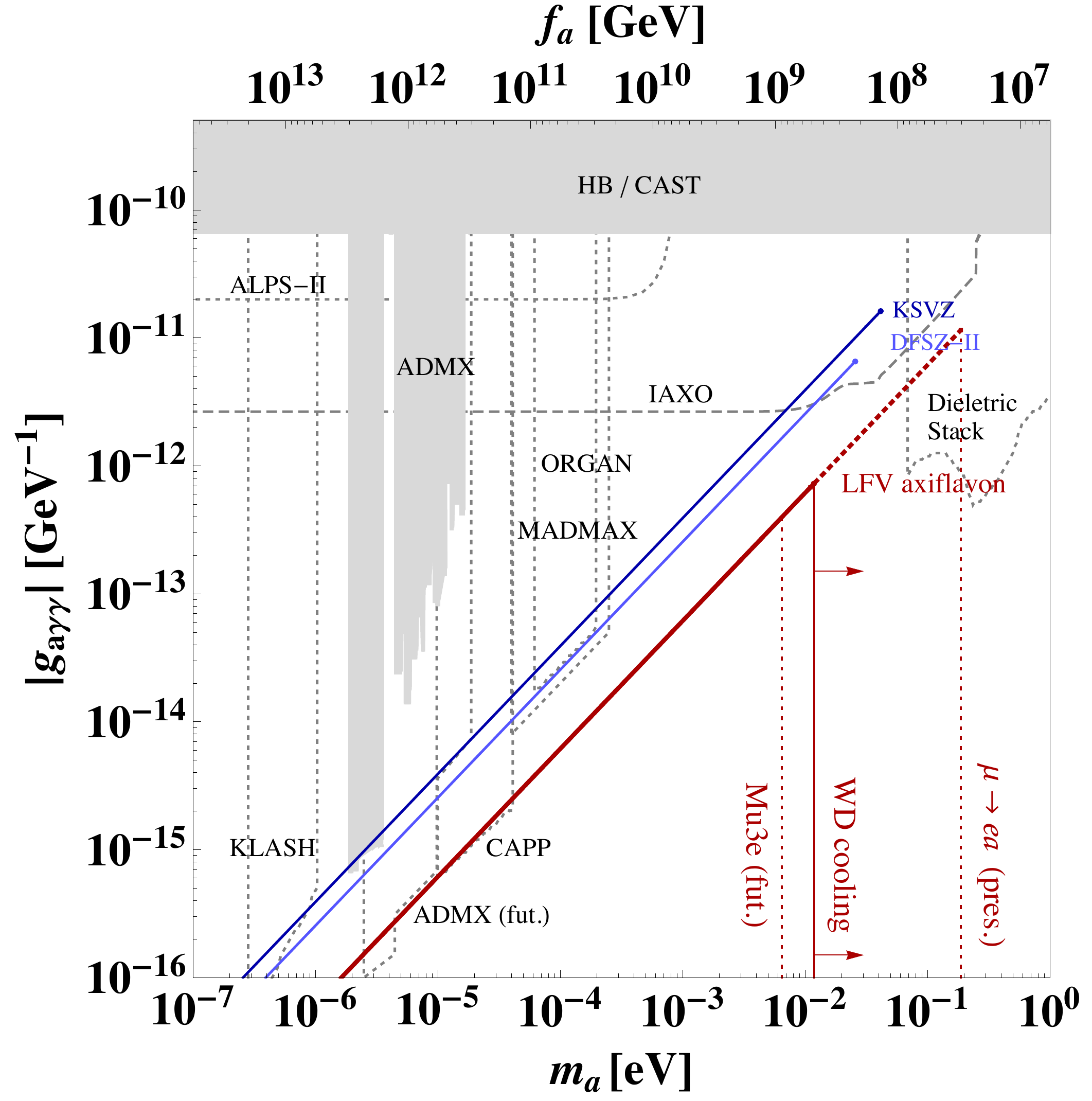}
	\caption{Present bounds and projected sensitivity for the LFV axiflavon model illustrated in Section \ref{sec:axiflavon}, with the LFV couplings given in Eq.~\eqref{eq:axiflavon:benchmark}. The {\bf red bold solid line} shows the predicted coupling to photons (as defined in Eq.~\eqref{eq:gagg}) of the LFV axiflavon model. The {\bf solid red} vertical line shows the upper bound on the axion mass from star cooling while the {\bf dotted red} vertical lines the present bound and future sensitivity from LFV axion searches. The {\bf grey shaded} and {\bf grey dotted} regions, and the  {\bf blue} and {\bf dark blue} lines are the same as in Fig. \ref{fig:axion-plot}.}
	\label{fig:axiflavon-plot}
\end{figure} 

The PNGB corresponding to the $U(1)_F$ factor is the LFV axiflavon, i.e., it acts as the QCD axion that solves the strong CP problem (see Ref.~\cite{Linster:2018avp} for details). The anomaly coefficients controlling the couplings of the LFV axiflavon  to gluons and to photons in Eq. \eqref{eq:La:QCDaxion} are given by
\begin{align}
N & = 9/2 \, , & E & = 10 \, .
\end{align}
The couplings of the LFV axiflavon to the SM fermions depend on the unitary rotations in Eqs.~\eqref{eq:VLu:LFVaxiflavon}, \eqref{eq:VLe:LFVaxiflavon}, as in Eq. \eqref{Cdef}. Because of the $SU(2)_F$ structure of the LFV axiflavon model it is useful to introduce the matrices ($f = u,d,e; P=L,R$)\footnote{Note the difference between $\epsilon_{ij}^{e_P}$ defined here and in Eq.~\eqref{eq:epsilon:eLR:QCDaxion}. For this reason we restrict the use of $\epsilon_{ij}^{f_P}$ symbols to the two respective sections.}
\begin{align}
\eps^{f_{P}}_{ij} \equiv (V^f_P)^*_{3i}  (V^f_P)_{3j} \, , 
\end{align}
 which satisfy
 \begin{align}
 0 & \le \eps^{f_P}_{ii} \le 1 \, , & \sum_i \eps^{f_P}_{ii}  & = 1 \, , &  |\eps^{f_P}_{ij}| & = \sqrt{\eps^{f_P}_{ii} \eps^{f_P}_{jj}} \, .
 \end{align}
They have the parametric structures
\begin{align}
\eps^{u_L}_{ij} \sim \eps^{u_R}_{ij} \sim \eps^{d_L}_{ij} \sim \eps^{e_R}_{ij} &   \sim \begin{pmatrix}
\lambda^6  & \lambda^5   & \lambda^3  \\ 
 \lambda^5 & \lambda^4 & \lambda^2  \\
\lambda^3 & \lambda^2 &1
\end{pmatrix} \, , & 
\eps^{d_R}_{ij} & \sim \eps^{e_L}_{ij}  \sim \begin{pmatrix}
\lambda^2  & \lambda   & \lambda  \\ 
 \lambda & 1 & 1  \\
\lambda & 1 &1
\end{pmatrix} \, .
\end{align}
 The axiflavon couplings to fermions are given in terms of charges and these parameters as \begin{align}
C^V_{f_i f_j}  & =  \frac{X_{f^c_a} - X_{f_a}}{2N} \delta_{ij} + \frac{X_{f^c_3} - X_{f^c_a}}{2N} \eps^{f_R}_{ij} - \frac{ X_{f_3} - X_{f_a} }{2N}  \eps^{f_L}_{ij}   \, , \\
C^A_{f_i f_j}  &  =  \frac{X_{f^c_a} + X_{f_a}}{2N} \delta_{ij} + \frac{X_{f^c_3} - X_{f^c_a}}{2N} \eps^{f_R}_{ij} + \frac{ X_{f_3} - X_{f_a} }{2N}  \eps^{f_L}_{ij} \, ,
\end{align}
where $f = u,d,e$ denotes the fermion sector and $X_{f^c_a}, X_{f_a}, X_{f^c_3}, X_{f_3}$ are the $U(1)_F$ charges in Table \ref{tab:charges}. 

The $sd$ couplings are strongly suppressed, $C^{V,A}_{sd} \sim \lambda^5/(2N)$, as a result of small rotations in the LH sector, $\eps^d_{L, 12} \sim \lambda^5$, and the fact that RH rotation do not lead to off-diagonal terms because of universal charges in the RH sector, $X_{D_a^c} = X_{D_3^c}$. The RH contributions to the $\mu e$ couplings are CKM-like suppressed, while the contribution from the LH rotations are large, since the corresponding charges are  non-universal (in contrast to Ref.~\cite{Linster:2018avp}), $X_{L_a} \ne X_{L_3}$,  giving $C^{V}_{\mu e} = - C^{A}_{\mu e} \sim  \sqrt{m_e/m_\mu}/N$. The axiflavon couplings to nucleons and electrons are identical to Ref.~\cite{Linster:2018avp} and are to good approximation given by 
\begin{align}
C^A_{ee} & = C^A_{uu}  = C^A_{dd} =  C^A_{cc} = C^A_{ss}  \approx \frac{2}{9} \, , & C^A_{tt} & \approx 0 \, , & C^A_{bb} &  \approx \frac{1}{9} \, .
\end{align} 

The ${\cal O}(1)$ coefficients in the rotation matrices can be fixed by  performing an explicit fit to all the observables -- the masses and mixings,  including the neutrino sector. Using the same procedure and the SM inputs as in Ref.~\cite{Linster:2018avp}, we find that a good fit is obtained by choosing $\eps_\phi = 0.023, \eps_\chi = 0.080$, $M = 4.8 \times 10^{11} \GeV$, with Yukawa couplings for the up quark sector
$\lambda^u_{\{12,22,23,32,33\}} = \{-2.0,1.0,  -3.1, -1.1, -0.79\}$, and for the down quark sector 
$\lambda^d_{\{12,22,23,32,33\}} = \{1.3,  1.1,  -0.76,  0.44,  -0.85\}$,  with the couplings $\lambda^u_{11,13, 31}$ and $\lambda^d_{11,13, 31}$  irrelevant because they give only subleading contributions to quark masses and mixings.
The couplings for the lepton Yukawa matrices are
$\lambda^e_{\{11,12,13,22,23,31,32,33\}} = \{1.0, 2.3,  1.0, 1.7,  0.40,  -0.30,  -1.3,  0.55\}$,
while  for the neutrinos they are
$\lambda^\nu_{\{11,12,13,22,23,33\}} = \{-1.0, -1.0, 0.38,  -1.2,  1.9,  -0.69  \}$. 
For this fit we find for the $sd$ axiflavon coupling $C^V_{sd}  = 1.6 \times 10^{-5} $, while the relevant couplings in the leptonic sector 
are
\begin{align}
\label{eq:axiflavon:benchmark}
 C^V_{\mu e} & = - C^A_{\mu e} = 0.043 \, , & C_e & = 0.21 \, , & C_{\tau e} & = 0.029 \, , & C_{\tau \mu} & = 0.12 \, ,
\end{align}
where $C_{\ell_i \ell_j}\equiv \sqrt{|C^{V}_{\ell_i \ell_j}|^2 + |C^{A}_{\ell_i \ell_j}|^2}$. We use the above benchmark values for axiflavon coupling to derive the sensitivities of different observables to axiflavon in Fig.~\ref{fig:axiflavon-plot}. 
Other phenomenologically viable choices of  parameters that differ by ``${\cal O}(1)$" factors can also give a good fit to the SM masses and mixings, so the constraints obtained in our benchmark should be viewed only as indicative, with  ${\cal O}(1)$ variations,  when this larger class of axiflavon parameters is considered.

The red line in Fig. \ref{fig:axiflavon-plot} shows the predicted coupling to photons, $g_{a\gamma\gamma}$, which for the LFV axiflavon is given by Eq. \eqref{eq:gagg} with $E/N=20/9$, as a function of axiflavon mass, Eq. \eqref{eq:axion:mass}. The bound on $f_a$ from WD cooling is denoted with a vertical solid red line. The next less stringent bound comes from the $\mu\to e a$ search by TWIST (dotted vertical red line). Fig. \ref{fig:axiflavon-plot} shows that the Mu3e future reach (dotted vertical red line) will exceed the WD cooling constraints. In order not to clutter Fig. \ref{fig:axiflavon-plot}, we do not show the less constraining present bound (future senstivity) on LFV axiflavon from $K^+\to \pi^+ a$ which is $f\gtrsim 5.4 \times 10^{6} {\rm~GeV}~(1.6 \times 10^7 {\rm~GeV})$ \cite{MartinCamalich:2020dfe}. The expected reach from Belle II is $f_a\gtrsim 3(1)\times 10^6$ GeV from $\tau \to \mu a (\tau \to ea)$ searches, which is outside the range plotted in Fig. \ref{fig:axiflavon-plot}. The other constraints, shown in grey with future sensitivities denoted with grey dashed lines, are as in Fig. \ref{fig:axion-plot}. Clearly, there is significant parameter space where the LFV axiflavon can be discovered in LFV experiments, especially considering the potential astrophysical uncertainties in the WD cooling bounds.

 \subsection{The Leptonic Familon}\label{sec:familon}
Our next example of an LFV ALP is the familon, i.e., the PNGB arising from the spontaneous breaking of a global horizontal symmetry, which we take to be the Froggatt-Nielsen (FN) flavor symmetry, $U(1)_{\rm FN}$ \cite{Froggatt:1978nt}. We consider the case where the $U(1)_{\rm FN}$ only acts on the leptonic sector so that the LFV ALP is the {\em leptonic familon}. Unlike the previous two examples, the leptonic familon does not solve the strong CP problem and, as a consequence, its mass is not determined by the QCD anomaly. The mass of the leptonic familon is therefore taken to be a free parameter, yet still small enough that it can be produced in tau or muon decays. The predictive power of the model is limited to the parametric prediction of the LFV coupling, which are related to the neutrinos mass texture. 

The couplings of the leptonic familon to the SM leptons are determined by the positive $U(1)_{\rm FN}$ charges $[L]_i$ and $[e]_i$ carried by the lepton doublets 
$L_i$ and singlets $e_i^c$, respectively. The $U(1)_{\rm FN}$ symmetry is broken by the vev $f_a$ of a scalar field $\Phi$ with charge $[\Phi]=-1$, 
\begin{align}
\Phi = \frac{f_a+\phi}{\sqrt{2}} e^{i a/f_a},
\end{align}
where $a$ is the familon, while $\phi$ is the radial mode with the mass ${\mathcal O}(f_a)$. Integrating out all the heavy fields the charged lepton Yukawa couplings and the Majorana neutrino mass matrix are given by (as determined through spurion analysis) 
\begin{align}
y^e_{ij} = a^e_{ij}  \left(\frac{\langle\Phi^*\rangle}{M}\right)^{[L]_i + [e]_j},\quad\quad m^\nu_{ij}=\kappa^\nu_{ij}\frac{v^2}{\Lambda_N} \left(\frac{\langle\Phi^*\rangle}{M}\right)^{[L]_i + [L]_j},
\end{align}
where $a^e$ and $\kappa^\nu$ are assumed to be flavour-anarchical matrices of $\ord{1}$ coefficients, $M$ is a cut-off scale with 
$$
\epsilon\equiv \frac{\langle\Phi^*\rangle}{M} <1,
$$ 
controlling the hierarchies among charged lepton masses, 
while $\Lambda_N$ is the lepton-number breaking scale suppressing the dimension 5 Weinberg operator.\footnote{Introducing instead the RH neutrinos such that the neutrinos have  Dirac masses would not change the following discussion and the resulting PMNS matrix.}
An anarchical PMNS matrix featuring $\ord{1}$ mixing angles \cite{Hall:1999sn} can be achieved by taking equal charges for the lepton doublets, 
\begin{align}
\label{eq:ana}
([L]_1,\,[L]_2,\,[L]_3)&~=~(L,\,L,\,L), & [\tt Pure~Anarchy]\,.
\end{align}
Good fits to the neutrino oscillation data can also be obtained for mildly hierarchical charges (at the price of somewhat larger values of $\eps\sim 0.3-0.4$) \cite{Altarelli:2012ia,Bergstrom:2014owa},
\begin{align}
\label{eq:mutau}
([L]_1,\,[L]_2,\,[L]_3)&~=~(L+1,\,L,\,L), & 
[\mu\tau~{\tt Anarchy}]\,,\\
([L]_1,\,[L]_2,\,[L]_3)&~=~(L+2,\,L+1,\,L), &
 [{\tt Hierarchy}]\,. \label{eq:hie}
\end{align}
The hierarchy of charged leptons is then reproduced by a suitable choice of the charges of the RH leptons.
Up to some freedom due to uncertainties relative to the expansion parameter $\epsilon$ and more importantly to the $\mathcal{O}(1)$ coefficients $a^e_{ij}$, 
a successful charge assignment is the following:
\begin{align}
&([e]_1,\,[e]_2,\,[e]_3)~=~(5-L,\,3-L,\,2-L),&\epsilon = 0.1\quad\quad & [\tt Pure~Anarchy]\,,\\
&([e]_1,\,[e]_2,\,[e]_3)~=~(9-L,\,5-L,\,3-L),&\epsilon = 0.3\quad\quad & [\mu\tau~{\tt Anarchy}]\,,\\
&([e]_1,\,[e]_2,\,[e]_3)~=~(11-L,\,6-L,\,4-L),&\epsilon = 0.4\quad\quad & [\tt Hierarchy]\,.
\label{eq:RHcharges}
\end{align}

The couplings of the  familon to the leptons are given by, 
\begin{align}
\mathcal{L} \supset \frac{\partial^\mu a}{2 f_a}\left(C^V_{\ell_i \ell_j}\, \bar{\ell}_i \gamma_\mu \ell_j + C^A_{\ell_i \ell_j}\, \bar{\ell}_i \gamma_\mu \gamma_5 \ell_j \right),
\end{align}
where 
\begin{align}
\label{eq:CVA:flavon}
C^{V/A} = V^{e\,^\dag}_R X^e_R\, V^e_R  \pm  V^{e\,^\dag}_L X^e_L \,V^e_L.
\end{align}
The unitary rotation matrices are defined as $V_L^{e\,\dag} y^e \,V^e_R \equiv y^e_{\rm diag}$,
while 
$(X^e_{L})_{ij} = [L]_i \delta_{ij}$, $(X^e_{R})_{ij} = -[e]_i \delta_{ij}$ are diagonal matrices of the FN charges.
Up to $\ord{1}$ coefficients the LH and RH rotations are given by $\big(V^e_{L/R}\big)_{ij} \approx \eps^{|[L/e]_i-[L/e]_j|}$. 

\begin{figure}[t]
	\centering
	\includegraphics[width=0.48\linewidth]{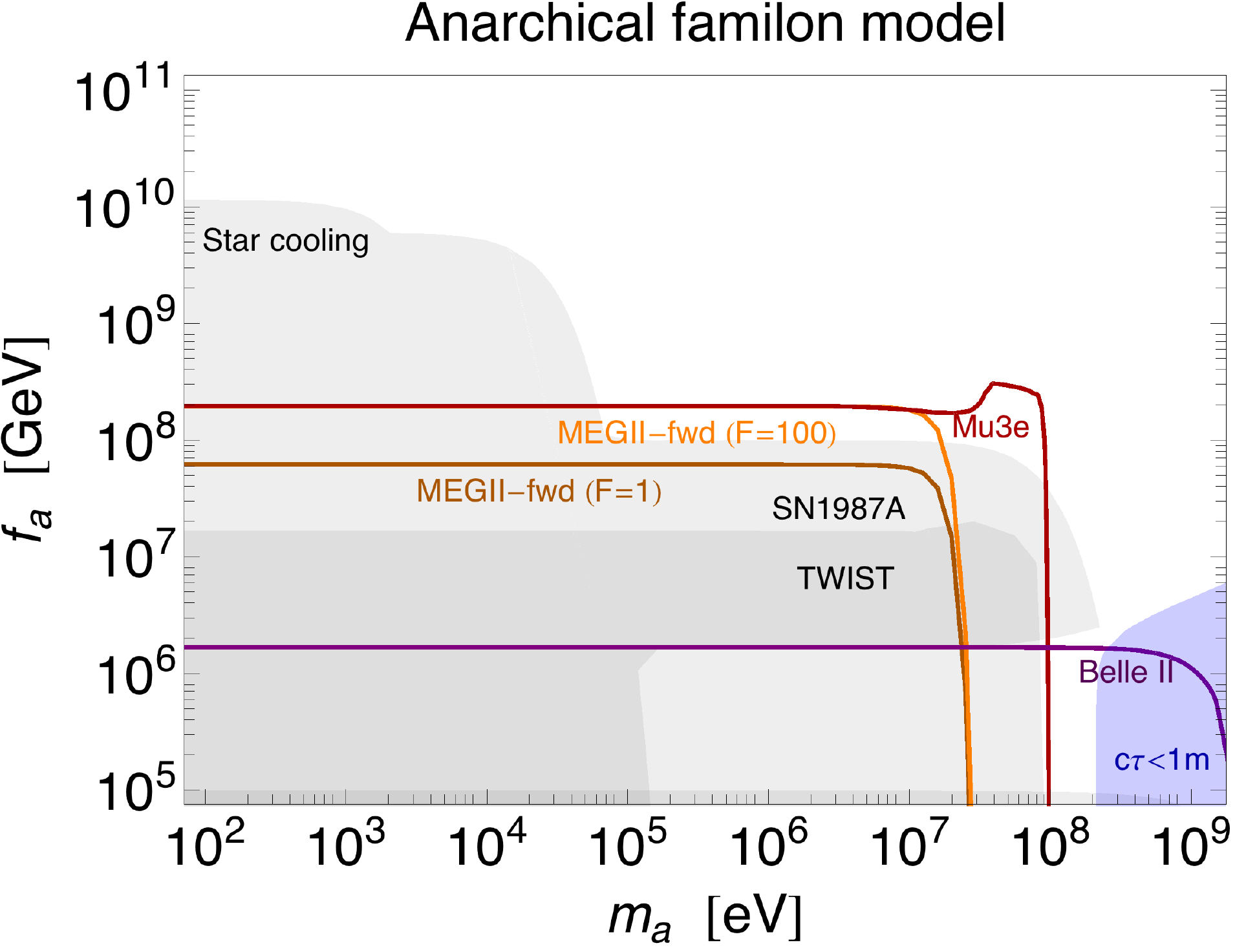}
		\hfill
		\includegraphics[width=0.48\linewidth]{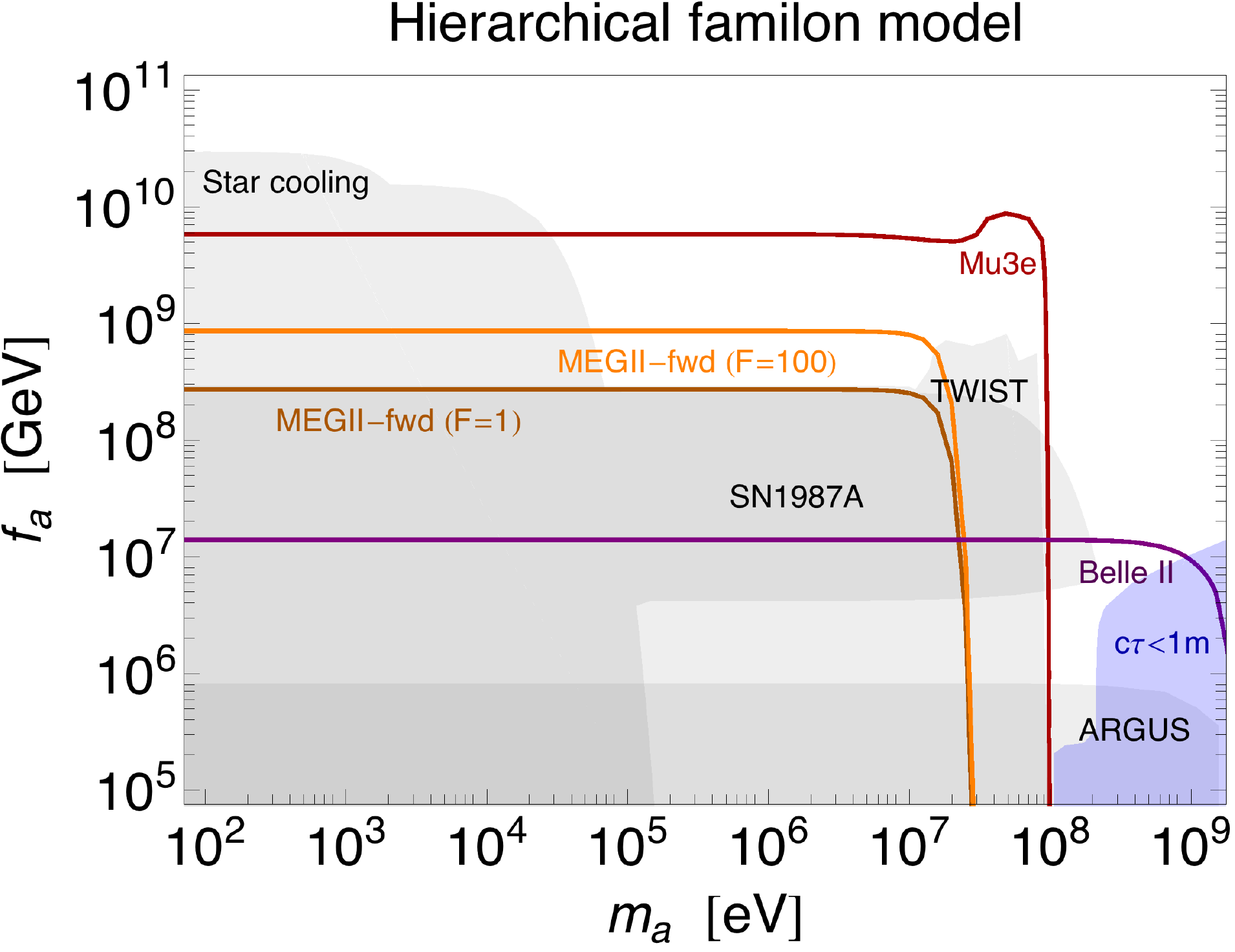}
	\caption{Present bounds and future sensitivities on the decay constant $f_a$ (in GeV) vs.~the  mass $m_a$ for two familon models. The {\bf blue shaded} regions indicate where the familon decays promptly into SM states. The {\bf grey shaded} regions are currently excluded by a combination of $\mu^+\to e^+ a$ experiments and star cooling bounds on the ALP-electron coupling. We show projected sensitivities of MEGII-fwd in {\bf orange}, Mu3e in {\bf red}, Belle II in {\bf purple}. {\bf Left:} Anarchical model, cf. Eq.~(\ref{eq:anarchy}), {\bf Right:} Hierarchical model, cf. Eq.~(\ref{eq:hie}).
 }
	\label{fig:familon-plot}
\end{figure}

The three charge assignments \eqref{eq:ana}-\eqref{eq:hie} lead to qualitatively different values of LFV familon couplings, Eq. \eqref{eq:CVA:flavon}. In the purely anarchical model the $[L]_i$ are universal, cf.~Eq.~\eqref{eq:ana}, so that the off-diagonal couplings in Eq.~\eqref{eq:CVA:flavon} are entirely due to the RH charges. The RH rotations are small, of the order of the ratios of lepton masses, $(V^e_{R})_{ij} \approx (V^e_{R})_{ji} \approx m_{\ell_i}/m_{\ell_j}$ ($i<j$), and thus the LFV couplings are suppressed,
\begin{align}
\label{eq:anarchy}
C^{V}_{\ell_i \ell_j} = C^{A}_{\ell_i \ell_j}  \approx ([e]_i-[e]_j) \frac{m_{\ell_i}}{m_{\ell_j}}~~~({i<j})\quad \quad [\tt Pure~Anarchy]\,.
\end{align}
The purely anarchical leptonic familon couples to the $V+A$ current. It is thus subject to constraints from the old Jodidio et al.~experiment and can be an important target for the proposed MEGII-fwd setup, despite the severe suppression of the LFV couplings, see Fig.~\ref{fig:familon-plot} (left).

For the hierarchical model the $[L]_i$ are non-universal, Eq. \eqref{eq:hie}, and the LH rotations give the dominant  contributions to the off-diagonal familon couplings in Eq.~\eqref{eq:CVA:flavon}. Even so, the $V+A$ couplings induced via RH rotations remain non-negligible, and are even enhanced compared to the anarchical case, 
 $\big(V^e_{R}\big)_{ij} \approx \big(V^e_{R}\big)_{ji} \approx  \eps^{[e]_i-[e]_j} \approx (m_{\ell_i}/m_{\ell_j})/\epsilon^{[L]_i-[L]_j}$ ($i<j$).
The LFV couplings of the hierarchical leptonic familon are thus, from Eq.~\eqref{eq:CVA:flavon},  
\beq
\begin{split}
\label{eq:hierarchy}
\begin{matrix}
C^{V}_{\ell_i \ell_j} &\approx  ([e]_i-[e]_j)\frac{1}{\epsilon^{[L]_i-[L]_j}} \frac{m_{\ell_i}}{m_{\ell_j}} - ([L]_i-[L]_j)\eps^{[L]_i-[L]_j},  
\\[3mm]
  C^{A}_{\ell_i \ell_j} &\approx   ([e]_i-[e]_j)\frac{1}{\epsilon^{[L]_i-[L]_j}} \frac{m_{\ell_i}}{m_{\ell_j}} + ([L]_i-[L]_j)\eps^{[L]_i-[L]_j} ,
\end{matrix}
\qquad  (i<j) \quad
   [\tt Hierarchy]\,.
\end{split}
\eeq
Because the hierarchical familon mostly couples to the $V-A$ current the dominant constraint comes from the bound on $\mu\to e\,a$ due to TWIST, Eq.~\eqref{eq:twistV-A}, even though this is otherwise the weakest bound on ${\rm BR}(\mu\to e\,a)$. In the future, {MEGII-fwd} can improve the reach on the hierarchical familon beyond the present bounds, because of the non-vanishing $V+A$ contributions, cf.~Fig.~\ref{fig:familon-plot} (right). 

The $\mu\tau$ Anarchy scenario is in between the above two cases. There is no LH mixing in the $2-3$ sector, while there is one in the case of $1-2$ and $1-3$ transitions. The RH mixings are suppressed, but with $1-2$ and $1-3$ mixing enhanced by $1/\epsilon$ over the pure anarchical case.

The familon coupling to electrons, relevant for assessing the star cooling bounds, is given by 
\begin{align}\label{ea:fam-cee}
C^{A}_{ee}  \approx -\big( [e]_1 +[L]_1\big)\,.
\end{align}
Finally, the coupling to photons is controlled by the anomaly contribution $E_\text{UV}$, Eq.~\eqref{eq:EUV}, which can be calculated in terms of FN charges in Eqs.~\eqref{eq:ana}-\eqref{eq:RHcharges} as 
\begin{align}
E_\text{UV} = \sum_i \left( [e]_i + [L]_i \right),
\end{align}
such that, for the three illustrative charge assignments (Anarchy, $\mu\tau$ Anarchy, Hierarchy), one gets respectively 
$E_\text{UV}=\{10,18,24\}$.

Fig.~\ref{fig:familon-plot} summarizes the present and expected future bounds  on the FN breaking scale $f_a$ for the purely anarchical (left panel) and for the hierarchical leptonic familon (right panel). 
  For the anarchical model, we set $[e]_1-[e]_2 = 2$ and $[e]_2-[e]_3 = 1$ in Eq.~(\ref{eq:anarchy}), and $[e]_1 + [L]_1 =8$ in Eq.~\eqref{ea:fam-cee}.  For the hierarchical case we estimated the LH rotations appearing in Eq.~(\ref{eq:hierarchy})  employing $\eps =0.4$ as in Refs.~\cite{Altarelli:2012ia,Bergstrom:2014owa}, while the RH couplings are obtained by setting 
 $[e]_1 - [e]_2 =5$ and  $[e]_2 - [e]_3 =2$.  In this case the diagonal coupling to electrons in  Eq.~\eqref{ea:fam-cee} follows from $[e]_1 + [L]_1 =13$. In all the cases the $\ord{1}$ coefficients were set  to $1$ exactly. The blue area in Fig.~\ref{fig:familon-plot} indicates where the familon proper decay length is shorter than 1\,m (due to a sizeable width to leptons, $a\to\ell_i\ell_j)$, so that searches for $\ell_i \to \ell_j +$ invisible lose sensitivity. 
Fig.~\ref{fig:familon-plot}  shows that in both representative models all future searches for LFV processes performed by MEG II, Mu3e and Belle II can probe well into the yet unexplored parameter space. 
 
 As far as DM is concerned, the leptonic familon resembles very much the generic ALP DM discussion in Sec.~\ref{fig:moneyDM}. We refer to that section for an extensive discussion of the implications of LFV searches on the ALP DM parameter space.

\subsection{The Majoron}\label{sec:majoron}
The majoron~\cite{Chikashige:1980ui, Schechter:1981cv} is the PNGB due to spontaneous breaking of the lepton number. A natural context where this kind of ALP arises are the seesaw models, where the breaking scale of lepton number is associated with the mass scale of heavy new fields. In type-I seesaw models at least two singlet fermions, the right-handed neutrinos $N_i$, are added to the SM (for concreteness we will assume below two RH neutrinos). 
The RH neutrinos couple to LH leptons via the Yukawa coupling matrix, $y_N$, while their masses are described by the Majorana mass matrix $M_N$,
\begin{align}
\mathcal{L} =\mathcal{L}_{\rm SM} + i \overline{N}\slashed{\partial}N -\left( y_N\overline{L} N \widetilde{H}+
\frac{1}{2}\overline{N^c} M_N  N + {\rm h.c.}\right).
\label{eq:Lseesaw}
\end{align}
In the seesaw limit,  $M_N\gg m_D \equiv y_N v $, the RH neutrinos are heavy and can be integrated out, while the light neutrinos are predominantly part of the SM doublets, $L_i$, with the Majorana mass matrix given by
\begin{equation}
\label{eq:seesaw}
m_\nu = - m_D M_N^{-1} m_D^T \,.
\end{equation}

The majoron $J$ arises when the mass matrix $M_N$, which breaks the lepton number by two units, is generated dynamically by the vev of a new SM singlet scalar field $\sigma$. In this case the RH neutrino mass matrix in Eq.~\eqref{eq:Lseesaw} is replaced by the Yukawa couplings of RH neutrinos to the scalar $\sigma$, i.e., $M_N\to \lambda_N \sigma$ in Eq.~\eqref{eq:Lseesaw}, where 
\begin{equation}
\label{eq:maj}
\sigma = \frac{f_N +  \hat{\sigma}}{\sqrt{2}} e^{i J/f_N} \, ,
\end{equation}
so that the RH neutrino mass matrix is given in terms of the matrix $\lambda_N$ as
\begin{equation}
M_N = \frac{\lambda_N f_N}{\sqrt{2}}\,. 
\end{equation}
The radial mode $\hat \sigma$ is heavy and can be integrated out, while the majoron $J$ is a PNGB and is light (we take its mass to be a free parameter). 

The majoron couples at tree level to neutrinos through the $\bar N^c\lambda_N \sigma N$ Yukawa terms.  
These Yukawa interactions then induce couplings of $J$ to charged leptons and quarks at loop-level~\cite{Chikashige:1980ui}, see Ref.~\cite{Pilaftsis:1993af} for complete expressions. Here we are interested only in the seesaw limit of these general expressions, which we match onto the effective Lagrangian~\eqref{couplings} upon identifying $a \to J, f_a \to f_N$. Using the results of Ref.~\cite{Garcia-Cely:2017oco,Heeck:2019guh}, we find for the majoron couplings to quarks and leptons, respectively,
\begin{align}
\label{eq:Cqq:Majoron}
C^V_{q_i q_j} & = 0 \, , & C^A_{q_i q_j} & =  - \frac{T^q_3}{16 \pi^2} \delta_{ij}  \, \Tr \left( y_N y_N^\dagger \right) \, , \\
C^V_{e_i e_j} & =  \frac{1}{16 \pi^2} \left( y_N y_N^\dagger \right)_{ij} \, , & C^A_{e_i e_j} & = \frac{1}{16 \pi^2} \left[   \frac{\delta_{ij} }{2}  \, \Tr \left( y_N y_N^\dagger \right) - (y_N y_N^\dagger)_{ij} \right]\, ,
\label{CeeMajoron}
\end{align}
where $T^{u,d}_3 = \pm 1/2$. Note that $F^V_{\mu e} = - F^A_{\mu e}$, i.e., the LFV couplings of type-I seesaw majoron have the  $V-A$ form (the couplings to the $V+A$ leptonic current are flavor conserving). The TWIST experiment is sensitive to such a majoron, while the more stringent bounds from the 1986 experiment by Jodidio et al. do not apply, see Section \ref{sec:pastmutoe} for details. 

In contrast to the other  ALP scenarios discussed above, in the case of a majoron, couplings to neutrinos are particularly relevant to assess the stability of the particle, hence whether it is a viable DM candidate. This is a consequence of the suppressed coupling to photons of the majoron, which in fact decays preferably to neutrinos when its mass is below the 2$m_e$ threshold. In the seesaw limit, the coupling to light neutrinos is diagonal and the decay width reads~\cite{Garcia-Cely:2017oco,Heeck:2019guh} 
\begin{align}
\Gamma(J\to \nu_i \bar{\nu}_i) = \frac{m_J}{16\pi f_N^2} m^2_i \sqrt{1-\frac{4 m_i^2}{m_J^2}}\,,
\end{align}
where $m_i$ are the light neutrino mass eigenvalues.

Another crucial  issue is whether the experiments are able to probe scales that are interesting for the neutrino mass generation. We can distinguish two limits: 
\begin{itemize}
\item In the \emph{standard seesaw setup}, sizeable entries of the Yukawa matrix $y_N$ are only compatible with the observed neutrino masses for an ultra-high seesaw scale. For instance, let us consider the case where elements of the Yukawa matrix are all of similar size, without any special structure, $|(y_N)_{ij}| \sim y$ (a hierarchical structure would not qualitatively change the argument). The light neutrino masses are thus of the size $m_\nu\sim y^2 v^2/M_N$, where $v=246$ GeV is the electroweak symmetry breaking scale. The effective scale suppressing the LFV $\ell_i\to \ell_j J$ decays is then given by $F \sim f_N/C \sim 16 \pi^2 f_N/y^2 \sim 16 \pi^2 v^2/m_\nu \gtrsim 10^{16} \GeV$.  The standard set-up therefore cannot be probed by the present nor any of the planned LFV experiments. 

\item In the \emph{low-scale seesaw setup}, the neutrino masses are additionally suppressed, such that large couplings in $y_N$ and a lower seesaw scale are compatible with the observed light neutrino masses. Indeed, since the neutrino masses $m_\nu \propto y_N M_N^{-1} y_N^T$ transform non-trivially under the lepton number (in contrast to $ y_N y_N^\dagger$), it is possible that light neutrino masses are parametrically suppressed in the presence of an approximate generalized lepton number. Such scenarios, usually referred to as the ``TeV scale seesaw mechanism", have been extensively studied in the literature~\cite{Broncano:2002rw,Raidal:2004vt, Kersten:2007vk, Abada:2007ux,Gavela:2009cd,Ibarra:2010xw,Ibarra:2011xn,Dinh:2012bp,Cely:2012bz,Alonso:2012ji}. In the following we use the results of Ref.~\cite{Ibarra:2011xn} to construct a concrete example of a majoron model with parametrically suppressed neutrino masses (and thus with enhanced majoron couplings to the SM leptons).  
\end{itemize}

In the simplest low-energy seesaw model one considers two right-handed neutrinos $N_{1,2}$ with a $2\times 2$ Dirac mass matrix, $M_N$,  and $3\times 2$ Dirac  Yukawa couplings, $y_N$,  
\begin{align}
M_N & = \begin{pmatrix}
0 & M \\
M & 0
\end{pmatrix} = \frac{\lambda}{\sqrt{2}} \begin{pmatrix}
0 & f_N \\
f_N & 0
\end{pmatrix} \, ,
& 
 y_N & = \begin{pmatrix}
y_{e1} & y_{e2}  \\
y_{\mu 1} & y_{\mu 2}  \\
y_{\tau 1} & y_{\tau 2}  
\end{pmatrix} \, ,
\end{align}
where $\lambda \equiv \sqrt{2} M/f_N$ is a real free parameter.
 In the $y_{\ell 1} \to 0$ limit the model has a global $U(1)$ symmetry, $M_N \to P M_N P$, $y_N \to e^{i \alpha} y_N P$ with $P = {\rm diag} \{e^{i \alpha}, e^{-i \alpha} \}$. The majoron couplings, which are proportional to $y_N y_N^\dagger$, are invariant under this symmetry, while the neutrino masses are not, $m_\nu \to e^{2 i \alpha} m_\nu$. This means that the neutrino masses are proportional to symmetry breaking parameters, $y_{\ell 1}$, which, if small, additionally suppress the neutrino masses compared to the majoron couplings. 
 
 Working in a basis where  the charged lepton matrix is diagonal, we can adjust the input parameters, $M$ and $y_{\ell i}$, such that all neutrino observables (2 mass differences + 3 mixing angles) are at the central experimental values. This leaves two free parameters, which we choose to be $M$, the mass scale of RH neutrinos, and the largest eigenvalue of the Dirac Yukawa matrix, $y = \max \left[ {\rm eig} ( y_N y_N^\dagger )\right]$. Using the results of Ref.~\cite{Ibarra:2011xn}, we obtain for the Normal Ordering (NO) in the seesaw limit,
 \begin{align}
 \label{eq:yNyNdagger}
 y_N y_N^\dagger & \approx y^2 \frac{m_3}{m_2 + m_3} A_i^* A_j \, , &
  \text{where~} A_i & = U_{i3} + i U_{i2} \sqrt{m_2/m_3} \, , 
 \end{align}
with $m_2 = \big({\Delta m_{21}^2}\big)^{1/2}$ and  $m_3 = \big({\Delta m_{31}^2}\big)^{1/2}$ the light neutrino masses (the lightest neutrino mass is $m_1 = 0$ as we introduced only two RH neutrinos). The $U_{ij}$ are the elements of the PMNS matrix, all of which are experimentally observable, while $y \approx \big({ \Tr \, ( y_N y_N^\dagger) }\big)^{1/2}$ is a free parameter, only  bounded by perturbativity, $y \lesssim 4$.  The result for Inverted Ordering 
 (IO) is obtained from \eqref{eq:yNyNdagger} by replacing $m_3 \to m_2 = \big(- \Delta m_{32}^2\big)^{1/2}, m_2 \to m_1 = \big(- \Delta m_{21}^2 - \Delta m_{32}^2\big)^{1/2}$ and $U_{i3} \to U_{i2}, U_{i2} \to U_{i1}$.

 For the numerical analysis we use the latest global neutrino oscillation fit results~\cite{Esteban:2018azc,nuFIT}, and set in Eq.~\eqref{eq:yNyNdagger} the mass differences, the mixing angles and the Dirac CP phase in PMNS matrix to their central experimental values. This gives the  $3\times 3$ Hermitian matrix $y_N y_N^\dagger$ that still depends on $y$ and one Majorana phases, $\alpha_m$. 
 For simplicity, we set the latter to zero, $\alpha_m = 0$. The effective suppression scales for the majoron couplings are in the NO case given by (similar results are obtained for IO)
 \begin{align}
  F^A_{ee} & = \frac{1.1 \times 10^{10} \GeV}{\lambda y^2}  \left(\frac{M}{10^7 \GeV} \right) \, , & 
  F_{\mu e} & = \frac{1.4 \times 10^{10} \GeV}{\lambda y^2}   \left( \frac{M }{10^7 \GeV} \right) \, ,\\
  F_{\tau e} & = \frac{1.6 \times 10^{10} \GeV}{\lambda y^2}   \left( \frac{M }{10^7 \GeV} \right) \, , &
  F_{\tau \mu} & = \frac{0.71 \times 10^{10} \GeV}{\lambda y^2}   \left( \frac{M }{10^7 \GeV} \right) \, ,
 \end{align}
where for the flavor-violating cases we quote the bound on combined $A$, $V$ effective scale, as defined in Eq.~\eqref{eq:FVA}.
For completeness we also show the results for the flavor diagonal couplings to muons and taus, although they are at the moment of little phenomenological relevance,
 \begin{align}
  F^A_{\mu\mu} & = -\frac{2.7 \times 10^{11} \GeV}{\lambda y^2}  \left(\frac{M}{10^7 \GeV} \right) \, , & 
  F^A_{\tau\tau} & = \frac{3.7 \times 10^{10} \GeV}{\lambda y^2}   \left( \frac{M }{10^7 \GeV} \right) \, .
  \end{align}
 The coupling to muons is comparatively suppressed due to an accidental cancellation between  the two contributions to  $C^A_{\mu\mu}$ in Eq.~\eqref{CeeMajoron}.
  
The majoron also couples to nucleons via its couplings to quarks \eqref{eq:Cqq:Majoron}. These couplings do not depend on PMNS elements and are given by 
 \begin{align}\label{eq:maj-nuclei}
  F_N & \approx \frac{0.88 \times 10^{10} \GeV}{\lambda y^2} \left(\frac{M }{10^7 \GeV} \right) \, .
 \end{align}
 
   \begin{figure}[t]
	\centering
		\includegraphics[width=0.75\linewidth]{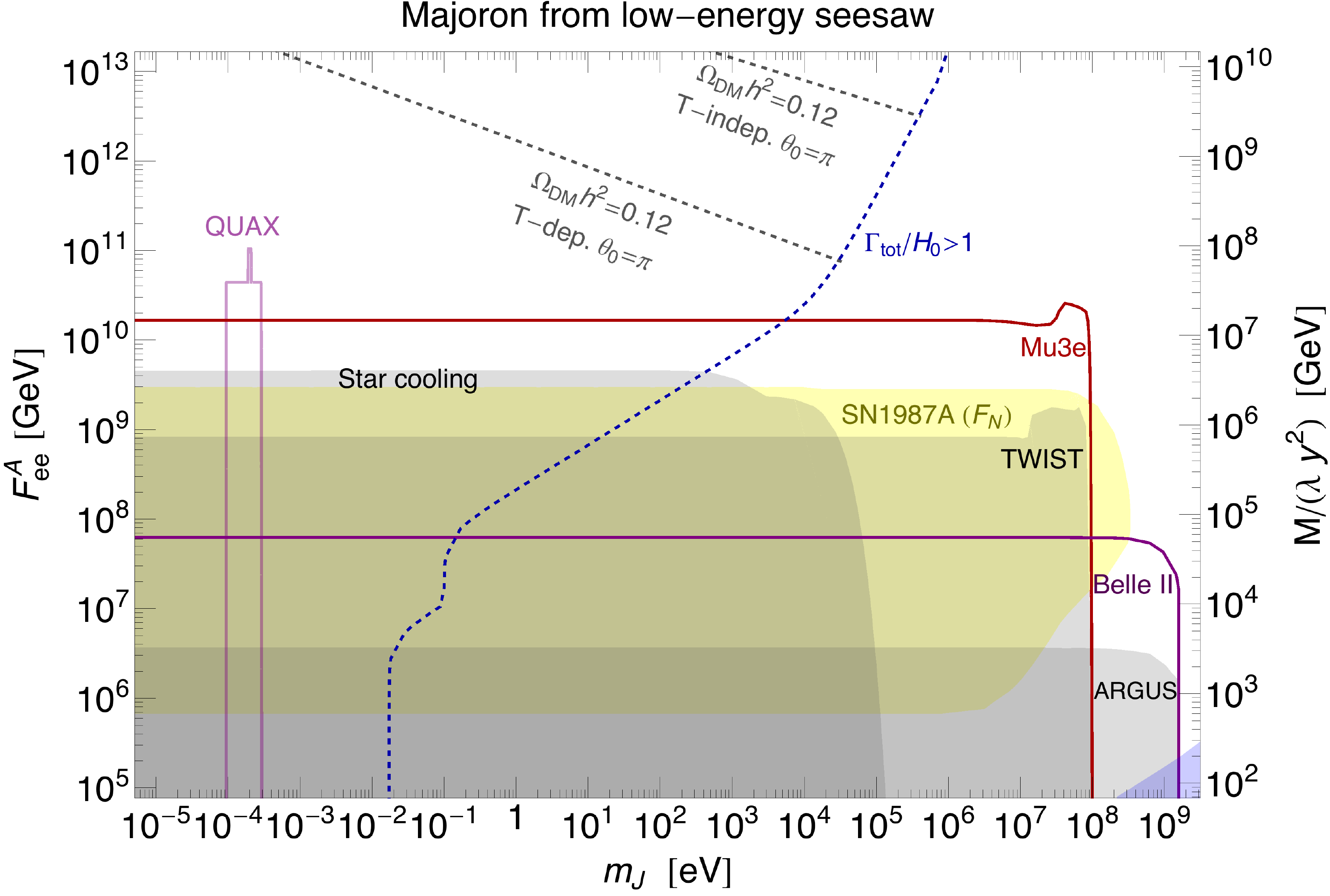}
	\caption{Present bounds and future sensitivities on the coupling to electrons $F^A_{ee}$ (in GeV) and the RH neutrino mass scale $M$ (normalized by the combination of free parameters $\lambda y^2$) vs.~the  majoron mass $m_J$ (in eV) for the low-energy seesaw majoron model. On the left of {\bf dashed blue} line the the majoron is stable on time scale of our Universe and could account for the DM abundance. The two dashed {\bf gray lines} are examples of misalignment production discussed in Sec~\ref{sec:ALP:DM}. The {\bf gray shaded} region is excluded by star cooling bounds on the majoron-electron coupling and $\mu^+\to e^+ a$ bounds. The {\bf yellow shaded} is excluded by SN1987A. We show in {\bf red} the future reach of Mu3e, in {\bf purple} the reach of Belle II and in {\bf light purple} the reach of QUAX which requires the majoron to be DM.}
\label{fig:majoron-plot}
\end{figure} 
 In Fig. \ref{fig:majoron-plot}, we summarize the current status and future prospect of the constraints on the parameter space 
 of the low-energy seesaw majoron model described above. Besides the present and future bounds from LFV experiments and the astrophysical limits discussed in Sec.~\ref{sec:astro} involving the coupling of the majoron to electrons, we display 
as a yellow-shaded area the region excluded by SN1987A, according to the study of Ref.~\cite{Lee:2018lcj}, due to the nucleon-majoron coupling of Eq.~\eqref{eq:maj-nuclei}. 

We also show as a dashed blue line where the majoron lifetime equals the lifetime of the universe. Anywhere to the left of this line, the majoron is a viable DM candidate. In fact, there is no strong constraint from observations of the extragalactic background light, since coupling of the majoron to photons are suppressed. The two representative lines explained in Sec.~\ref{sec:ALP:DM} show where the misalignment mechanism accounts for the whole DM abundance today with $\theta_0\sim1$. Below these lines the majoron is a sub-component of the total DM abundance unless new dynamical mechanism or a tuning of the initial condition is put into place. Notice that here we ignore further mechanisms of production that could arise from Higgs portal-type of couplings (see Ref.~\cite{Frigerio:2011in} for a discussion). 

Finally, the light purple line denotes the expected reach of the QUAX experiment~\cite{Barbieri:1985cp,Barbieri:2016vwg,Crescini:2020cvl},  which is directly sensitive to the coupling of the majoron to electrons.
The shown sensitivity is as assessed in Ref.~\cite{Chigusa:2020gfs}, under the assumption that the majoron is responsible for 100\% of the observed DM abundance.
 
Fig.~\ref{fig:majoron-plot} shows the remarkable reach in terms of the effective scale $F^A_{ee}$ and the RH mass neutrino mass scale $M$ of the Mu3e experiment, which, by searching for the $\mu\to e J$ decay, will be able to probe for the first time the uncharted territory beyond the star cooling and SN1987A bounds. In particular, Mu3e will be able to probe that part of the parameter space where the majoron can be a substantial component of DM.
 
Stringent constraints on low-energy seesaw models arise also from LFV decays, $\mu \to e \gamma, eee$, and $\mu \to e $ conversion, mediated by the heavy neutrinos at the scale $M$, if this is not much above the TeV scale. Following again Ref.~\cite{Ibarra:2011xn}, one  for instance finds for the branching ratio of $\mu\to e\gamma$
\begin{align}
{\rm BR} (\mu \to e \gamma) & = \frac{3 \alpha_{\rm em}}{32 \pi}
\frac{|(m_D m_D^\dagger)_{12} |^2}{M^4} \left| H \left(X \right) \right|^2 \, , 
\end{align}  
where $X = M^2/ M_W^2$ and 
\begin{align}
H(x) = \frac{x (1-6x+3x^2+2x^3 - 6 x^2 \log x)}{(1-x)^4} = \begin{cases} x & x \ll 1\\ 2 + (11- 6 \log x)/x & x \gg 1 \end{cases} \, .
\end{align}
This gives, in the $M \gg M_W$ limit,
\begin{align}
{\rm BR} (\mu \to e \gamma) & = \frac{2.4 \times 10^{-13}}{\lambda^2}  \left( \frac{ 10^3 \GeV}{  M} \right)^2 \left( \frac{ {\rm BR} (\mu \to e a)}{5.8 \times 10^{-5}} \right) \, .
\end{align}
This should be compared to the current 90\% CL  limit set by the MEG experiment, ${\rm BR} (\mu \to e \gamma) <4.2\times 10^{-13}$ \cite{TheMEG:2016wtm}.
The LFV  decays to majorons therefore typically provide stronger bounds on the seesaw scale $M$ than the LFV  processes, $\mu\to e\gamma, \mu \to 3 e,$ and $\mu N\to eN$, unless either $M\approx f_N \approx 1$ TeV, or $\lambda \ll 1$, such that the mass scale of the RH neutrinos is much lower than the $L$-breaking scale $M\ll f_N$.
Apart from these limits, $\mu\to e J$ tends to provide the dominant constraint, as a consequence of different scalings with the heavy scale (the rate of $\mu\to e J$ is $\propto M^{-2}$, while the rates of $\mu\to e\gamma, \mu \to 3 e,$ and $\mu N\to eN$, are $\propto M^{-4}$). See also the detailed discussion in Ref.~\cite{Heeck:2019guh}. 

Finally, we comment on the chiral structure of the LFV majoron couplings which in our minimal implementation are entirely of $V-A$, given that LFV is mediated by the $W$-loops. This will not be the case in a two Higgs doublet model, where the majoron LFV couplings can be induced by loops of the charged Higgs. The $V+A$ LFV couplings are then going to be suppressed by $\sim y_H^2 v^2/m_{H^+}^2$, where $m_{H^+}$ is the mass of the charged Higgs and $y_H$ its coupling to leptons. Given the strong bounds on $H^+$ coming from indirect searches in $b\to s\gamma$~\cite{Katz:2014mba} and direct collider searches~\cite{Craig:2015jba} we expect the right-handed LFV couplings of the majoron to be generically suppressed with respect to the left-handed ones. It would be  interesting to explore further the robustness of this statement in models where a light charged Higgs is still unconstrained~\cite{Nierste:2019fbx}.

\section{Conclusions}\label{sec:conclusions}
Generically, axion like particles (ALPs) can have flavor violating couplings to the SM fermions. In this manuscript we explored the phenomenological consequences of such couplings; 
in the first part we took a model-independent  perspective and summarized  present constraints and future sensitivities on generic lepton flavor violating couplings of light ALPs, see Fig.~\ref{fig:money}. 
 Here an important feature is our proposal for a new experimental set-up at MEG II, \emph{MEGII-fwd}, 
which consists of a forward calorimeter to be installed in front of the MEG II beamline, see Sec.~\ref{sec:futuremutoe} for details. The major benefit of such an experimental set-up is that the irreducible SM background from $\mu^+\to e^+ \nu\bar\nu$ is reduced at the highest positron momentum in the forward region.  Since the SM decay amplitude is controlled by left-handed couplings, it vanishes for an exactly forward positron of $p_{e^+}=m_\mu/2$ if produced from a muon that is completely negatively polarized. {MEGII-fwd} can be used to search for an effectively massles ALP produced in $\mu^+\to e^+ a$ with some amount of right-handed LFV coupling to the SM leptons. The signal will appear as a sharp  line in positron energy distribution, in contrast to the smoothly falling SM background. The final reach of {MEGII-fwd} depends on how well depolarization effects can be controlled, on the positron momentum resolution, and on whether or not magnetic focusing is applied in order to increase the positron luminosity in the forward direction. Assuming realistic estimates for these parameters, we expect that a two week run at MEG II in MEGII-fwd configuration will allow to explore new ALP parameter space well before Mu3e and beyond the current astrophysical limits from star cooling. 

Exploring this new region of parameter space will provide new insights on the couplings of an ALP DM produced non-thermally in the early Universe as discussed in Sec.~\ref{sec:ALP:DM}. A possible signal in an LFV experiment could be cross-correlated with the future experimental campaigns of axion haloscopes and future intensity mapping searches. This would give us more information about the light ALP mass beyond the expected resolution of MEGII-fwd.

In the second part of this paper we discussed several UV models where ALPs with flavor-violating couplings to leptons arise when addressing various shortcomings of the SM, such of the Strong CP Problem or the Flavor Puzzle, see Sec.~\ref{sec:models}. These models explicitly demonstrate that the LFV couplings, which will be tested by future laboratory experiments, are correlated to the flavor-diagonal lepton couplings, which are strongly constrained by star cooling. 
Despite these constraints we find encouraging prospects for the expected sensitivity of MEGII-fwd to test these theoretically well-motivated models. We believe that this warrants a more systematic exploration of the feasibility of MEGII-fwd, including proper detector simulations. In particular, it would be interesting to investigate in more detail the required rearrangement of the MEGII magnetic field and its interplay with the achievable momentum resolution.  We hope that this proposal will be explored further by the experimental collaboration. The MEGII-fwd proposal should also be carefully compared with other possibilities of improving the MEGII reach on LFV decays, for example the possibility of a dedicated trigger for $\mu^+\to e^+ a\gamma$ decays as discussed in Sec.~\ref{sec:futuremuegammaa}. We hope to return  to these issues in the future. 

Beyond the new proposed experimental ideas, the paper includes several novel theoretical results. First of all, in Sec.~\ref{sec:astro} we derived the astrophysical bounds from stellar cooling on leptonic couplings of an ALP with arbitrary mass. While the bounds on couplings of massless ALPs to electrons were readily available in the literature, this was not the case, to the best of our knowledge, for massive ALPs coupling to $ee$, as well as for the couplings to $\mu e$  and $\mu \mu$ (recently, Ref.~\cite{Bollig:2020xdr} treated the case of massless ALP coupling to muons). 
Second, we showed how the chiral structure of the LFV couplings emerges in various explicit (and mostly novel) models, in which the presence of a light ALP is motivated by addressing the strong CP Problem (the LFV QCD Axion and the LFV Axiflavon), the hierarchical structure of SM fermion masses (the LFV Axiflavon and the Leptonic Familon) or the origin of neutrino masses (the Majoron). 
For the LFV Axiflavon and the majoron the ALP couples mainly to left-handed leptons, and thus improvements in the reach on $\mu^+\to e^+ \gamma a$ at MEGII (or ultimately the inclusive on-line bump hunt proposed at Mu3e) will be needed in order to test these scenarios. On the other hand, for the case of LFV QCD axion and the leptonic familon the right-handed couplings are sufficiently large to allow MEGII-fwd to be the first experiment to discover the flavor violating ALP.

To conclude, we hope that this work may boost a renovated interest in the experimental possibilities at near future LFV experiments. The unprecedented luminosity of these facilities has been conceived to probe extremely rare lepton decays mediated by heavy new physics. The same data could also be used to probe decays with an (invisible) light new particle in the final state.
 We believe that we merely started to scratch the surface of the possible experimental improvements and theoretical motivations to search for light new physics in rare lepton flavor violating decays.

\paragraph*{Acknowledgements.} 
We thank Ryan Bayes, Marco Francesconi, Art Olin, Angela Papa, Ann-Kathrine Perrevoort, Francesco Renga, Andre Sch\"oning and Giovanni Signorelli for many useful discussions and precious explanations on the different experimental setups. We thank Simon Knapen and Gustavo Marques-Tavares for discussions. We especially thank Angela Papa and Giovanni Signorelli for extensive feedback on our work and Matthias Linster for providing the numerical fit used in Section \ref{sec:axiflavon}. 
We thank Galileo Galilei Institute for kind hospitality and for providing the collaborative atmosphere that initiated this work. 
The work of RZ  is supported by project C3b of the DFG-funded Collaborative Research Center TRR 257, ``Particle Physics
Phenomenology after the Higgs Discovery". JZ acknowledges support in part by the DOE grant de-sc0011784.

\bibliographystyle{JHEP}

\bibliography{FV_lepton}

\end{document}